\documentclass[aps,twocolumn]{revtex4-1}

\usepackage[margin=1.5cm,footskip=1cm]{geometry}
\usepackage{amsmath,amssymb,hyperref,graphicx,float,xcolor,enumitem}
\usepackage[english]{babel}
\usepackage{dsfont}

\graphicspath{{./figures/}}

\newcommand{\beq}{\begin{equation}}
\newcommand{\eeq}{\end{equation}}
\renewcommand{\vec}{\mathbf}
\newcommand{\erf}{\operatorname{erf}}
\newcommand{\Tr}{\operatorname{Tr}}
\newcommand{\Corr}{\operatorname{Corr}}
\newcommand{\Oh}{{\cal O}} 

\newcommand{\EE}{\operatorname{E}} 
\newcommand{\Std}{\operatorname{Std}} 
\newcommand{\Var}{\operatorname{Var}}
\newcommand{\FF}{{\cal F}}
\newcommand{\GG}{{\cal G}}
\renewcommand{\AA}{G}
\newcommand{\BB}{F}
\newcommand{\RR}{R} 
\newcommand{\rr}{{\cal R}} 
\newcommand{\VV}{V} 
\newcommand{\DD}{D} 
\newcommand{\LL}{L} 
\renewcommand{\SS}{S} 
\renewcommand{\ss}{s} 
\newcommand{\pinv}{p_{\it inv}}
\renewcommand{\aa}{I} 
\newcommand{\bb}{b} 
\newcommand{\TT}{T} 
\newcommand{\KK}{K} 
\newcommand{\UU}{U} 
\newcommand{\JJ}{J} 
\newcommand{\Smax}{\hat\SS} 
\newcommand{\Lmax}{\hat\LL} 
\newcommand{\Ymax}{\hat Y}
\newcommand{\Upsmax}{\hat\Upsilon}
\newcommand{\CF}{C} 
\newcommand{\PM}{W} 
\newcommand{\dd}{d} 
\newcommand{\tsurv}{{\cal T}}
\newcommand{\Comm}{{\cal C}} 
\newcommand{\Ext}{{\cal E}} 
\newcommand{\Lyap}{\Lambda} 
\newcommand{\tinv}{\mathbb{T}} 
\newcommand{\dens}{\Pi} 

\newcommand{\Ktot}{K_{\rm tot}}
\newcommand{\sigzeta}{\sigma_\zeta} 
\newcommand{\assembsig}{\sigma_\zeta} 
\newcommand{\Wtilde}{\tilde W}
\newcommand{\sdist}{p}
\newcommand{\NN}{{\cal N}}
\newcommand{\YY}{Y} 
\newcommand{\locsusc}{\chi} 
\newcommand{\globsusc}{X} 
\newcommand{\loceta}{\eta} 
\newcommand{\globeta}{H} 

\begin{document}
\title{Continual Evolution in Nonreciprocal Ecological Models }

\author{Aditya Mahadevan}
\affiliation{Department of Physics, Stanford University, Stanford, CA 94305, USA}

\author{Daniel S.~Fisher}
\affiliation{Department of Applied Physics, Stanford University, Stanford, CA 94305, USA}

\date{\today}

\begin{abstract}
Feedbacks between evolution and ecology are ubiquitous, with ecological interactions determining which mutants are successful, and these mutants in turn modifying community structure. We study the evolutionary dynamics of several ecological models with overlapping niches, including consumer resource and Lotka-Volterra models. Evolution is assumed slow, with ecological dynamics producing a stable community (with some strains going permanently extinct) between successive introductions of an invader or mutant. When new strains are slowly added to the community, either from an external pool or by large-effect mutations, the ecosystem converges, after an initial evolutionary transient, to a diverse eco-evolutionary steady state. In this ``Red Queen'' phase of continual evolution, the biodiversity continues to turn over without the invasion probability of new variants getting any smaller. For resource-mediated interactions, the Red Queen phase obtains for any amount of asymmetry in the interactions between strains, and is robust to ``general fitness" differences in the intrinsic growth rates of strains. Via a dynamical mean field theory framework valid for high-dimensional phenotype space, we analytically characterize the Red Queen eco-evolutionary steady state in a particular limit of model parameters. Scaling arguments enable a more general understanding of the steady state and evolutionary transients toward it. This work therefore establishes simple models of continual evolution in an ecological context {\it without} host-pathogen arms races, and points to the generality of Red Queen evolution. However, we also find other eco-evolutionary phases in simple models: For generalized Lotka-Volterra models with weakly asymmetric interactions an ``oligarch'' phase emerges in which the evolutionary dynamics continually slow down and a substantial fraction of the community's abundance condenses into a handful of slowly turning-over strains. \end{abstract}

\maketitle
\tableofcontents

\section{Introduction}

Ecological and evolutionary processes are inextricably linked. In diverse microbial communities with large population sizes, there is constant interplay between ecological interactions and new genetic variants. Evolution continually creates new variants that interact with the community differently from their parent strains, and these ecological interactions determine the community's properties and the fate of the new mutants, thereby influencing future evolution. As the increasing body of data from natural microbial communities shows, microbes live in close proximity with one another, with multiple strains and substrains of the same species coexisting in many natural environments~\cite{sunagawa2015structure,human2012framework,fierer2017embracing}. Understanding how ecological interaction webs influence evolution---and vice versa---is crucial for developing a richer picture of microbial evolution in both experimental and natural settings. A major puzzle is the ubiquity of closely related microbial strains coexisting, as evinced by the broad distribution of genomic divergences found within bacterial species apparently occupying the same ecological niche~\cite{kashtan2014single,rosen2015fine}. How does such diversity evolve and persist without clear ecological differences stabilizing the coexistence of strains? Theory is needed to develop distinguishable scenarios for the evolution and maintenance of fine-scale biodiversity in large microbial populations.

Even in controlled laboratory experiments with initially-clonal microbial populations, the development of ecology is inevitable. Although {\it apparent} selection pressures in the laboratory can be held constant, evolution in ``simple'' environments still generically results in the diversification and coexistence of different strains~\cite{rosenzweig1994microbial,friesen2004experimental,maharjan2006clonal}. 
Long after increases in fitness (as measured by competition with the ancestor) have slowed and populations seem to be well adapted to laboratory conditions, it has been observed, e.g. in the {\it E. coli} experiments of Richard Lenski, that evolution continues without further decrease in the rate of successful mutations~\cite{good2017dynamics}. Furthermore, diversification and coexistence of lineages is observed even in such experiments {\it designed} to be simple enough to eliminate ecological effects~\cite{lenski2017experimental,good2017dynamics}. Remarkably, even in laboratory experiments on bacterial populations without an external nutrient supply---which might be expected to exhibit severe resource limitation and diversity collapse---diversification and phenotypic turnover is found to continue over a timescale of years~\cite{finkel1999evolution}. 

With interacting bacteria and phage populations, there is clear evidence---both experimental and observational---for the continual turnover and diversification of microbial lineages~\cite{ignacio2020long,marston2012rapid,buckling2002antagonistic}. The ``Red Queen hypothesis''~\cite{van1973new} is often invoked to describe such arms-race evolutionary dynamics of hosts and pathogens. In the current work we use ``Red Queen dynamics'' more generally to describe continual turnover of biodiversity in a community without successful invasions of new mutants becoming rarer, and we look for such dynamics in other ecological contexts. 

Inspired by observations of microbial populations, we aim to understand what generic scenarios can emerge in the long term dynamics of evolution of ecological communities. We are particularly interested in robust behaviors that obtain across a range of parameters and models---such ``phases'' of eco-evolutionary dynamics present a set of qualitative possibilities that one can hope to compare with experiment or observation. 
The phases that we study emerge in the limit of large biological complexity: {\it high-dimensional phenotype and interaction spaces} upon which evolutionary and ecological processes play out. Looking for phases that are robust in the limit of high phenotypic dimensionality $\DD$---roughly the number of phenotypic properties of organisms that affect the ecological dynamics---increases the likelihood that the conclusions we draw are somewhat universal and relevant to natural systems, which are more complex---in multiple ways---than the simple models we study.

In order to understand what long term behaviors emerge in high-dimensional eco-evolutionary models, it is useful to first think separately about the dynamics of evolution and ecology, initially in a well-mixed population with {\it no spatial structure}. 
A further simplification is natural for large microbial populations: that the dynamics are essentially deterministic, with demographic fluctuations negligible except close to extinctions.

{\it Evolution without ecology:}
In a static environment without ecology, gradual evolution of a large almost-clonal population can be caricatured as gradient ascent on a complex fitness landscape. The fitness landscape is a map from some high-dimensional organismal phenotype to the ``fitness'': a misleading but canonical term for a measure of reproductive success that is surely not a single quantity except in the simplest non-ecological contexts. The emergence and fixation of beneficial mutations drives the population to high-fitness regions of phenotype space and evolution slows down as the population approaches one of the (typically) many local maxima of the fitness landscape~\cite{de2014empirical}. However in high dimensions, this simple intuition is only part of the story: the phenotype will generally spend a long time wandering around saddles before ``committing" to a particular maximum, which is likely to be only marginally stable~\cite{fisher2021inevitability}. 

 In large populations with an abundant supply of mutations, the evolutionary dynamics of the population of genotypes is complex, with simultaneous exploration of multiple phenotypic directions: this regime has not been much studied theoretically. However if there is no ecology except resource limitation capping the total population size, eventually the supply of beneficial mutations will become so limited that the population will again become almost clonal. By contrast, in an extrinsically changing environment, the fitness landscape depends on time, and is better caricatured as a ``seascape.'' Not surprisingly, a continually changing environment can lead to evolutionary dynamics which proceed indefinitely without slowing down~\cite{agarwala2019adaptive} and produce continually turning-over diversity: again, such phenomena have not been much studied for large populations in high phenotypic dimension.

{\it Ecology without evolution:}
Just as evolutionary dynamics of large populations can display rich behavior even without ecology, ecological dynamics with no evolution can exhibit various behaviors, with globally stable, multistable, and chaotic states all possible in simple models~\cite{bunin2017ecological,opper1992phase,blumenthal2024phase}. The most-studied contexts are ecological dynamics of {\it top-down assembled} communities, which consist of a large number of species brought together with their abundances obeying some prescribed ecological dynamics. The simplest ecological phase that obtains at long times is a state in which species abundances reach a globally attracting stable fixed point with some number of species (the stable community) having nonzero abundance and the rest going extinct, unable to reinvade even if reintroduced. As the diversity of this assembled community increases, the stable phase can yield to a more complicated behavior exhibiting multiple attractors and transient chaos, which remains only partially understood~\cite{bunin2017ecological,blumenthal2024phase,ros2023generalized}. 
Chaotic fluctuations in the high-diversity unstable regime lead to many extinctions and the system settles down to one of many possible stable communities or limit cycles, but these fixed points are generically unstable to invasion by some of the extinct strains. A population of dormant spores that germinate and resuscitate extinct populations can sustain the chaos for times exponentially long in the lifetimes of the spores, but as a matter of principle such dynamics do not constitute a true ecological phase~\cite{pearce2020stabilization}.

{\it Interplay of ecology and evolution:} To bring evolution and ecology together, we focus here on the slowly evolving limit in which the ecological dynamics are much faster than the evolutionary dynamics and spatial mixing is fast enough to be ignored. Motivated especially by the puzzle of fine-scale diversity, we are primarily interested in multiple strains of a single species with overlapping niches. This regime is the hardest in which to find sustained diversity and continual evolution because it does not allow short-term transient coexistence driven by continual invasions of new mutants, which was studied in refs.~\cite{xue2017coevolution,huisman2001towards,yan2019phylodynamic}. 

We study models in which the ecological dynamics are simplest, converging to a (usually) globally attracting fixed point with some strains permanently extinct. Under slow evolution, do the resulting eco-evolutionary dynamics remain diverse and continually turn over in a ``Red Queen'' steady state, or does evolution get progressively slower and slower? Do generalist mutants take over and cause the diversity to crash? Are there multiple different phases of eco-evolutionary dynamics for different parameter regimes?

It has been shown that Red Queen dynamics emerge generically in models of an almost-clonal population evolving in a high-dimensional fitness landscape with a small amount of environmental feedback~\cite{fisher2021inevitability}. In these models, the fitness landscape---now better caricatured as a ``snowscape''---changes as the evolving population affects its environment. For even a small amount of such feedback, the population wanders through phenotype space without settling down to a stable fixed point, due to selective forces that arise from the gradient of its continually-modified fitness landscape. Universality emerges in these models in the limit of large phenotype dimension, with different microscopic models displaying similar phase diagrams, including a phase of sustained evolutionary chaos~\cite{fisher2021inevitability}.

In the eco-evolutionary models we study here, ecology acts loosely as a manifestation of feedback between organism and environment, with the ``environment'' of any particular strain set by its interactions with the rest of the community, and thus subject to change every time a new mutant arises in the community. Can natural models of this process result in a Red Queen phase of evolution? If so, what sets the diversity of the community in this phase?

Some previous work has found that ecological communities, although capable of sustaining extensive diversity through standard mechanisms of resource competition~\cite{armstrong1980competitive} lose much of their diversity when they start to gradually evolve either by small mutations or by the introduction of independent species~\cite{shoresh2008evolution,schippers2001does}. These results were argued to deepen the puzzle of the extensive diversity observed in nature. However, the models in which such results have been reported are complicated enough that analytical understanding is difficult, and generality of the claims is not clear---indeed other work has suggested that phenotypic complexity {\it allows} the gradual evolution of high diversity~\cite{doebeli2010complexity}.

{\it Evolutionary dynamics}: The evolutionary process we study involves adding strains one by one to an ecological community, with extinctions permanent when they occur. Once extinct, a strain will never be able to re-emerge, even if conditions become favorable for its growth. Therefore the order in which strains are introduced influences the fate of the community. However, the {\it statistical} properties of the evolving community for large phenotypic dimension $\DD$ depend only on the pool from which invaders are drawn, rather than on the specific properties of the invaders.

Previous work has studied the process of adding strains into ecological models {\it without} permanent extinctions~\cite{tikhonov2018innovation,servan2021tractable,capitan2009statistical}, where re-invasion of previously-extinct strains is possible if conditions later become favorable for them. This process eliminates the important difference between communities assembled all at once (top-down) and gradually (bottom-up): the latter is our focus.

A large amount of work by theoretical ecologists and statistical physicists has focused on ecological models that are nongeneric because they admit a Lyapunov function which is always increasing under the eco-evolutionary dynamics~\cite{altieri2021properties,tikhonov2017collective,cui2020effect,biroli2018marginally}. The existence of such a Lyapunov function depends on the symmetry of ecological interactions~\cite{marsland2020minimum}, e.g. through perfect efficiency in the conversion of nutrients to biomass, which does not hold in biological systems. Therefore the generality of conclusions drawn from such models is unclear. In the current work we focus on generic behavior, drawing conclusions away from the special slice of parameter space with a Lyapunov function. Indeed we will show that, under evolution, the behavior of models with a Lyapunov function is qualitatively different from the general case.
Several other studies~\cite{bonachela2017eco,law1997evolution,nordbotten2016asymmetric} have noted the importance of {\it nonreciprocal} or {\it asymmetric} ecological interactions for reaching a state of continual evolution---we will provide further evidence for this conclusion in the high-diversity limit, which complements the more often-studied low-diversity case~\cite{bonachela2017eco,law1997evolution}

{\it Models}: We are interested in dynamics of closely related strains, where differences between strain interactions with one another are sums of many positive and negative effects. To caricature this biological complexity, we parameterize the strain-to-strain differences in phenotypic properties as {\it random variables} drawn from some distribution. This approximation of unknown complexity by randomness is a useful way to get a sense of general patterns of behavior that might apply across a range of ecological-evolutionary situations~\cite{may1972will}.

In the simplest models, strains interact with one another {\it directly}, with all pairwise interactions chosen from some specified ensemble. These models, which include generalized Lotka-Volterra interactions~\cite{bunin2017ecological}, suffer from a lack of strain phenotypes which, when present, dictate how each strain interacts with any other prospective strain. As a result, the effective phenotypic dimensionality in Lotka-Volterra models is infinite. Problematically, such models do not generally admit a straightforward way of drawing random mutants correlated with a parent strain in the community~\cite{mahadevan2023spatiotemporal}. Nonetheless we will study the Lotka-Volterra eco-evolutionary dynamics in the case of independent invaders added to the community. This simple limit is instructive for understanding evolution driven by mutant strains correlated with their parents. 

To explicitly study the effects of parent-mutant correlations we introduce structure into the ecological interactions, considering the case where they are mediated by a collection of chemicals (or resources) in the environment. Each strain has a phenotype which describes its effect on the resources and the resources' effect on its growth rate---these phenotypes consequently determine the inter-strain interactions. Small mutations correspond to having highly correlated parent and mutant phenotypes, with the amount of similarity between mutant and parent playing an important role.

The phenotype of each strain in such {\it consumer resource} models is given by its growth and consumption rates for each of the resources. The number of resources---or more generally, interaction-mediating chemicals---is the key quantity $\DD$. This number sets the {\it dimensionality} of the environment and of the interactions between consumers. It also sets a natural scale for the number of strains that can coexist: in the models we study here, the diversity is bounded above by $\DD$, and ``high-diversity" means that the number of coexisting strains is $\Oh(\DD)$. 
The limit of large $\DD$---surely biologically relevant~\cite{doebeli2010complexity}---will allow us to analyze statistical (hopefully universal) properties of the resulting eco-evolutionary dynamics. 

When mutational effects are sufficiently large, we find that key aspects of the evolutionary dynamics are captured by the dynamics of {\it bottom-up assembly} (with extinctions), wherein phenotypically {\it independent} invaders enter the community serially. The understanding from this simplified regime gives us qualitative insight into differences between a top-down-assembled and an evolved community, along with a framework for understanding the Red Queen phase that occurs for correlated mutants.

In order for the Red Queen phase to be robust, it must be stable to the presence of ``general fitness'' mutations which increase the growth rate of a strain {\it regardless} of the composition of the community. These general fitnesses appear naturally in models of resource competition, where they correspond to the growth rates of strains in isolation. Metabolic tradeoffs, which constrain the width of the general fitness distribution, are sometimes invoked to sustain high levels of diversity in ecological settings. These tradeoffs can be strict~\cite{posfai2017metabolic,caetano2021evolution} or---more realistically---allow some variation in the total metabolic budget~\cite{cui2020effect}. Without strict tradeoffs, can consumer resource and Lotka-Volterra models evolve towards a diverse state in which the distribution of extant general fitnesses is narrow? Or does evolution constantly push into the tail of the general fitness distribution and fail to reach a steady state? Do generalist species emerge and reduce diversity by outcompeting other strains?

Our main result is that there exists a robust high-diversity Red Queen phase of eco-evolutionary dynamics characterized by continual evolution and turnover of biodiversity, and that this phase occurs across a range of parameters and models.
Importantly, the Red Queen phase persists in the presence of general fitness differences, as long as their distribution decays fast enough. The tail of the general fitness distribution controls the probability of successful invasion and therefore the rate of turnover in the steady state, but otherwise does not limit the diversity.

\section{Consumer resource model}

\subsection{Ecological dynamics and stable communities}

We first study a standard model of resource competition with $\DD$ externally-supplied resources (or more generally, interaction-mediating chemicals), and an assembled population of $\SS$ different strains that consume these resources (see refs.~\cite{cui2020effect,marsland2020minimum,tikhonov2017collective,posfai2017metabolic}). The ecological dynamics of this consumer resource (CR) model are 
\begin{subequations}
\label{eq:CR_dyn}
\begin{align}
\frac{dn_i}{dt} &= n_i\left(\sum_{\beta=1}^\DD \GG_{i\beta} \rr_\beta-\dd\right) \\
\quad \frac{d\rr_\alpha}{dt} &= \KK_\alpha-\rr_\alpha \sum_{j=1}^\SS \FF_{j\alpha}n_j-\omega \rr_\alpha,
\end{align}
\end{subequations}
where the $\{n_i\}$ are the populations of strains, labeled by Roman indices $i\in\{1,2,\dots\SS\}$ and the $\{\rr_\alpha\}$ are resource abundances labeled by Greek indices $\alpha \in\{ 1,2,\dots\DD\}$. The resource supply vector $\vec \KK$ has components $\KK_\alpha$, and $\dd$ and $\omega$ are respectively the death rate of the consumers and the decay (or dilution) rate of the resources. Without substantial loss of generality, we take all the $\KK_\alpha$ to be equal and given by a constant $\KK$, and we set $\omega=0$ to reduce the number of parameters without changing the qualitative behavior. The $\SS\times\DD$ matrices $\GG$ and $\FF$ respectively parameterize the {\it growth} and {\it feeding} rates of each consumer on each resource: the phenotype of strain $i$ is the corresponding row of $\GG$ and $\FF$ taken together, so that the dimensionality of phenotype space is $2\DD$. The elements of $\GG$ and $\FF$ consist of some positive mean---an average growth or feeding rate---in addition to a heterogeneous part that varies between strains and resources. Ref.~\cite{cui2020effect} studied the ecological dynamics of a {\it symmetric} version of this model with $\FF=\GG$. Although we expect $\GG$ and $\FF$ to be positively correlated, some amount of asymmetry---arising from, e.g., varying efficiencies in utilization of resources---is generic, and we will show that this asymmetry is important for the dynamics of interest to us.

For closely related strains, the differences between interactions will be sums of positive and negative effects: thus we treat $\GG_{j\alpha}$ and $\FF_{j\alpha}$ as random variables drawn from some distribution. With the interpretation of consumption and growth, one expects the elements of $\FF$ and $\GG$ to be positive, though a small fraction being negative (corresponding loosely to crossfeeding or inhibitory effects) does not change the behavior. For simplicity we draw the elements from gaussian distributions with means $\mu_g$, $\mu_f$ and variances $\sigma_g^2$, $\sigma_f^2$ respectively. The choice of a gaussian distribution is convenient as we will later take advantage of this structure to evolve strains by drawing from the ensemble of interactions conditional on inequality constraints. However, as we show in Section~\ref{sec:s_dist}, the upper tails of the distributions of the elements of $\GG$ and $\FF$ play an important role in the evolution.

We parameterize elementwise correlations between the gaussian entries of $\GG$ and $\FF$ by $\Corr[\GG_{j\beta},\FF_{j\alpha}] = \kappa\delta_{jk}\delta_{\alpha\beta}$, where $\Corr[A,B]$ is the Pearson correlation coefficient between $A$ and $B$, and the {\it symmetry parameter} $\kappa\in[0,1]$ parameterizes the similarity between growth and consumption rates. The ratios $\sigma_g/\mu_g$ and $\sigma_f/\mu_f$ determine what fraction of the entries of $\GG$ and $\FF$ are positive; we consider $\sigma_f=\mu_f/4$ and $\sigma_g=\mu_g/4$ so that most entries are positive. With the choice of $\omega=0$ there are only three important parameters other than $\SS$ and $\DD$: $\kappa$, $\sigma_g/\mu_g$ and $\sigma_f/\mu_f$, with $\KK$ setting a scale for the consumer abundances (Appendix~\ref{app:parameters}). Without loss of generality we will set $\mu_f = \mu_g =1$, $\dd=1$ and $\KK=1$.

The degree of correlation, $\kappa$, between the feeding and growth rates plays an important role: $\kappa=1$ presumes identical efficiencies of conversion between resources and biomass between strains~\cite{cui2020effect}. This symmetry guarantees the existence of a Lyapunov function (see Appendix~\ref{app:parameters}) that is always increasing along the ecological (and later evolutionary) dynamics of the CR model~\cite{marsland2020minimum}. However for generic $\kappa<1$ the dynamics do not have a Lyapunov function, which makes a big difference for the evolutionary behavior as we will see shortly.

One must supplement the deterministic dynamics of Equation~\ref{eq:CR_dyn} with a small abundance threshold below which strains go {\it extinct}: their abundances are then set to zero and {\it cannot recover}. In large populations the dynamics and extinctions are driven deterministically by growth rate differences, so we neglect demographic noise. 
Starting from positive initial conditions, the strain abundances in the CR model generally reach a fixed point with some fraction of strains going extinct and others persisting at nonzero abundance. We denote by $\Comm$ the {\it community of extant strains}: those with positive fixed point abundances. For small enough $\SS$---less than some multiple of $\DD$---the fixed point is generally stable and cannot be invaded by any of the extinct strains. For large $\DD$, in the parameter ranges of interest, the stable uninvadable community is highly likely to be unique~\cite{marsland2020minimum}. The breakdown of this uniqueness above a critical value of $\SS$ (discussed for a related model in ref.~\cite{blumenthal2024phase}), will not play a role in the evolving communities of interest here.

At the ecological fixed point of the CR dynamics, the abundances $\{n^*_i\}$ of the extant strains and the resources $\{\rr_\alpha^*\}$ are determined (with $\omega=0$) by
\begin{subequations}
\begin{align}
\rr^*_\alpha &= \frac{\KK_\alpha}{\sum_{j\in\Comm} \FF_{j\alpha} n^*_j}\\
0&=\sum_{\alpha=1}^\DD \GG_{i\alpha}\rr^*_\alpha-\dd\ \text{ for }\ i\in\Comm\\
n^*_j &=0 \text{ for } j \notin\Comm 
\end{align}
\end{subequations}
with the strains $j\notin \Comm$ extinct and unable to reinvade. Whether a strain can invade from low abundance is determined by the sign of its {\it invasion fitness} 
\beq
\xi_i = \sum_{\alpha=1}^\DD \GG_{i\alpha} \rr^*_\alpha-\dd.\eeq
If $\xi_i$ is negative, strain $i$ cannot invade.

The total population size, $N \equiv \sum_i n^*_i$, is of order $\mu_g \sum_\alpha\KK_\alpha/\mu_f$, which is roughly $\DD$ for our choice of parameters. (We have kept the dependence on $\vec \KK$ explicit because it carries a physical meaning and characterizes the environment. In our simulations all of its components are $1$: the effects of changing $\vec \KK$ are discussed briefly in Section~\ref{sec:discussion}.)

We define the {\it extant strain diversity}, $\LL = |\Comm|$, as the number of strains {\it left} in the community at the ecologically stable fixed point. In accordance with the competitive exclusion principle~\cite{armstrong1980competitive}, the maximum $\LL$ that can stably coexist at a fixed point is $\DD$, as seen from the insufficiency of the fixed point conditions otherwise. For natural choices---including ours---of the ensemble of interactions, the typical number of strains that can stably coexist at a fixed point scales linearly with $\DD$ with a coefficient less than $1$. This linear relationship between the diversity of the community and the number of resource-mediating chemicals is expected to be generic across a range of consumer resource models with heterogeneous interactions, with the statistics of the interactions affecting the prefactor in the scaling~\cite{cui2020effect}. In ecological language, the number of {\it potential niches} is $\DD$ but not all of these can be filled. 

{\it Measures of diversity}: In the ecological literature many different measures of diversity of communities have been introduced~\cite{daly2018ecological}. We will use $\LL$, the total number of species in the community, which is sometimes called the {\it richness}. Other popular measures are the Shannon diversity: the weighted average of $\log(N/n_i)$, and the {\it inverse Simpson index} ({\it inverse participation ratio} in physics terminology), given by $N^2/(\sum_i n_i^2)$.
If there is a broad distribution of abundances in a community, these measures can be quite different. However in our models, as we shall see, the distribution of abundances is not very broad so which measure we use does not affect our conclusions. For simplicity we refer to $\LL$ as the ``diversity.''

\subsection{Evolutionary processes}
\label{sec:CR_evo}

{\it Evolution by independent invaders:} 
We now focus on slow evolution of stable communities in the CR model, assuming a clean separation of timescales between evolution and ecology. If the resource dynamics are fast enough, they can be integrated out to yield equations for the strain dynamics which reach the same fixed point irrespective of the resource timescale---we can therefore assume fast resource dynamics to simplify our dynamical equations without affecting the evolutionary behavior. The ecological dynamics of the community run until a fixed point is reached. Then a single new strain is introduced, and the ecological dynamics again run until a fixed point is reached. The ecological time scale thus does not affect the evolutionary dynamics of primary interest. 

We first study the case of {\it independent invaders} into the community. Although this process is sometimes referred to as ``bottom-up assembly," it will give us insight into evolutionary dynamics with large-effect mutations. We start with a single strain of consumer and a collection of $\DD$ resources, and repeatedly add an unrelated consumer strain drawn from the gaussian phenotype ensemble. We define an {\it epoch} as the unit of time corresponding to a {\it successful} invasion followed by the relaxation of the ecological dynamics to a fixed point. Although the length of an epoch in ecological time varies between epochs, in our case of slow evolution the ecological timescale is not important.

With each new strain added, we find a stable uninvadable ecological fixed point of the community either by integrating the dynamical equations with a small extinction threshold, or by the more efficient iterative algorithm described in Appendix~\ref{app:numerics}. Once $\LL$ is sufficiently large, we expect that, with high probability, this fixed point is the {\it unique} stable uninvadable fixed point for the collection of extant strains in each epoch---this intuition is based on analogy with the assembled community in which the uninvadable fixed point is unique as long $\SS$ and $\LL$ are of similar size~\cite{bunin2017ecological}, as is the case in our evolving system where the size of the extant community {\it plus} the new invader is analogous to $\SS$.

{\it Invaders}: In order for a new strain, labelled $\aa$, to be able to invade, it must have positive invasion fitness $\xi_\aa$. Furthermore, if it successfully invades, its fixed point abundance $n_\aa^*$ will be approximately proportional to $\xi_\aa$, with the constant of proportionality determined by the properties of the community into which $\aa$ invades, as can be seen by a perturbative analysis for large $\LL$. The dot product between the growth vector $\boldsymbol\GG_\aa$ (the $\aa$th row of $\GG$) and the resource vector $\boldsymbol \rr^*$ of the community determines the invasion fitness of strain $\aa$. 
Therefore when we draw the phenotype of $\aa$---consisting of $\DD$-dimensional vectors $\boldsymbol\GG_\aa$ and $\boldsymbol\FF_\aa$---from their gaussian ensemble, we condition on $\xi_\aa>0$, which drastically speeds up numerics when the evolution pushes the extant distribution of interactions to have large mean, making invaders with $\xi_\aa>0$ very rare~\cite{vrins2018sampling}.

The resource availability vector $\boldsymbol\rr^*$ can be thought of as a modified version of the supply vector $\vec \KK$, where the availability of resource $\alpha$ is reduced from $\KK_\alpha$ by a factor of $\sum_j \FF_{j\alpha}n_j^*$. Therefore alignment between $\boldsymbol\GG_{\aa}$ and $\vec\KK$ results in a contribution to the invasion fitness of strain $\aa$ that is independent of the other strains in the community---this contribution to the invasion fitness is a form of {\it general fitness}, here defined as $\UU_i \equiv\sum_\alpha \GG_{i\alpha}\KK_\alpha$, which captures the average performance of strain $i$ against all of the $\DD$ resources at the rate that they are supplied by the environment, and thus scales with the magnitude of the resource supply vector.
 Therefore $\xi_\aa$ for an invader has a community-independent contribution proportional to the general fitness $\UU_\aa$ as well as a community-dependent contribution from the ecological interactions. We refer to the community-dependent part of the invasion fitness as the {\it drive}, which will be defined more precisely for the models we introduce in Section~\ref{sec:simple_models}.

It is useful for analysis to extend the definition of the invasion fitness to strains already in the community: for $i\in\Comm$ we define $\xi_i$ as the growth rate strain $i$ would have if it were removed from the community and then reintroduced at low abundance. This quantity arises naturally in the mean field description of ecological models (Section~\ref{sec:analysis} and Appendix~\ref{app:cavity}) and we will sometimes refer to it as the {\it bias}, consistent with previous work~\cite{pearce2020stabilization}. The bias is therefore an invasion fitness appropriately defined for strains both inside and outside the community. Differences between biases of different strains come from both the general fitnesses and the drives, and such differences are responsible for the differential growth and extinction of strains.

As we show in Appendix~\ref{app:consistency}, in a randomly assembled community the {\it selective differences}---related to differences between general fitness terms $\UU_i$---are {\it first order} in the phenotypic differences, proportional to $\frac{1}{\Ktot}\sum_\alpha\GG_{i\alpha}\KK_\alpha-\langle\GG\rangle$, while the variations in the drives are second order, proportional to both $\frac{1}{\Ktot}\sum_\alpha\GG_{i\alpha}\KK_\alpha-\langle \GG \rangle$ and $\frac{1}{N}\sum_i\FF_{i\alpha}n^*_i-\langle \FF \rangle$, where we have defined $\Ktot\equiv\sum_\beta\KK_\beta$, and weighted averages $\langle\GG\rangle = \frac{1}{N\Ktot}\sum_{i,\alpha}\GG_{i\alpha}n^*_i\KK_\alpha$ and similarly for $\langle \FF\rangle$. Therefore the selective differences comprise the dominant contribution to the width of the bias distribution, larger by a factor of $\sqrt \DD$ than the contribution from the drives. For large $\DD$ the selective differences determine both the ecological and early-stage evolutionary dynamics and can, at least initially, strongly limit the diversity.
Previous work~\cite{cui2020effect} has forced the contributions to be comparable by scaling the mean and variance of $\GG$ and $\FF$ as $1/\DD$, while in fact it is their mean and {\it standard deviation} that should be the same order in a biologically-motivated model. We will later show that in a long-evolved community, the variations in the general fitnesses become much smaller, indeed comparable to the variations in the drives. However---importantly---this feature must be shown to emerge rather than assumed. 

\subsection{Eco-evolutionary dynamics}

\begin{figure}
\centering
\includegraphics[scale=.6]{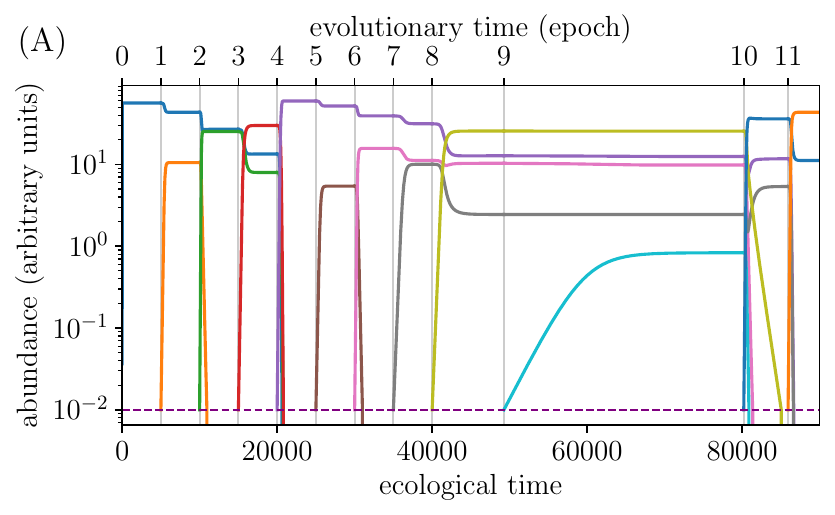}
\includegraphics[scale=.6]{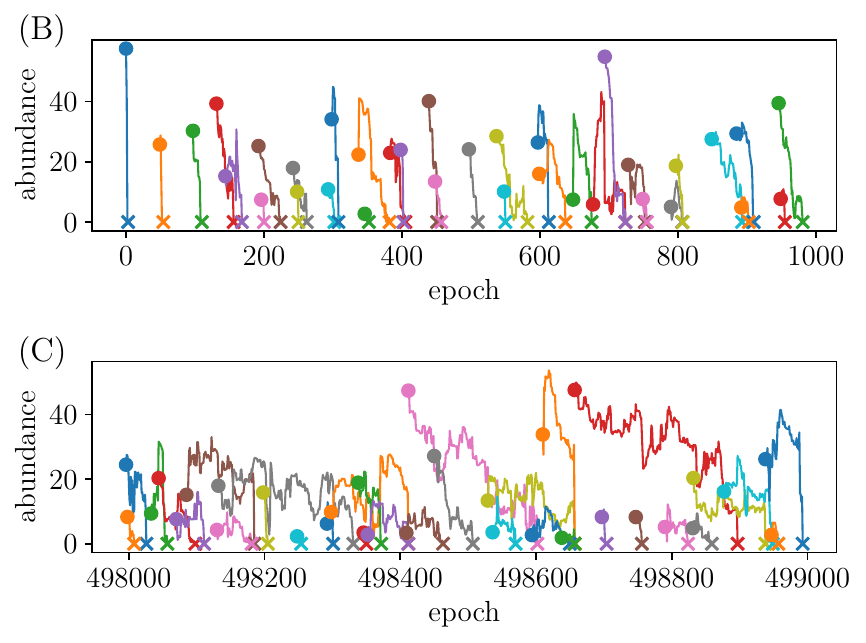}
\caption{Dynamics of strain abundances on ecological and evolutionary timescales in the consumer resource (CR) model for $\kappa=0.4$ and $\DD=50$ resources. Starting with a single strain, ecological dynamics reach a fixed point between each new strain invasion which begins a new epoch (vertical lines). Ecological time on lower axis; evolutionary time, in epochs, on upper axis. (B)--(C) Selected strain abundance trajectories on evolutionary timescales. (B) first $1000$ epochs. (C) after a long period of evolution. Turnover of diversity is slower than during initial stages of evolution, but has converged to a steady average rate.}
\label{fig:CR_dynamics}
\end{figure}

\begin{figure}
\centering
\includegraphics[scale=.6]{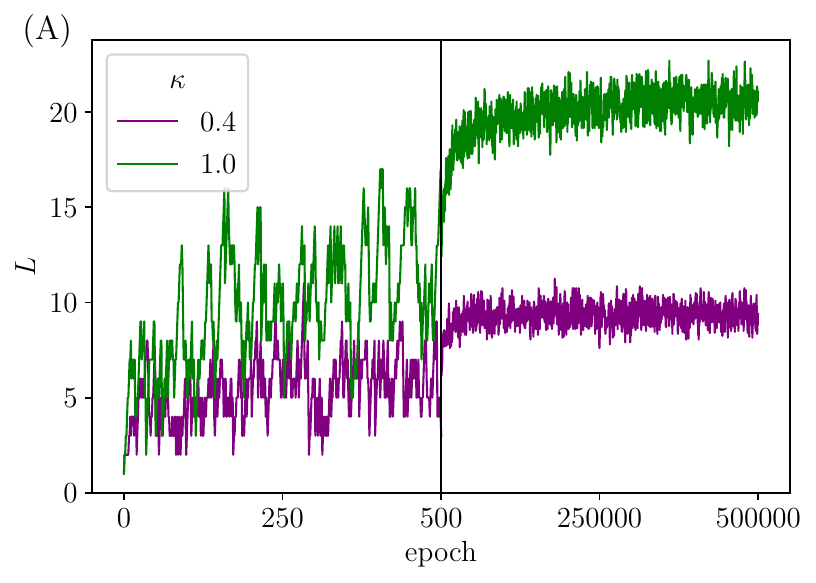}
\includegraphics[scale=.6]{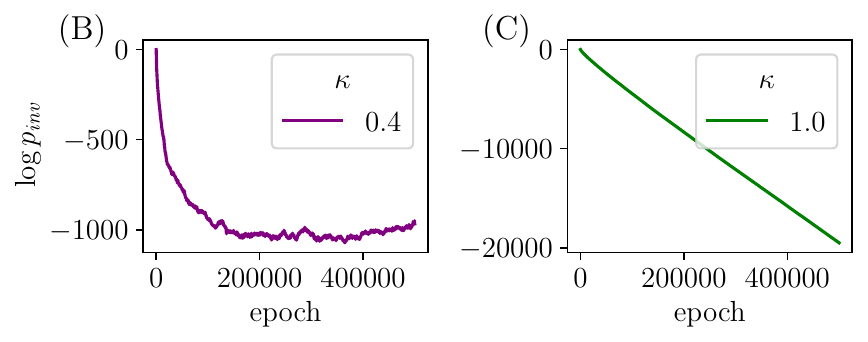}
\caption{The CR model reaches a steady state for symmetry parameter $\kappa=0.4$, but not for perfect symmetry, $\kappa=1$. (A) Community diversity $\LL$ as function of evolutionary time, starting from a single strain with $\DD=50$. First $500$ epochs on stretched scale; thereafter as moving average over windows of $1000$ epochs. $\LL$ increases, then fluctuates around steady state for $\kappa=0.4$, but creeps up for $\kappa=1$. (B)--(C) The invasion probability decreases initially, then reaches steady state for $\kappa=0.4$, but continues to decrease for $\kappa=1$. The absurdly small $\pinv$ is an artifact of the gaussian distribution of phenotypes (see Section~\ref{sec:s_dist}).}
\label{fig:CR_L}
\end{figure}

\begin{figure}
\centering
\includegraphics[scale=.6]{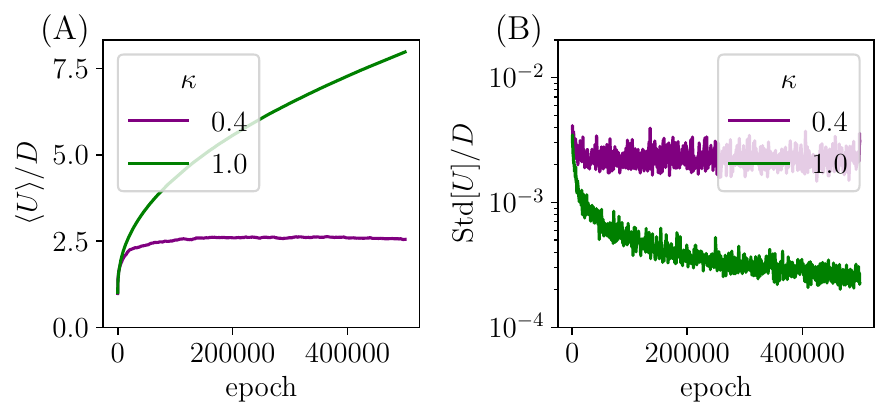}
\caption{(A)--(B) Mean and standard deviation (weighted by abundances) of extant distribution of generalized fitness, $\UU_i$, over evolutionary time for $\kappa=0.4 $ and $1$ with $\DD=50$. Distribution stabilizes for $\kappa=0.4$. For $\kappa=1$, mean continues increasing with standard deviation decreasing.}
\label{fig:CR_U}
\end{figure}

To explore the eco-evolutionary dynamics, we choose $\DD=50$ resources with $\kappa=0.4$ well away from the special value $\kappa=1$ for which there is a Lyapunov function: we investigate that special case later. Starting from a single strain ($\LL=\SS=1$), Figure~\ref{fig:CR_dynamics}A shows the ecological dynamics of the strain abundances in Equation~\ref{eq:CR_dyn} (with resources integrated out), punctuated by well-spaced evolutionary events when an invader comes in at the start of each new epoch. When the system has reached an ecological fixed point, a new strain is introduced, drawn randomly from the ensemble of independent strains conditioned on being able to invade (the conditioning speeds up numerics, since new strains which cannot invade will not affect the community). The new strain joins the community and the ecological dynamics find a new stable fixed point at which the abundances of the other strains are perturbed, with some strains possibly going extinct. Thereafter another strain is introduced. 
 
 From now on, we focus on the evolutionary dynamics and ignore the ecological transients at the beginning of each epoch that drive some strains extinct and others to nonzero abundances. 
The changes caused by invading strains can be visualized over longer times by coarse-graining the time resolution from ecological to evolutionary epochs and tracking the fixed point abundances of strains (calculated with the algorithm of Appendix~\ref{app:numerics}). Such strain abundance trajectories are shown near the beginning of the evolution in Figure~\ref{fig:CR_dynamics}B and, after much evolution, in Figure~\ref{fig:CR_dynamics}C. We show only a few of the strains for ease of visualization.

Primary quantities of interest are the extant strain diversity $\LL$, the population mean general fitness $\langle \UU\rangle \equiv \frac{1}{N}\sum_{i} n_i \UU_i$ (with angular brackets denoting weighted averages over the community), and the probability that a randomly chosen strain can invade, denoted $\pinv$. In addition to the community level properties, we are interested in the statistical properties of strains including the distributions and temporal correlations of their abundances $n^*_i$, their general fitnesses $\UU_i$ and their evolutionary lifetimes which we call $\tsurv_i$ (see Section~\ref{sec:dyn_cavity} for more details).

The evolution of the diversity for $\kappa=0.4$ is shown in purple in Figure~\ref{fig:CR_L}A. Starting from a single strain, $\LL$ initially grows fast and then slows down, appearing to saturate at $\LL\approx10$ at long times, for which we plot a moving average. As shown in Appendix~\ref{app:L_fluct} for a related model that we will introduce shortly, the fluctuations in $\LL$ are $\Oh(\sqrt\LL)$ resulting from rough independence of strains along with robustness of the community to perturbations (Section~\ref{sec:fragility}). 
For $\kappa=0.4$, it thus appears that the evolution converges to a Red Queen steady state. Note that although for these parameters $\LL$ remains rather small, we later show that it is expected to be of order $\kappa^2 \DD$, here only $8$, in the large $\DD$ limit of primary interest. The modest $\kappa^2\DD$ used here was needed for convergence to steady state in reasonable time.

Further evidence for the Red Queen state for $\kappa=0.4$ is shown in Figures~\ref{fig:CR_L}B, where the invasion probability of a new strain, $\pinv$, after initially decreasing rapidly, saturates at long times (albeit to an extremely small value). This behavior can be understood through the general fitnesses. Figures~\ref{fig:CR_U}A and B show in purple the mean and width, respectively, of the extant $\UU_i$, distribution. Initially the mean increases, but then saturates at a value of order $\DD$ with the strains all having similar $\UU_i$. The evolution therefore pushes far out into the tail of the general fitness distribution, making it very hard for other strains to invade. We shall show in Section~\ref{sec:s_dist} that the width of the distribution of extant general fitnesses, scaled by the mean consumption rate of the resources---which is what enters in the growth rate of the strains (Appendix~\ref{app:consistency})---shrinks during evolution until it is similar to the width of the drive distribution. At this point the intrinsic (general fitness) and community-derived (drive) parts of the bias contribute similarly to bias differences between strains. Once this occurs, the distribution of extant $\UU_i$ stops pushing higher and we reach a Red Queen evolutionary steady state with a roughly steady rate of turnover. 

In an evolutionary steady state (when it occurs) we denote {\it averages over time} by overbars. Because of the turnover of strains, in steady state temporal averages should be equal to averages over the ensemble of strains.

{\it Turnover of strains}: In steady state, one new strain invades in each epoch and on average, one extant strain goes extinct. This means that a strain remains in the community for an average of $\overline{\LL}$ epochs and the rate of turnover of strains is $1/\overline{\LL}$ per epoch. 
This is seen in Figure~\ref{fig:CR_dynamics} where at early times when $\LL$ is small, strains do not survive long, while in the long-time steady state with larger $\LL\approx \overline{\LL}$ they survive longer on average even though there are still strains that go quickly extinct, as we discuss in Section~\ref{sec:dyn_cavity}. 

The continual turnover of strains together with the plateauing of $\pinv$ mean that in the Red Queen phase, strains that successfully invade and later go extinct will have a substantial chance of being able to reinvade once the community has sufficiently turned over and lost memory of its state at the extinction time of the strain. Each strain keeps ``memory" of its general fitness which, because it was able to invade earlier, must be similar to that of the other strains in the steady state. 

While these simulations provide strong evidence for a Red Queen steady state for $\kappa=0.4$, quantitatively they are problematic. It is apparent from Figure~\ref{fig:CR_L}B that for the ensemble of strains we have used, the invasion probability of a new invader is unrealistically small in the steady state ($\pinv\approx 10^{-400}$). In Section~\ref{sec:lin_model} we will introduce a family of models in which $\pinv$ at steady state can be much larger, and in Section~\ref{sec:s_dist} we analyze how the choice of the $\GG$ and $\FF$ ensembles affects the invasion probability in steady state, and show that the very small $\pinv$ observed above is an artifact of the gaussian phenotype ensemble we have used for numerical convenience.

{\it Totally symmetric ensemble:}
We now show that the perfectly-symmetric case, $\kappa=1$, which possesses a Lyapunov function $\Lyap$ (see Appendix~\ref{app:parameters}), behaves very differently from the $\kappa=0.4$ ensemble for which there is no Lyapunov function. 
From the green trajectory of the diversity in Figure~\ref{fig:CR_L}A, it is difficult, based only on the dynamics of $\LL$, to know whether a steady state obtains by the end of the simulation. Invading strains must increase $\Lyap$ and there is no absolute upper bound of $\Lyap$ in the ensemble of all possible invaders. Therefore even if $\LL$ saturates, there cannot be a steady state. The larger $\Lyap$ gets, the more unlikely that adding new strains can increase it further---thus the probability of invasion, $\pinv$ must steadily decrease, as indeed seen in Figure~\ref{fig:CR_L}C. Concomitantly, the mean general fitness pushes further and further into the tail of its distribution as shown by the green trajectory in Figure~\ref{fig:CR_U}A, while the standard deviation of the $\UU_i$ decreases (green curve Figure~\ref{fig:CR_U}B) due to the steepening of the distribution around $\langle \UU\rangle$ (see Section~\ref{sec:s_dist}). 

Because of the Lyapunov function, we conclude that there is no Red Queen steady state for $\kappa=1$. It is interesting to note that, in spite of the gradual increase of $\Lyap$, $\LL$ stays far away from its upper bound of $\DD=50$. Indeed, $\LL$ stays below its maximum value of $\DD/2$ for a randomly {\it assembled} community with $\kappa=1$~\cite{cui2020effect}. However it is not clear whether under evolution $\LL$ might eventually reach or exceed this bound.

{\it Evolution via mutations:} 
So far we have considered independent invaders in the CR model. In Appendix~\ref{app:CR_corr} we study the evolutionary dynamics when invaders are correlated with a parent strain chosen with probability proportional to its abundance in the community. Indexing the parent by $P$ and the mutant by $M$, we introduce a parameter $\rho\in[0,1)$: the elementwise Pearson correlation coefficient between $\boldsymbol\GG_P$ and $\boldsymbol\GG_M$ and between $\boldsymbol\FF_P$ and $\boldsymbol\FF_M$. Given a parent, we draw mutants from this ensemble (while also preserving correlation $\kappa$ between $\boldsymbol\GG_M$ and $\boldsymbol\FF_M$): therefore the correlation between $\UU_P$ and $\UU_M$ is also $\rho$. Our main finding is that for $\rho$ sufficiently far from $1$---i.e. mutations of substantial effect---the eco-evolutionary phenomenology is similar to the independent invader limit $\rho=0$.This finding allows us to base qualitative conclusions for true evolution on results from independent invaders.

\begin{table}
\centering
\begin{tabular}{c || l}
CR&Consumer resource model (Equation~\ref{eq:CR_dyn})\\\hline
LV & Lotka-Volterra (Equation~\ref{eq:LV_model})\\\hline
$\SS$ & number of strains in initial pool\\\hline
$\DD$ & phenotypic dimension; number of resources\\\hline
$\LL$ & number of extant strains at fixed point\\\hline
$n_i$ & abundance of strain $i$\\\hline
$\UU_i$ & general fitness in CR model; $\UU_i = \sum_\alpha \GG_{i\alpha}\KK_\alpha$ \\\hline 
$\xi_i$ & invasion fitness = bias\\\hline
$\nu_i$ & fractional abundance of strain $i$; $\sum_i\nu_i=1$\\\hline
$\zeta_i$ & drive = community-dependent component of bias\\\hline
$\sigzeta$ & width of drive distribution\\\hline
$\ss_i$ & general fitness; intrinsic component of bias\\\hline
$\sdist(\ss)$ & distribution from which $\{\ss_i\}$ are drawn\\\hline
$\Sigma$ & scale of $\ss$ distribution\\\hline
$\langle A\rangle$ & community average of $A$ (unless otherwise stated)\\\hline
$\hat A$ & value of $A$ in the saturated assembled community\\\hline
$\vec\KK$ & vector of resource supply rates\\\hline
$\GG$ & growth rate matrix\\\hline
$\FF$ & feeding rate matrix\\\hline
$\AA$ & matrix of growth rate differences; mean $0$\\\hline
$\BB$ & matrix of feeding rate differences; mean $0$\\\hline
$\Upsilon$ & community mean ``fitness"; $\Upsilon = \sum_i \nu_{i}(\ss_i+\sum_j \VV_{ij}\nu_{j})$\\\hline
$\VV$ & pairwise strain interaction matrix\\\hline
$\kappa$ & correlation between growth and feeding; $\kappa\in[0,1]$\\\hline
$\gamma$ & symmetry parameter for LV model; $\gamma\in[-1,1]$ \\\hline
$Q$ & niche parameter of LV model; $\EE[\VV_{ii}] = -Q$\\\hline

$\pinv$ & probability of a new variant successfully invading\\\hline
$\rho$ & correlation between parent and mutant\\\hline
$\overline{A}$ & evolutionary time average of $A$\\\hline
$\Delta(\Upsilon)$ & scale of $\ss$ distribution around $\ss\approx\Upsilon$ (Equation~\ref{eq:Delta_ups})
 
\end{tabular}

\caption{Definitions of variables and terms.}
\label{table:notation}
\end{table}

\section{Simplified models}
\label{sec:simple_models}

The saturation of the community-mean general fitness in the Red Queen eco-evolutionary steady state for $\kappa<1$ is a manifestation of continual evolution in an ecological system. We seek simpler---but still robust---models that display such a Red Queen steady state, but which allow us to separate out and independently tune the importance of ecological interactions and general fitnesses. The two related models that we shall study both have a generalized Lotka-Volterra form, but with different ensembles for the interaction matrix between strains. The first model is a linearized version of the CR model with the resource dynamics integrated out, yielding a maximum rank $\DD$ for the strain interaction matrix. The second model is of a more standard Lotka-Volterra type, with direct interactions drawn from some ensemble and diagonal ``niche" interactions enabling substantial diversity to coexist. For some ranges of parameters, both of these models exhibit a Red Queen phase characterized by constant turnover of biodiversity. In addition they allow us to isolate the effects of the general fitnesses on the eco-evolutionary dynamics.

\subsection{Linearized resource model}
\label{sec:lin_model}

In the Red Queen steady state of the CR model, the mean of $\sum_j \FF_{j\alpha}n_j^*$, the total consumption rate of resource $\alpha$, is large compared to its variations (Figure~\ref{appfig:resource_width}A), suggesting an expansion in these variations. In addition, the total population of the ecosystem $N=\sum_j n^*_j$ is roughly constant on evolutionary timescales (Figure~\ref{appfig:resource_width}B), due to the large mean of $\GG$ and $\FF$ relative to their variations---this motivates us to separate out the mean part of these matrices from their variations, and introduce zero-mean residual interaction matrices $\AA$ and $\BB$. We work in terms of fractional abundances $\nu_i=n_i/N$ rather than absolute abundances, and integrate out the resource dynamics since these do not affect the fixed point.

Expanding in the {\it differences} between resources and between strains yields general fitnesses at first order and interactions at second order (Appendix~\ref{app:consistency}).
The simplified model that we introduce, and refer to hereafter as the {\it linearized resource model}, or simply the linearized model, is
\begin{subequations}
\label{eq:eco_dyn}
\begin{align}
\frac{d\nu_i}{dt} &= \nu_i\left(\ss_i + \sum_{\alpha=1}^\DD\AA_{i\alpha}\RR_\alpha-\Upsilon(t)\right) \\
\RR_\alpha &= -\sum_{j=1}^\SS \BB_{j\alpha}\nu_j .
\end{align}
\end{subequations}
The term $\Upsilon(t)$ is a Lagrange multiplier, equal to $\sum_{i}\nu_i(\ss_i+\sum_\alpha\AA_{i\alpha}\RR_\alpha)$, which incorporates the average effects of growth, death, and resource depletion, and keeps the population constant with $\sum_{i} \nu_i=1$. Here, the per capita growth rate of each strain is a linear function of the abundances of all the other strains, with a quadratic part from $\Upsilon$ that is the same across all the strains and includes the average general fitness. Differences in general fitness are parametrized by $\ss_i$, the {\it selective differences} between strains, which play the role of the $\UU_i$ in the CR model. 
As the means have been subtracted off, the residual part of the resource vector $\vec\RR$ can have components which are negative. $\vec\RR$ then measures {\it deviations} of resource availability away from some positive baseline.

The structure of Equation~\ref{eq:eco_dyn} is that of a generalized Lotka-Volterra model or system of replicator equations (discussed further in Section~\ref{sec:LV_def}), where the interaction matrix between $\SS$ strains, denoted by $\VV$, has a structure inherited from the $\AA$ and $\BB$ matrices via $\VV = -\AA\BB^\intercal$, where $\AA$ and $\BB$ are $\SS\times \DD$ matrices. We take each element of $\AA$ and $\BB$ to be gaussian distributed with mean zero and unit variance, with no correlations within $\AA$ or $\BB$. Elementwise correlations between $\AA$ and $\BB$ are again parameterized by $\kappa\in[0,1]$ via $\EE[\AA_{i\alpha}\BB_{j\beta}]=\kappa\delta_{ij}\delta_{\alpha\beta}$.

The off-diagonal entries of $\VV$ have mean zero and standard deviation $\sqrt\DD$ while the diagonal entries have mean $-\kappa\DD$ and standard deviation $\sqrt{(1+\kappa^2)\DD}$. The parametrically-larger (for large $\DD$) diagonal entries thus endow the interactions with sufficient {\it niche} structure to stabilize the dynamics. (Note that $\kappa<0$ would result in unstable behavior with the diagonal entries of $\VV$ having positive mean.) As in the CR model, the number of resources $\DD$ sets the dimensionality of phenotype space, with the phenotype of strain $i$ comprised of $\ss_i$ along with the $i$th rows of $\AA$ and $\BB$. The maximum number of coexisting strains at a stable fixed point of the linearized model is equal to $\DD$ and will generally be proportional to $\DD$ with a prefactor less than $1$. 
 
When $\kappa=1$, $\Upsilon$ is a convex Lyapunov function for Equation~\ref{eq:eco_dyn}. If additionally $\ss_i=0$, we have $\Upsilon =-\sum_\alpha(\sum_i\AA_{i\alpha}\nu_i)^2$ which is nonpositive. For nonzero $\ss_i$, the Lyapunov function gets an extra piece $\langle\ss\rangle\equiv\sum_i\nu_i\ss_i$ and can be of either sign, though it remains a convex function of the $\{\nu_i\}$.

We will take the general fitnesses $\ss_i$ to be drawn from a distribution $\sdist(\ss)$, which for most of this paper is gaussian with mean $0$ and variance $\Sigma^2$. In Section~\ref{sec:s_dist} we show that the value of $\Sigma$ determines the early evolutionary behavior and the value of $\pinv$ at steady state, with the shape of the large-$\ss$ tail of $\sdist(\ss)$ playing a controlling role. As the community evolves and the distribution of extant $\ss_i$ pushes into the tail of $\sdist(\ss)$, the effective $\Sigma$---the variance of the extant $\ss_i$---decreases. Therefore tuning $\Sigma$ to be small {\it a priori} mimics the effects of {\it prior evolution} and allows us to study communities that have been evolving for a long time, without the long initial evolutionary transient. 

As discussed previously in Section~\ref{sec:CR_evo}, for the CR model in Equation~\ref{eq:CR_dyn}, the width of the $\UU_i$ distribution is a factor of $\sqrt\DD$ larger than the width of the drive distribution. 
In the linearized model, this corresponds to $\Sigma = \Oh(\sqrt\DD)$, since the distribution of the drives in a diverse community has $\Oh(\sqrt{\DD/\LL}) = \Oh(1)$ width. However, in the Red Queen steady state, as we shall see, the first and second order---in $1/\sqrt\DD$---components of the growth rate of each strain become comparable to one another. Thus we mostly consider $\Sigma = \Oh(1)$, and sometimes set $\Sigma=0$, which shortens the transient period of evolution before the steady state is reached---and enables analysis---but does not change our conclusions. In Section~\ref{sec:s_dist} we also discuss non-gaussian distributions of the $\ss_i$, which affect quantitative properties but not the basic phenomenology.

\subsection{Lotka-Volterra model}
\label{sec:LV_def}

In addition to the linearized model of Equation~\ref{eq:eco_dyn}, we have studied the generalized Lotka-Volterra (LV) model (or replicator equations)~\cite{diederich1989replicators} which take the form
\beq \frac{d\nu_i}{dt} = \nu_i\left(\ss_i+\sum_{j=1}^\SS V_{ij}\nu_j-\Upsilon(t)\right),\label{eq:LV_model}\eeq
with $\Upsilon(t) = \sum_i \nu_i(\ss_i + \sum_j \VV_{ij}\nu_j)$.
The LV model has similar structure to the linearized model of Section~\ref{sec:lin_model}; the difference lies in the statistics and parameterization of the interaction matrix $\VV$. Instead of the $\VV = -\AA\BB^\intercal$ with random $\AA$ and $\BB$ of the linearized resource model, in this LV model the $\VV_{ij}$ are themselves gaussian random variables with $\EE[\VV_{ij}]=-Q\delta_{ij}$, $\EE[\VV_{ij}^2] = 1+\gamma \delta_{ij}$, and $\EE[\VV_{ij}\VV_{ji}] = \gamma$ for $i\neq j$, with other covariances zero. Here $\gamma\in[-1,1]$ is the symmetry parameter tuning between predator-prey-like interactions for $\gamma<0$ and competitive interactions for $\gamma>0$: a rough correspondence with the linearized model for positive $\gamma$ is $\gamma \sim \kappa^2$. The negative diagonal term $-Q$ is a ``niche'' parameter that measures the strength of self-interaction relative to inter-strain interaction, playing a similar role to $-\kappa\sqrt\DD$ in the linearized model. Our parameterization of the LV equations differs from a common parameterization in which interactions have nonzero mean~\cite{bunin2017ecological}: here the $\VV$ matrix captures {\it variations} in interaction magnitudes, with $\Upsilon$ taking the role of the mean interaction strength~\cite{pearce2020stabilization}. As in the linearized model, when the interactions are symmetric ($\gamma=1$), $\Upsilon$ is a Lyapunov function of the dynamics, though in contrast to the linearized model~\cite{tikhonov2017collective}, it can become a nonconvex function of the $\{\nu_i\}$ for large enough $\SS$~\cite{biroli2018marginally}.

The LV model has a similar ecological phase diagram to the CR model and linearized models, with a globally attracting fixed point of the dynamics when $\SS$ is sufficiently small. The maximum $\LL$ at a stable uninvadable fixed point scales as $Q^2$ with a $\gamma$-dependent coefficient~\cite{opper1992phase}.
Note that the scale of the drive distribution is $\Oh(1/\sqrt\LL) = \Oh(1/Q)$ in the LV model, as opposed to $\Oh(\sqrt{\DD/\LL})=\Oh(1)$ in the linearized model, since the overall scale of the entries in the LV model is smaller by a factor of $\sqrt\DD$ than in the linearized model.

 In the LV model, unlike in the linearized and CR models, strains cannot be characterized by finite-dimensional phenotypes. Although one can attempt to define a phenotype of strain $i$ by $\ss_i$ along with its corresponding row and column of $\VV$, this phenotype depends on the rest of the community and thus changes---or grows in dimension---as the community turns over.

\subsection{Properties of invaders}
\label{sec:linearized_evo_dyn}

Having defined the simplified ecological models of interest, we will now specify the process by which they evolve. We first discuss general facts about invaders introduced into the community, and then define the specific parameterization of invaders that we have used for simulations.

As in the CR model, whether a new strain $\aa$ successfully invades the extant community or not is determined by its invasion fitness or bias, now written as $\xi_\aa = \ss_\aa + \zeta_\aa-\Upsilon$ with $\zeta_\aa = \sum_j \VV_{\aa j}\nu_{j}$ in the LV model and $\zeta_\aa = \sum_\alpha \AA_{\aa\alpha} R_{\alpha}$ in the linearized model. Splitting the bias up into the general fitness piece $\ss_\aa$, the Lagrange multiplier $\Upsilon$ and the drive $\zeta_\aa$ is useful since it separates the intrinsic and extrinsic contributions to the bias. Strains with positive bias can invade the community and reach abundance proportional to their bias, whereas strains with negative bias go extinct. As in the CR model, we sample properties of the invader $\aa$ conditional upon $\xi_\aa>0$~\cite{vrins2018sampling}. A strain's extinction occurs when its abundance goes to $0$ in the ecological dynamics or, equivalently, when its bias goes negative.

In the LV model, the lack of phenotypes makes attempts to parameterize correlations between parent and mutants {\it ad hoc} for general $\gamma$~\cite{mahadevan2023spatiotemporal}, so we confine our study to the case of independent invaders. An invader strain $\aa$ is generated by drawing a new row and column of $\VV$ from the original gaussian ensemble with $\Corr[\VV_{\aa i},\VV_{j\aa}]=\gamma\delta_{ij}$, and by drawing general fitness $\ss_\aa$ from a gaussian distribution with mean $0$ and variance $\Sigma^2$. Therefore the invasion probability of an independent invader is $\pinv = 1-\Phi[\Upsilon/\sqrt{\sum_i\nu_i^2 + \Sigma^2}]$ where $\Phi$ is the standard normal CDF.

In models with phenotypes, such as the CR or linearized model, there is a natural way to generate mutants through small changes to the phenotype of an extant parent strain. This procedure creates a mutant whose interactions with all other strains are similar to those of its parent.  In our gaussian ensemble, the probability of any parent generating a mutant that can invade is the same---therefore we can, in an unbiased way, simply select a parent with probability proportional to its abundance. 

The parameter $\rho\in[0,1)$ describes the correlation between parent $P$ and mutant $M$.
The mutant phenotype is drawn from an ensemble that preserves $\EE[\AA_{M\alpha}\AA_{P\beta}] = \EE[\BB_{M\alpha}\BB_{P\beta}] = \rho\delta_{\alpha\beta}$, $\EE[\AA_{M\alpha}\BB_{M\beta}]=\kappa\delta_{\alpha\beta}$, and $\Corr[\ss_M,\ss_P]=\rho$. We fix the desired correlations by drawing the phenotype of $M$ as
\begin{subequations}
\label{eq:mut_ensemble}
\begin{align}
\AA_{M\alpha} &= \rho\AA_{P\alpha} + \sqrt{1-\rho^2}Z_\alpha \\
\BB_{M\alpha} &= \rho \BB_{P\alpha} + \sqrt{1-\rho^2}(\kappa Z_\alpha+\sqrt{1-\kappa^2}Z_\alpha'),\\
\ss_M &= \rho \ss_P + \Sigma\sqrt{1-\rho^2} Z''
\end{align}
\end{subequations}
where the $\{Z_\alpha\}$, $\{Z_\alpha'\}$ and $Z''$ are uncorrelated standard normal random variables. This choice ensures that the elements of the interaction matrix $V=-\AA\BB^\intercal$ have correlations $\Corr[\VV_{Mi},\VV_{Pj}] = \Corr[\VV_{iM}\VV_{jP}]=\rho\delta_{ij}$ for $i,j\notin\{ M,P\}$.
The invasion probability of a mutant drawn from this ensemble is given by
\beq \pinv = 1-\Phi\left[\sqrt{\frac{1-\rho}{1+\rho}}\frac{\Upsilon}{\sqrt{|\vec \RR|^2+\Sigma^2}}\right]\label{eq:pinv}\eeq
where $\Phi$ is the standard normal CDF.
In this paper we restrict study to large effect mutations, with $\sqrt{1-\rho^2}=\Oh(1)$. The behavior as $\rho\to1$, corresponding to small effect mutations, is both interesting and biologically relevant: we will discuss it in depth in future work.

\begin{figure}
\centering
\includegraphics[scale=.6]{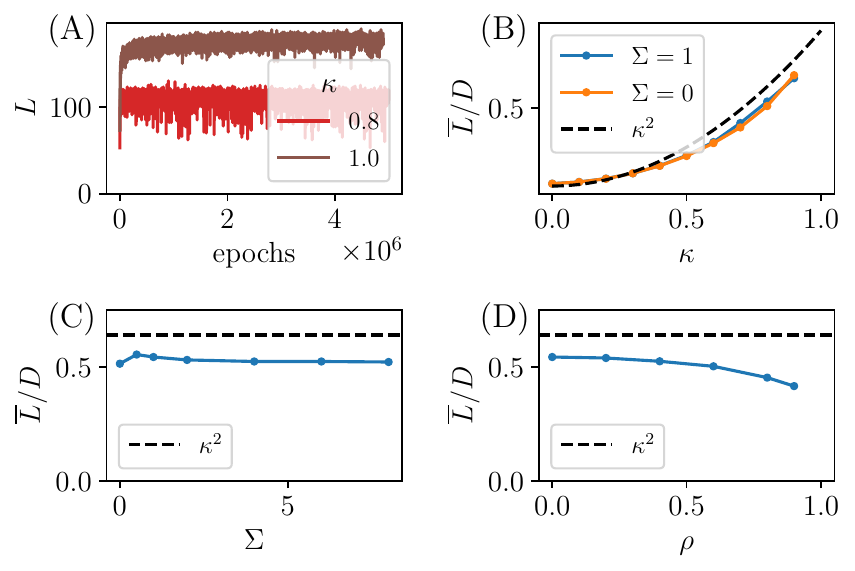}
\caption{Dynamics and steady-state of diversity in linearized model as function of $\kappa$, $\Sigma$ and $\rho$. Default values $\DD=200$, $\kappa=0.8$, parent-mutant correlation $\rho=0$, width of distribution of selective differences $\Sigma=1$, with each varied as specified per panel. (A) Dynamics of $\LL$ over evolutionary time for $\kappa=0.8,1$. Although $\LL$ appears to be reaching steady state for both parameters, for $\kappa=1$, $\Upsilon$ and $\pinv$ indicate it is not (Figures~\ref{fig:LR_Ups} and~\ref{fig:LR_pinv}). (B) Steady state diversity. $\overline\LL$ normalized by its maximum, $\DD$, as function of $\kappa$, for $\Sigma=0$ and $1$. Dotted line shows maximum diversity in {\it top-down assembled} community; in Red Queen phase, diversity is lower. (C) $\overline\LL$ depends only weakly on $\Sigma$. (D) $\LL$ depends weakly on $\rho$ for $\rho$ not too near unity.}
\label{fig:LR_L}
\end{figure}

\begin{figure}
\centering
\includegraphics[scale=.6]{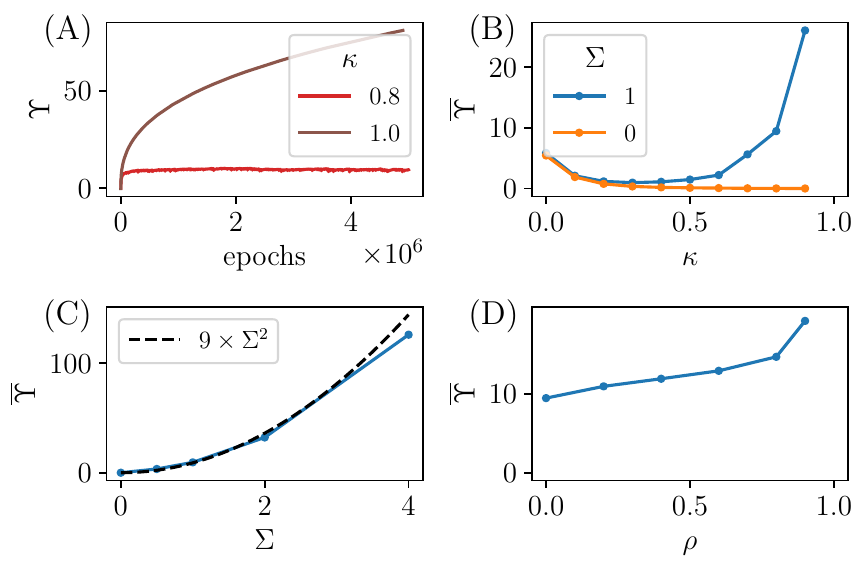}
\caption{Dynamics and steady state of Lagrange multiplier $\Upsilon$ ---loosely community ``fitness"---as a function of parameters for linearized resource model with default values as in Figure~\ref{fig:LR_L}. (A) Dynamics: for $\kappa=1$, $\Upsilon$ (Lyapunov function in this case) is ever-increasing. For $\kappa=0.8$, $\Upsilon$ reaches steady state $\overline\Upsilon$. (B) $\overline\Upsilon$ as function of $\kappa$ for $\Sigma=0$ and $1$. Only for $\Sigma>0$ does the phenotype push into the tail of its distribution indicated by large $\Upsilon$. Simulations are less reliable for small $\kappa$ because $\LL$ is smaller (see Appendix~\ref{app:numerics}). For $\kappa=0.9$, $\Sigma=1$, steady state not yet reached by end of simulation. (C) Dependence on $\Sigma$. Dotted line is rough fit to predicted form, $\overline\Upsilon\sim\Sigma^2$. At $\Sigma=4$, $\Upsilon$ has not yet saturated. (D) Weak dependence of $\overline\Upsilon$ on $\rho$.}
\label{fig:LR_Ups}
\end{figure}

\begin{figure}
\centering
\includegraphics[scale=.6]{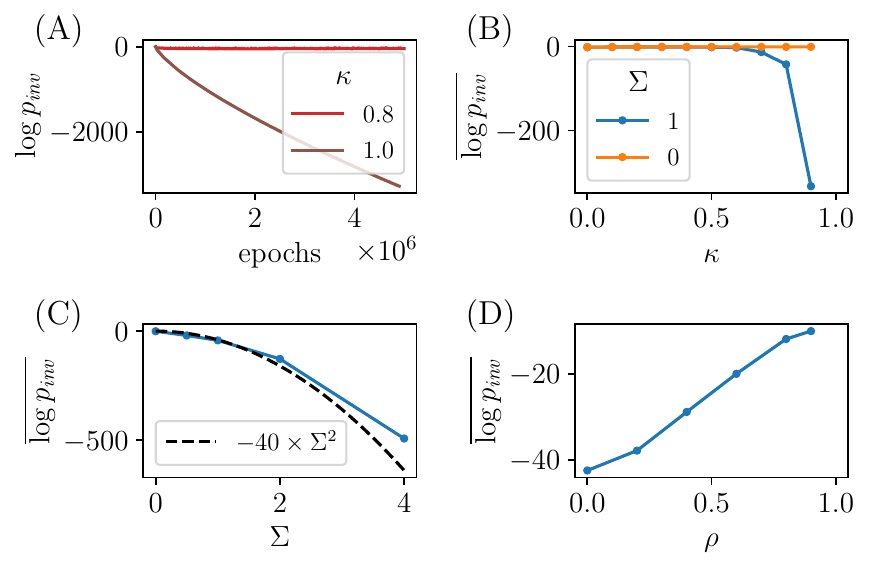}
\caption{Dynamics of invasion probability, and mean logarithm of invasion probability in steady state, $\overline{\log\pinv}$, as function of parameters with default values as in Figure~\ref{fig:LR_L}. (A) Dynamics: $\pinv$ saturates for $\kappa=0.8$ but continues to decrease for $\kappa=1$. (B) Dependence on $\kappa$.  (C) Dashed line is rough fit to predicted $-\overline{\log\pinv}\sim\Sigma^2$, by analogy with assembled community. (Equation~\ref{eq:S_hat}).  Simulations for $\Sigma=4$ have not reached steady state and $\overline{\log\pinv}$ is overestimated (see Appendix~\ref{app:evo_transient}). (D) Invasion probability increases with $\rho$ for $\rho$ far from $1$. }
\label{fig:LR_pinv}
\end{figure}

\subsection{Evolution in the linearized resource model}

The motivation for introducing the linearized model is to capture the phenomenology of the CR model---a Red Queen state with continual turnover but roughly-constant mean general fitness and $\pinv$---in a simple setting where the effects of the general fitnesses can be separated out from the interactions, and with a shorter transient period before the steady state is reached.
We begin by describing the basic phenomenology of evolution in the linearized model of Equation~\ref{eq:eco_dyn}, with the evolutionary dynamics of Equation~\ref{eq:mut_ensemble}. In general, the evolutionary dynamics are parameterized by $\DD$, $\kappa$, $\rho$, $\Sigma$ and $\SS$, the size of the initial pool of strains.
However $\SS$ only plays a role in the transient behavior: the evolution converges to the same state independent of the initial diversity as illustrated in Section~\ref{sec:lv_evo} for the LV model.

 In Figure~\ref{fig:LR_L}A we plot the dependence of $\LL$ on evolutionary time for two values of $\kappa$: $0.8$ and $1$. (We have here chosen $\kappa=0.8$ to yield higher diversity than the $\kappa=0.4$ used for the CR model.) Starting from a single strain, there is a transient during which the diversity builds up, after which both trajectories flatten out.
Figure~\ref{fig:LR_Ups}A shows the trajectories of the Lagrange multiplier, $\Upsilon$---analogous to the mean general fitness $\langle\UU\rangle$ in the CR model, which clearly showed the difference between the behavior with and without a Lyapunov function. For $\kappa=0.8$, saturation of $\Upsilon$ indicates a Red Queen steady state and results in saturation of $\pinv$. By contrast, for $\kappa=1$, $\Upsilon$ continues to increase and $\pinv$ concomitantly decreases: there is no Red Queen phase.

To investigate the Red Queen phase in the absence of a Lyapunov function, we study the long-time averages $\overline{\LL}$, $\overline{\Upsilon}$ and $\overline{\log\pinv}$ for ranges of the parameters $\kappa$, $\Sigma$, and mutant-parent correlation $\rho$. (Note that the {\it typical} $\pinv$ of interest is better characterized by $\overline{\log\pinv}$ than $\log\overline\pinv$ as $\pinv$ can fluctuate by large amounts especially when $\Sigma$ is large as discussed in Section~\ref{sec:s_dist}.) 
Figures~\ref{fig:LR_L},~\ref{fig:LR_Ups} and~\ref{fig:LR_pinv} display these quantities across a range of $\kappa$, $\Sigma$, and $\rho$. The important conclusion from these simulations is that the Red Queen phase obtains across a range of parameters for $\kappa<1$, and that parameter choices, though affecting the length of the pre-steady state transient and the qualitative properties of the steady state, do not eliminate the Red Queen phase. Therefore the Red Queen state can be qualitatively understood by studying just a few values of the relevant parameters. The value of $\DD$ sets the overall scale of $\LL$ but, as long as it is large, does not matter much---we therefore choose a large $\DD=200$ and $\kappa=0.8$ for our simulations. This choice yields a large number (roughly $\DD/2$) strains surviving in the Red Queen phase yet keeps away from the non-generic $\kappa=1$ behavior. 

In Figure~\ref{fig:LR_L}A, the observed $\overline\LL$ is compared with the curve $ \kappa^2\DD$ which is the maximum $\LL$ in a top-down assembled community with $\BB$ and $\AA$ drawn from the gaussian ensemble for the linearized model (see Section~\ref{sec:assembled_comm}). We see that $\overline\LL$ from simulations is less than this naive upper bound, as will be discussed in Section~\ref{sec:fragility}. In Section~\ref{sec:s_dist} we will give scaling arguments for how $\overline\Upsilon$ and $\overline{\log{\pinv}}$ at steady state depend on $\kappa$ and $\Sigma$: some of these scalings are compared with simulations in Figures~\ref{fig:LR_Ups} and~\ref{fig:LR_pinv}.

\subsection{Evolution in the LV model}
\label{sec:lv_evo}

\begin{figure}
\centering
\includegraphics[scale=.6]{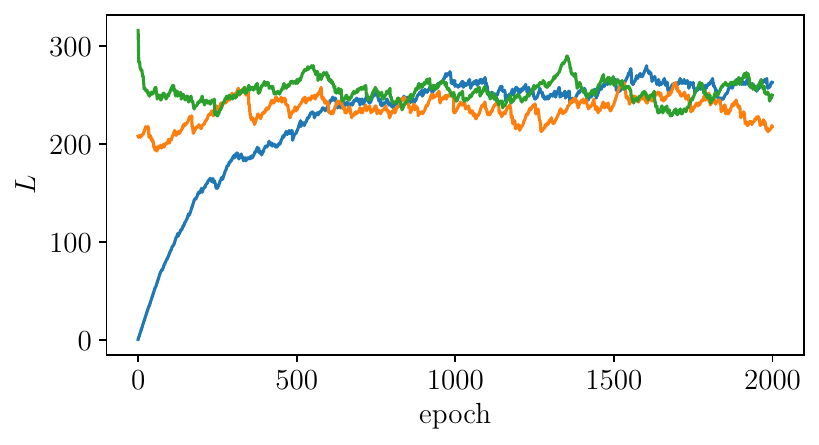}
\caption{Same steady state is reached in LV model for a range of initial strain pool sizes $\SS$ which determine initial diversity. All initial conditions tend toward same Red Queen phase after short initial transient. Parameters of LV model $Q=20$, $\gamma=0$, $\rho=0$, $\Sigma=0$.}
\label{fig:S_trans}
\end{figure}

Given the apparent similarities between the CR and linearized models, it is natural to ask how the eco-evolutionary phenomenology in the random Lotka-Volterra model resembles or differs from these: in particular whether the difference in the ensemble of the interaction matrix between the LV and linearized models matters. 
 In Appendix~\ref{app:lv_evo} we investigate evolution---by independent invaders---in the LV model across a range of $\gamma$ and $Q$, the symmetry and niche parameters defined in Section~\ref{sec:LV_def}. For a range of $\gamma$ we find a Red Queen phase analogous to those of the CR and linearized model, with the product of $Q^2$ and a function of $\gamma$ determining the number of strains in the steady state; we have also checked that a Red Queen phase obtains for nonzero $\Sigma$.
 Similarly to in the linearized model, $\LL$ in the Red Queen phase fluctuates around a mean value $\overline\LL$ that is less than the maximum $\LL$ from a top-down assembled community. The gap between the realized $\overline\LL$ and the assembled upper bound as a function of $\gamma$ is shown in Figure~\ref{appfig:LV}A. This gap diverges as $\gamma\to-1$, meaning that in this regime, the Red Queen steady state is much less diverse than the largest possible community assembled top-down. In contrast, in the linearized model, the gap is bounded over the whole range of $\kappa$. The distinction is not surprising: there is no regime of the linearized model that corresponds to the negative cross-diagonal correlations in $\VV$ that characterize the LV model with negative $\gamma$. 

The fact that a top-down assembled community can be substantially larger than the evolved community in the LV model allows us to check that the Red Queen phase obtains across a range of initial number of strains, $\SS$. In particular, since the top-down assembled community can have $\LL>\overline\LL$, there are situations in which there is a naturally-findable stable ecological community that is unstable to evolution, with $\LL$ decreasing as the evolving community converges to the Red Queen phase. Figure~\ref{fig:S_trans} shows this behavior in the LV model for $\gamma=0$ and $\Sigma=0$, with $\LL$ approaching $\overline\LL$ from both above or below, depending on the $\SS$ of the initial community. Therefore, even in simple models, the question of whether evolution increases or decreases diversity from a top-down assembled initial condition has no general answer~\cite{shoresh2008evolution,doebeli2010complexity}.

{\it Breakdown of Red Queen phase:}
Although we find a robust Red Queen phase for a range of $\gamma$ and $\Sigma$ in the LV model, closer investigation of the effect of $\gamma$ reveals an interesting feature. The evolutionary dynamics in the LV model change sharply with $\gamma$, appearing to exhibit a discontinuous transition around $\gamma_c\approx0.65$ for $\Sigma=0$. For $\gamma<\gamma_c$ a Red Queen phase obtains, while for $\gamma>\gamma_c$ we find a phase in which the rate of successful invasion is ever-slowing and a few strains rise to high abundance. We therefore name this phase the ``oligarch'' phase of eco-evolutionary dynamics---a counterpart to the Red Queen's monarchy.
 The oligarch phase is discussed further in Section~\ref{sec:discussion}, and we leave its fuller analysis for future work. For now, we note that there does not appear to be an oligarch phase in the linearized or CR models: the Red Queen steady state obtains for all $\kappa<1$, albeit with $\overline\Upsilon$ and $\overline{\langle \UU\rangle}$ respectively diverging as $\kappa$ approaches $1$ and evolutionary transients correspondingly lengthening.

\section{Analysis }
\label{sec:analysis}

In order to understand the phenomenology of the Red Queen phase described above, it is useful to first discuss the {\it top-down assembled} community of the linearized model in which all strains are brought together at once with dynamics determined by their ecological interactions. An understanding of the ecological phase diagram in this setting is informative for analyzing the Red Queen phase, and gives us an idea of how $\overline\Upsilon$ and $\overline{\log\pinv}$ scale with various parameters in the Red Queen phase.

To analyze the eco-evolutionary dynamics, we will introduce a particular limit of the linearized model which reduces to the LV model with $\gamma=0$. With the further simplifications of independent invaders and $\Sigma=0$---yielding the simplest ensemble of the LV model---we can exactly solve for the steady state of the evolutionary dynamics in the limit of large diversity. Key quantities are the distribution of strain abundances in an evolved community---strikingly different than a top-down assembled community---and temporal correlation functions of strain abundances in evolutionary time. 

We then add in selective differences, presenting general scaling arguments for the dependence of $\overline\LL$, $\overline\Upsilon$ and $\overline{\log\pinv}$ on $\Sigma$, which build on the intuition gleaned from the assembled communities. We discuss subtleties associated with long-lived strains with anomalously large $\ss_i$, generalize to different distributions, $\sdist(\ss)$, and show how the primary pathologies of the gaussian distribution can be alleviated.

\begin{figure}
\centering
\includegraphics[scale=.6]{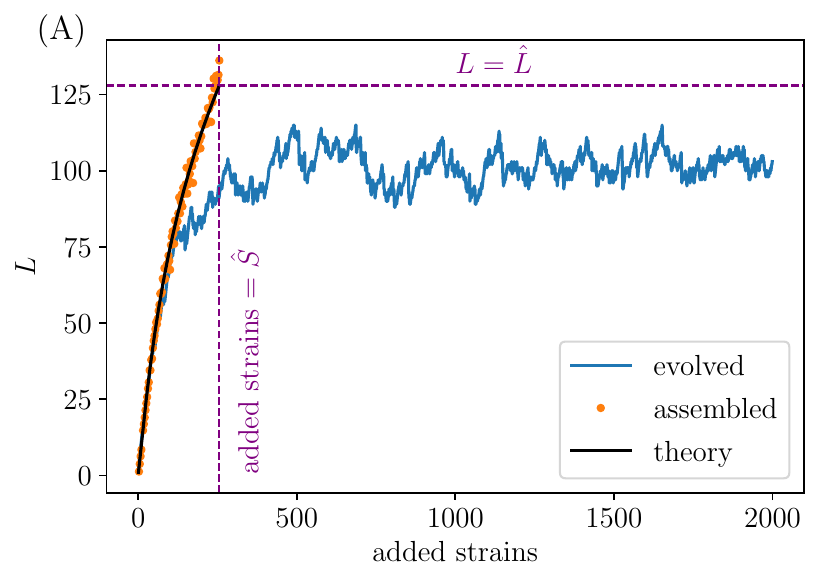}
\includegraphics[scale=.6]{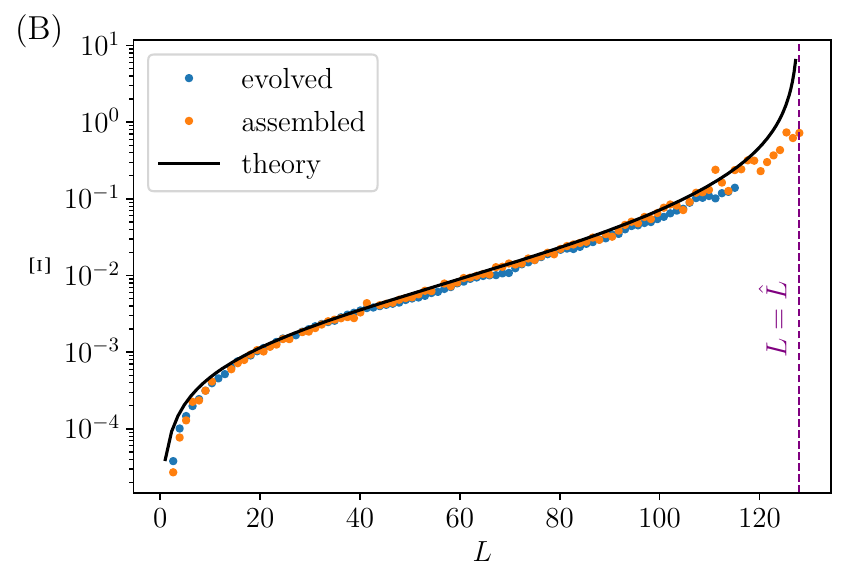}
\caption{Maximum diversity and stability to evolutionary perturbations comparing top-down assembled (yellow, black) with evolved (blue) communities, with data for small $\LL$ taken during build-up from a single strain. Linearized model with $\kappa=0.8$, $\DD=200$, and $\Sigma=0$. (A) Number of surviving strains as function of number of strains added. 
For assembled community, strains added all at once; for evolved community, one at a time. Maximum diversity in evolved case is less than saturated diversity, $\Lmax$ (horizontal dashed line) of assembled communities, reached with $\Smax=2\kappa^2 \DD$, (vertical dashed line). (B) Fragility, $\Xi$, a community-wide measure of susceptibility to perturbations, versus $\LL$. Fragility diverges at transition between stable and unstable phases (vertical dashed line), which is not reached in evolved communities. }
\label{fig:fragility}
\end{figure}

\subsection{Assembled communities}
\label{sec:assembled_comm}
In a top-down assembled community of the linearized model, all the strains are initialized at positive abundance, and are governed by the deterministic dynamics of Equation~\ref{eq:eco_dyn} without a sharp extinction cutoff. The cavity method~\cite{roy2019numerical} can be used to analyze both the statics and dynamics of the assembled community with $\SS$ initial strains and $\DD$ resources, in the limit of large $\SS$ and $\DD$ with the ratio $\SS/\DD$ constant. The spirit of the cavity method---in the static case---is to consider a large community to which one adds, separately, one new strain and one new resource: these each act as small perturbations to the fixed point of the community. The abundance of the new strain is determined by its invasion fitness, modified by the community-wide response to its introduction, and the new resource is treated similarly. 
One then enforces self-consistency conditions that the statistics of the strains and resources in the community are the same as the statistics of the newly added strain and resource. These conditions rely on the community being assembled with independent interactions, since then the independently-drawn invading strain and resource are from the same ensemble as the rest of the community. By solving the self-consistency equations one can extract abundance distributions of $\nu_i$ and $\RR_\alpha$, and solve for $\LL/\DD$ and $\Upsilon$ as a function of $\SS/\DD$, $\kappa$ and $\Sigma$. Similar results can be obtained for the CR model~\cite{cui2020effect} and---more simply---for the LV model~\cite{opper1992phase}.
 
 The cavity method also allows one to derive self consistent equations for the ecological {\it dynamics}~\cite{roy2019numerical}, in addition to their fixed point---we use a variant of this method to understand the {\it evolutionary} dynamics in Section~\ref{sec:dyn_cavity}.

Details of the static cavity analysis for top-down assembled communities with gaussian $\sdist(\ss)$ are in Appendix~\ref{app:cavity}, for both the linearized and LV models. We find that generically the linearized model exhibits two phases as a function of the ratio $\SS/\DD$. For $\Sigma>0$, when $\SS/\DD$ is small enough, there is a unique stable globally attracting fixed point, uninvadable by the strains that are going extinct: this fixed point is described by the cavity solution and the community diversity is $\LL/\DD<\kappa^2$. The abundances of the strains in the stable community are distributed as a truncated gaussian so that there is a constant density of abundances near zero. (Figure~\ref{fig:abund_dist} shows this feature---and its agreement with the result of the cavity calculation---for the LV model which is similar to the linearized model in this respect.) The strains destined for extinction have negative invasion fitnesses with a distribution that also has a constant density near zero. Thus the assembled community has many close-to-marginal strains. 

When the number of the strains assembled, $\SS$, becomes larger than a ($\Sigma$-dependent) critical value $\Smax$, at which $\LL = \hat{\LL}=\kappa^2 \DD$, the cavity solution becomes unstable and chaotic dynamics or multi-stability ensues~\cite{blumenthal2024phase}. We use a hat to denote the values of quantities at this phase transition, and refer to the community at this transition as {\it saturated}. The number of initial strains needed to achieve this saturated state, $\Smax$, diverges as $\kappa\to1$ as long as $\Sigma>0$. Therefore for $\kappa=1$ and $\Sigma>0$ the cavity solution is always stable.
When $\Sigma=0$ the picture for $\kappa<1$ remains the same, while at $\kappa=1$ there is a transition (for $\SS>2\DD$) to a nongeneric ``shielded'' phase~\cite{tikhonov2017collective} in which $\LL/\DD=1$---the maximum allowed by competitive exclusion---and this state is highly marginal.

With an extinction threshold, even for $\SS>\Smax$ the ecological dynamics will generally reach a stable fixed point with $L\lesssim\kappa^2\DD$. However the resulting community will generally be unstable to re-invasion by some of the strains that had gone extinct. Furthermore, in this super-saturated regime there are multiple possible stable communities, almost all of them invadable by some of the extinct strains. Which community occurs will thus depend on the initial conditions. 
The question of whether or not stable uninvadable communities are likely to exist for large $\DD$ is unresolved except for the special $\kappa=1$ case where there is an absolute maximum of the convex Lyapunov function. However it is believed that even if such fully-stable fixed points exist for $\kappa<1$, they may attract only an exponentially small fraction of initial conditions~\cite{cugliandolo1997glassy,fisher2021inevitability}.

 We can glean insights about the properties of an evolved community with $\rho=0$ from an analysis of the near-saturated assembled community, since after long period of evolution, a large number of strains---many more than $\Smax$---will have had a chance to invade the evolved community. Naively, one might think that the end result will be similar to assembling the whole community at once. However as we shall see, the evolved community never becomes saturated in the way that an assembled community can be. 
 
 We would like to understand the behavior for $\Sigma\gg 1$ which is what occurs ``naturally" (e.g. from the CR model): we thus focus for now on this limit. In the linearized model the saturated assembled community has $\hat\LL=\kappa^2\DD$ for all $\Sigma$ and, for $\Sigma\gg 1$,
\begin{subequations}
\label{eq:saturated_scalings}
\begin{align}
\Upsmax&\approx \frac{\Sigma^2\kappa}{1-\kappa^2} \label{eq:Upsilon_hat}\\
\Smax&\sim\exp\left[\frac{\kappa^2\Sigma^2}{2(1-\kappa^2)^2}\right].\label{eq:S_hat}
\end{align}
\end{subequations}
The reciprocal of $\Smax$ scales with the fraction of surviving strains, and therefore the invasion probability of a single independent invader is of order $\DD/\Smax$. 

We conjecture that the {\it forms} of the scalings of $\Upsmax$ and $\log\Smax$ with $\Sigma$ in the assembled community also describe the time-averaged quantities in the Red Queen steady state, i.e. that $\overline\Upsilon\sim\Upsmax$ and $\overline{\log\pinv}\sim -\log\Smax$, albeit with unknown coefficients. The scalings in Equation~\ref{eq:saturated_scalings}, along with the guess that $\overline\LL\sim\Lmax$, motivate the approximate theory curves in Figures~\ref{fig:LR_L},~\ref{fig:LR_Ups} and~\ref{fig:LR_pinv}: these are consistent with numerical data from the Red Queen phase. In Section~\ref{sec:s_dist} we give an argument that the conjectured scalings of $\overline\Upsilon$ and $\overline{\log\pinv}$ with $\Sigma$ should indeed be expected on evolutionary grounds. 

\subsection{Fragility of assembled and evolved communities}
\label{sec:fragility}

In the eco-evolutionary Red Queen phase, ecological communities remain stable, but are different from top-down assembled communities since their interaction statistics are conditioned on evolution and, crucially, they would be {\it unstable to invasion} by some fraction of previously extant strains that have gone extinct. An indication of this distinction, shown in Figures~\ref{fig:LR_L} and~\ref{fig:fragility}A, is the diversity of the Red Queen phase: $\overline{\LL}$ is smaller than in saturated assembled communities, $\overline\LL<\Lmax=\kappa^2 \DD$, across the range of $\kappa$. When a community is assembled one strain at a time with {\it permanent} extinctions, it is not possible to reach as high diversity as reached by top-down assembly. What is responsible for this gap between $\overline\LL$ in the Red Queen phase and $\Lmax$ from the saturated assembled community? The answer is associated with the relative fragility of the two states. 

The transition between phases in the assembled community is marked by the onset of an instability in the cavity solution. This instability entails the divergence of a community-level property that we call the {\it fragility} and denote by $\Xi$. The fragility quantifies the variance of the changes in strain abundances in response to a random perturbation to the growth rates of all the strains.
Concretely, if for each strain $i$ we add a small independently random term $h_i$ to its growth rate, with $\EE[h_i]=0$ and variance $\EE[h_i^2]$ (loosely like applying random fields in magnetic systems), the fragility is defined by $\sum_i\EE[\delta\nu_i^2] = \Xi\, \EE[h_i^2]$. The fragility can also be written as $\Xi = \Tr[\chi^\intercal \chi]$ where the $\chi_{ij} = \frac{d\nu_i}{d\xi_j}$ are elements of the susceptibility matrix $\chi$ of the community.
Deep in the stable phase the fragility is small, as small perturbations to strains' growth rates do not cause abundances to reshuffle by much. However as the transition is approached and $\LL$ nears $\Lmax=\kappa^2 \DD$, the fragility diverges, indicating a breakdown in the stability of the cavity solution. In Figure~\ref{fig:fragility}B we plot the fragility, computed directly from the interaction matrix, for both an evolving and top-down assembled community across a range of $\LL$. In both cases, $\Xi$ increases with $\LL$, and in the evolving community it plays an important role in the Red Queen phase, as we now discuss.

Each new invader adds a small random part to the biases of the other strains. Invader $\aa$, which reaches a stable abundance $\nu_\aa$, contributes $\VV_{i\aa}\nu_\aa$ to the bias of each strain $i$. The variance of this random perturbation is $\Var[\VV_{i\aa}\nu_\aa]=\nu_\aa^2\DD$ for the linearized model. This is multiplied by a factor of $\Xi/\LL\sim 1/\LL^2$ to yield $\Var[\delta\nu_i]$ of order $1/\LL^3$ and therefore $\delta\nu_i\sim \LL^{-3/2}$. For a typical strain with $\nu_i\sim1/\LL$, this perturbation will be small. However in a top-down assembled community, the lowest abundance strains will have $\nu_i$ smaller than typical by a factor of $1/\LL$, and $\Oh( \sqrt{\LL})$ of these will have abundance less than the magnitude of the perturbation. This implies that the invasion will drive the bias of $\Oh( \sqrt{\LL})$ of the lowest abundance strains negative, causing their extinction. Concomitantly, if we do not enforce permanent extinctions, the perturbation will increase the bias of a similar number of barely-extinct strains and these will enter the community.
As long as $\Xi$ is not large, this number of extinctions and invasions will be small. As $\SS\nearrow\Smax$, $\Xi$ diverges and the assembled community will be shaken a lot by a single additional strain. 

The behavior of an {\it evolved} community---with permanent extinctions---in response to a small random perturbation is quite different. When the community has just been assembled, if it is close to saturated, a single invader will drive a fraction of the strains with $\nu \leq \Oh(\LL^{-3/2})$ extinct. As this is $\Oh(\sqrt\LL)$ strains, $\LL$ will initially decrease (Figure~\ref{fig:S_trans}). Under the perturbations from further invasions, the density of low abundance strains will be hollowed out by this process which is like diffusion with an absorbing boundary condition at zero: This results in a density of abundances that goes linearly to zero as $\nu\to0$ (see Figure~\ref{fig:abund_dist}). 
The diffusion coefficient of the abundances in evolutionary time is proportional to the fragility of the community, so if $\Xi$ is large, then the perturbations are amplified and more extinctions result, on average, from each successful invader. As the diversity grows, we expect that (as in the top-down assembled communities), the fragility also grows. When it is large enough that on average one extinction is driven by each invasion, the diversity stabilizes. This must thus occur while the fragility is not too large suggesting---although not implying because of the complexities on conditioning on the evolutionary history---that the Red Queen state will be less diverse than the maximal assembled diversity, $\hat{\LL}=\kappa^2\DD$, as observed in Figure~\ref{fig:fragility}A. Note that the ratio $\overline\LL/\kappa^2\DD$ depends on the value of $\kappa$ (Figure~\ref{fig:LR_L}B); this is presumably due to the dependence of $\Xi$ on $\kappa$.

In the special case of perfect symmetry ($\kappa=1$ for the linearized model and $\gamma=1$ for the LV model), $\LL$ approaches $\Lmax$ at long times despite the lack of a Red Queen steady state. We do not have an explanation for this apparent saturation of the diversity through gradual assembly.

In recent work, de Pirey and Bunin~\cite{arnoulx2024many} have observed and studied a similar phenomenon of depletion of low-abundance strains due to continual perturbations to the community from invading strains. These authors looked at a purely {\it ecological} model but with small migration from a mainland that prevents total extinction of any strain. Species rising up from the brink of extinction play the role of invaders in our evolutionary model. The similarities between this ecological model and our work are discussed in Appendix~\ref{app:dPB}. A primary difference is due to the migration-induced boundary condition: As there is a flux of strains both in and out, the steady state distribution of $\nu$ does not vanish as $\nu\to0$ and the boundary condition must be determined self-consistently from the ecological dynamics.
In the evolutionary problem we study, more analytical progress is possible due to the simplicity of the extinction boundary condition in contrast to that which arises from migration.

\subsection{Exact solution in simplest case} 
\label{sec:dyn_cavity}

\begin{figure}
\centering
\includegraphics[scale=.6]{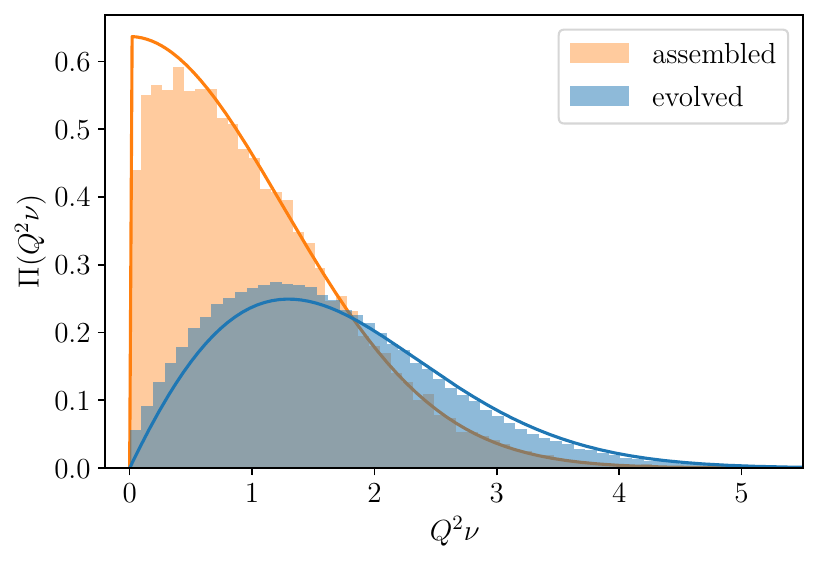}
\caption{Number density $\dens(z)$ of the scaled abundance $z = Q^2\nu$ in the $\gamma=0$, $\Sigma=0$ LV model with $Q=20$: for saturated assembled community (yellow) and evolved Red Queen steady state (blue), with normalization $\int d z \dens(z) =\overline\LL/Q^2$. Histograms are numerical data and solid lines predictions from, respectively, assembled cavity calculation (Appendix~\ref{app:cavity}) and evolutionary cavity calculation (Equation~\ref{eq:FP_rescaled}).}
\label{fig:abund_dist}
\end{figure}

\begin{figure}
\centering
\includegraphics[scale=.6]{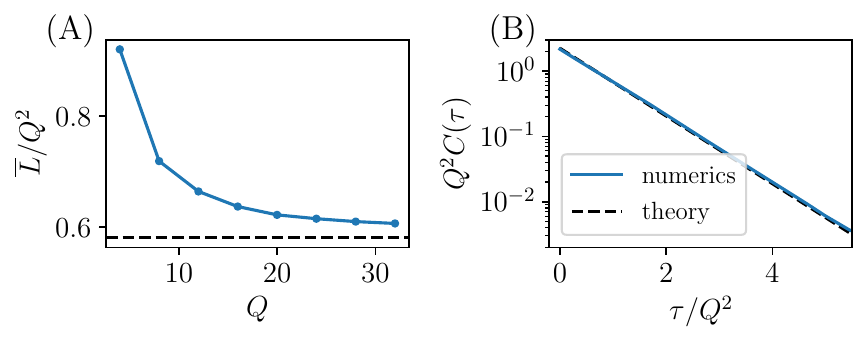}
\includegraphics[scale=.6]{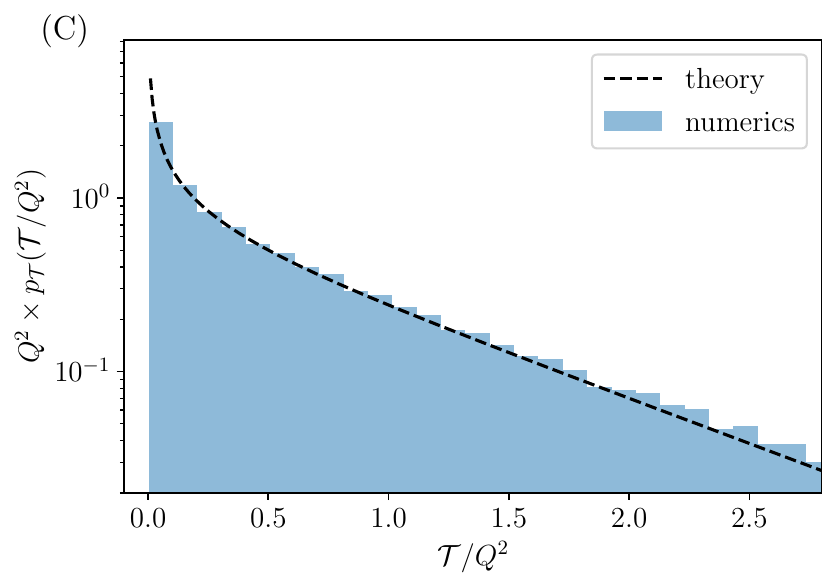}
\caption{Predictions of evolutionary cavity calculation and simulations for Red Queen state of $\gamma=0$, $\Sigma=0$, LV model with independent invaders. Niche parameter that sets time and diversity scales is $Q=20$ except in first panel. (A) Number of strains as function of $Q$ converges to predicted $\overline\LL\approx 0.58\times Q^2$ (dotted black line) as $Q\to\infty$, Equation~\ref{eq:Lbar_RQ}.
(B) Temporal correlation function $\CF(\tau) = \sum_j \nu_j(\TT)\nu_j(\TT+\tau)$ agrees with Equation~\ref{eq:corr_func} with no fit parameters. (C) Distribution of strain lifetimes $\tsurv$, normalized by basic timescale $Q^2$, agrees with prediction of Equation~\ref{eq:lifetime_dist}. }
\label{fig:theory}
\end{figure}

Having understood some features of the eco-evolutionary phenomenology by comparison with the saturated assembled community, we now proceed to directly analyze the evolutionary dynamics for a particularly simple choice of parameters.

With $\kappa>0$, there are correlations $\kappa^2$ between the cross-diagonal elements of the interaction matrix $\VV$, as in the case of the $\gamma>0$ LV model. These correlations induce complex memory into the evolutionary dynamics since the effect of an invader $\aa$ on a strain $j$ in the community via $\VV_{j\aa}$ feeds back via $\VV_{\aa j}$ onto $\aa$. This feedback on $\aa$ is coherent across the extant strains with magnitude proportional to $\kappa^2$ (or $\gamma$). Thus the properties of a strain in the community depend on its effects on the other strains over all the epochs since its invasion. This fact makes the evolutionary dynamics difficult to analyze (like the ecological dynamics in ref.~\cite{pearce2020stabilization}), as the form of this memory kernel needs to be self-consistently determined along with the dynamics of the strains abundances in evolutionary time. We set this analysis up in Section~\ref{sec:general_cavity} but do not carry it through.

However, there is a limit of the linearized model that is simple: if we take $\DD\to\infty$ and $\kappa\to0$ while keeping the product $\kappa^2 \DD$ finite, we can rescale the interaction matrix by a factor of $\sqrt \DD$ to obtain the LV model with $\gamma=0$ and $Q = \kappa \sqrt \DD$, implying $\overline\LL=\Oh(Q^2)$ in the Red Queen phase. (Note that in this limit, the elements of $\VV$ become gaussian and the off-diagonal elements are $\Oh(1)$.) The fact that for $\gamma=0$ the interaction matrix $\VV$ has no correlations across the diagonal makes the feedback effects smaller by a factor of $1/\sqrt{\LL}$ (and of random sign) and hence negligible---therefore the analysis that we undertake becomes tractable. We first treat the case of $\gamma=0$, $\Sigma=0$, and then set up the general structure of the cavity calculation for nonzero $\gamma$ and $\Sigma$, before heuristically analyzing the effects of $\Sigma>0$.

Our analysis of the Red Queen phase proceeds by means of a dynamical cavity method, which is one way of carrying out dynamical mean-field theory (DMFT) as used in systems with quenched randomness~\cite{roy2019numerical}. The idea is to consider the evolutionary-time dynamics of a focal ``cavity strain'' labelled $0$, that could potentially enter the community, persist for some number of epochs, and later go extinct. For $\gamma=0$, the abundance of the cavity strain while it is in the community is determined by its bias, if positive. In this simple case the strain's feedback on itself averages to zero so that $\nu_0=\xi_0/Q$ in the limit of large $\LL$ (Appendix~\ref{app:cavity}). However the bias, $\xi_0(\TT)=\ss_0+\zeta_0(\TT)-\Upsilon(\TT)$, changes in evolutionary time $\TT$. If the strain attempts to invade at epoch $\tinv_0$, then its abundance in the community at all later times is
\beq
\nu_0(\TT)=\frac{\xi_0(\TT)}{Q }\mathds{1}\left[\xi_0(\TT')>0 ~ \forall ~ \TT'\in (\tinv_0,\TT)\right]
\label{eq:zero_gamma_nu}
\eeq
where $\mathds{1}[E]$ is the indicator function which is unity when its argument is true and zero otherwise. The cavity strain's abundance is only nonzero as long as its bias remains positive---and it goes irrevocably extinct once its bias goes negative. Therefore the statistics of the cavity strain abundance are determined by the statistics of its bias together with the permanent extinction condition.

To carry out the analysis, one needs to make an {\it Ansatz} for the statistics of the cavity bias, and then calculate the statistics of the cavity abundance as filtered through the extinction condition. One then enforces self consistency, which relates the dynamics of strain abundances in the community to the original statistics of the cavity bias. 
This is possible since the cavity drive is simply $\zeta_0(\TT)=\sum_{j\in\Comm(\TT) } \VV_{0j}\nu_j(\TT)$, which has gaussian correlations for large $\LL$. Since the $V_{0j}$ are independent and not conditioned on properties of the community, we have, for the LV model with the $\VV_{0j}$ being $\Oh(1)$, 
\beq
\CF(\tau)\equiv \overline{\zeta_0(\TT)\zeta_0(\TT+\tau)} = \sum_j \overline{\nu_j(\TT)\nu_j(\TT+\tau)}
\label{eq:cav_self_con}
\eeq
where the sum is over {\it all} strains that have invaded before time $\TT$---most of which will be extinct at $\TT$ or $\TT+\tau$ and therefore not contribute to the correlation function. For large $\overline\LL$, the average over $\TT$, denoted by the overbar, is not needed: we expect $\CF(\tau)$ to be ``self-averaging" and depend only on the time-difference $\tau$ (up to subdominant corrections). A particular role is played by the equal-time correlations: 
 $\sigzeta^2\equiv\Var[\zeta_\aa] = \CF(0)=\sum_i\nu_i^2$, with the last equality for the LV model where $\sigzeta\sim 1/\sqrt{\LL}$. (In the linearized model $\sigzeta$ carries an additional factor of $\sqrt\DD$.)

The self-consistency condition at the heart of the cavity method is that the autocorrelation of the cavity strain, averaged over trajectories $\{\zeta_0(\TT)\}$ and invasion time $\tinv_0$, is the same as the autocorrelation function of strains $i$ in the community conditioned on having the same general fitness $\ss_i=\ss_0$. The correlation functions of strains in the community, weighted-averaged over the $\ss$ distribution, determine $\CF(\tau)$ via Equation~\ref{eq:cav_self_con}. The Lagrange multiplier
$\overline\Upsilon$ has to be adjusted to fix $\sum_i\nu_i=1$ (just as $\Upsilon$ is determined for the static ecological cavity analysis of Appendix~\ref{app:cavity}).

We now specialize to the simplest case where all $\ss_i$ are zero. Simulations of the $\gamma=0$, $\Sigma=0$ Lotka-Volterra model (Figure~\ref{appfig:LV}B) suggest that in this case $\overline\Upsilon=0$ and we make the {\it Ansatz} that this is exact. 
Hoping for some luck, we then make the simplest possible {\it Ansatz} for the statistics of $\zeta_0(\TT)$: namely that it has an exponentially decaying correlation function $\CF(\tau) = \sqrt{\JJ/\bb}\  e^{-b\tau}$. This {\it Ansatz} for the correlation function can be translated into a stochastic process that determines $\zeta_0$:
\beq
\frac{d\zeta_0}{d\TT} = -b\zeta_0+\sqrt{2\JJ}\eta_0(\TT),
\label{eq:drive_sde}
\eeq
with $\TT$ the time measured in epochs and $\eta_0(\TT)$ white noise satisfying $\langle \eta_0(\TT)\eta_0(\TT')\rangle = \delta(\TT-\TT')$ (angular brackets here denote averages over the stochasticity caused by invaders). The white noise represents the changes in the community from adding independent invaders which cause the diffusive changes in abundances discussed in Section~\ref{sec:fragility}. The deterministic term is, roughly, regression to the mean $\zeta_0$ of zero, which can be seen to arise from the correlation between $\zeta_0$ and $\delta\zeta_0$ due to the same elements of $\VV$ appearing in both quantities~\cite{mahadevan2023spatiotemporal}.
The quantities $\bb$ and $\JJ$ are (for now) phenomenological parameters that must be fixed later by self-consistency. The advantage of having a Markovian stochastic 
differential equation for $\zeta_0$, along with $\overline\Upsilon=0$, is that the method of images can be used to find the statistics of $\nu_0$ with an absorbing boundary condition at zero, conditioned on its invasion at time $\tinv_0$. 

In order to calculate the right hand side of the self-consistency condition Equation~\ref{eq:cav_self_con}, we need to calculate the two-time correlation function of $\nu_0$. Since (with all $\ss_i=0$) the strains are statistically equivalent, we can calculate $\langle\nu_0(\TT)\nu_0(\TT+\tau)\rangle$ for the cavity strain entering the community with the initial condition of $\zeta_0(\tinv_0$), averaged over the stochasticity in Equation~\ref{eq:drive_sde}. We then average this correlation function over $\tinv_0$ and $\nu_0(\tinv_0)$.
We know that $\nu_0(\TT)$ is determined by Equation~\ref{eq:zero_gamma_nu} with $\xi_0(\TT) = \zeta_0(\TT)$ since $\ss=\overline\Upsilon=0$.
From our {\it Ansatz}, the initial $\nu_0(\tinv_0)$ is distributed as a gaussian random variable with standard deviation $\sigzeta/Q=\frac{1}{Q}\sqrt{\JJ/\bb}$, conditioned to be positive (as epochs are measured per successful invasion). In order to calculate $\CF(\tau)$ from the strains in the community, we integrate $\langle \nu_0(\TT)\nu_0(\TT+\tau)\rangle$ over all initial times $\tinv_0$---equivalent to fixing $\tinv_0=0$ and integrating over all $\TT$---and over all starting abundances $\nu_0(\tinv_0)$ weighted by the thalf-gaussian measure. We work with all times in evolutionary epochs, so that the integral of the distribution of $\nu_0(\tinv_0)$ for $\nu_0(\tinv)>0$ is the rate of successful invasions per unit time, which is $1$ in these units.


A self-consistent solution is possible if the result of this calculation gives an exponentially decaying correlation function that can be equated with the correlation function of the cavity drive $\CF(\tau) = \sqrt{\JJ/\bb} \ e^{-b\tau}$. Fortunately, this turns out to be the case. An additional self-consistency condition fixes $\sum_i \nu_i = 1$. Put together, these conditions allow us to solve for $\JJ$ and $\bb$. We then compute the diversity $\overline\LL=\sum_i \Theta(\nu_i) $. 
These calculations are carried out in Appendix~\ref{app:dyn_cav}. The key results are the self-consistent exponential solution for the correlation function
\beq \CF(\tau) = \frac{2\pi\lambda^2}{Q^2} \exp\left[-\frac{2\lambda}{Q^2}\tau\right]\quad\text{with}\quad \lambda = \frac{1+2\log2}{4},\label{eq:corr_func}\eeq
 and the steady state diversity given by
\beq\overline\LL = \frac{2\log2}{1+2\log 2} Q^2\cong 0.58\times Q^2.\label{eq:Lbar_RQ}\eeq
The timescale of the decay of the correlation function is $\Oh(Q^2)=\Oh(\overline\LL)$, consistent with the community turning over every $\Oh(\overline\LL)$ successful invasions. We additionally find that $\bb = 2\lambda/Q^2$ and $\JJ = 4\pi \lambda^3/Q^4$.

Figure~\ref{fig:theory} compares the analytical results to numerics on the $\gamma=0$, $\Sigma=0$ LV model. As $Q$ increases in Figure~\ref{fig:theory}A, the numerics get closer to the large-$\overline\LL$ regime in which the theory holds, and $\overline\LL/Q^2$ in steady state approaches $\frac{2\log 2}{1+ 2\log 2}$ from above. Figure~\ref{fig:theory}B shows the scaled correlation function $Q^2 \CF(\tau)$ versus scaled time $\tau/Q^2$, which agrees well with Equation~\ref{eq:corr_func}. 

From the self consistent solution we can also calculate the distribution of strain {\it lifetimes}, where the lifetime of a strain $i$, denoted $\tsurv_i$, is the number of evolutionary epochs for which it has nonzero abundance. In Appendix~\ref{app:dyn_cav} we find a distribution
\beq p_{\tsurv}(\tsurv) = \frac{2\bb}{\pi\sqrt{e^{2\bb\tsurv}-1}}.\label{eq:lifetime_dist}\eeq
Figure~\ref{fig:theory}C shows good agreement between this analytical result and the numerics: $p_{\tsurv}(\tsurv)$ decays exponentially at large $\tsurv$ as expected, and has an integrable divergence as $\tsurv\to0$, due to the distribution of invaders having a constant density at low abundance. These invaders go extinct very quickly as discussed in Section~\ref{sec:fragility}. Note that, as successful invaders must balance extinctions, $ \langle\tsurv\rangle = \frac{\log2}{\bb} = \overline\LL$: the mean diversity is equal to the average strain lifetime in epochs. 

The dynamic cavity method also enables us to determine the distribution of strain abundances in steady state, as the solution to a Fokker Planck equation with appropriate boundary conditions and with a source term for the invading strains (Equation~\ref{eq:FP_drive}). This solution is plotted in Figure~\ref{fig:abund_dist}, in reasonable agreement with the numerical result for the distribution of abundances in the Red Queen phase. As anticipated in Section~\ref{sec:fragility}, there is a marked difference between the abundance distribution in the eco-evolutionary steady state and the saturated assembled community abundance distribution with $\LL=\Lmax$. In the assembled community the abundance distribution is a truncated gaussian---a half-gaussian at saturation. By contrast, in the Red Queen steady state, the density of strains vanishes at low abundance due to the constant jostling of the community by invaders which knock low-abundance strains extinct: the effective diffusion coefficient capturing this process is proportional to the fragility.

For the LV model with $\gamma=0$, the fragility can be calculated directly due to the simple relationship between $\nu_i$ and $\xi_i$. For each strain with positive bias, $\nu_i=\xi_i/Q$ so the susceptibility to a change in growth rate is $1/Q$ (Appendix~\ref{app:cavity}). However the change in each $\nu_i$ causes a small random change of the biases of all the other strains, which increases the effective variance of the random field, and so on. Summing up these effects~\cite{mahadevan2023spatiotemporal} yields a total effect of random $h_j$'s given by $\sum_i (\delta\nu_i)^2 = \Xi \Var[h_j]$ with the fragility $\Xi=\frac{\LL/Q^2}{1-\LL/Q^2}=\frac{\LL}{Q^2-\LL}$. This result does not depend on how the community is formed, or on the selective differences if they are present. It shows right away the maximum size of a stable community is $\hat{\LL}=Q^2$, which is saturated for an assembled community. For the evolved community, $\Xi$ cannot be large---otherwise multiple strains would be driven extinct by each invader. Thus $\overline\LL<\hat{\LL}$, as we find in numerics.

\subsection{General cavity formulation}
\label{sec:general_cavity}

We now generalize the cavity analysis away from $\gamma=0$, $\Sigma=0$. For $\gamma\neq 0$ in the LV model the analysis is particularly complicated. (For $\kappa>0$ in the linearized model it is even more so as both resources and strain dynamics have to be analyzed as in the ecological cavity calculation of Appendix~\ref{app:cavity}.) 
The bias and abundance of a focal strain are no longer simply related as in the $\gamma=0$ case: due to the feedback of the strain on itself via its influence on the other strains, $\nu_0(\TT)$ will depend both on its bias $\xi_0(\TT)$ and an additive term arising from the feedback integrated over all times since its invasion. Therefore the analog of Equation~\ref{eq:zero_gamma_nu} for $\gamma\neq 0$ is
\beq
\nu_0(\TT)=\frac{\phi_0(\TT)}{Q }\mathds{1}\left[\phi_0(\TT')>0 ~ \forall ~ T'\in (\tinv_0,\TT)\right]
\eeq
where 
\beq
\phi_0(\TT)=\xi_0(\TT)+\gamma \int_{\tinv_0}^T dT' M(T-T')\nu_0(\TT') \label{eq:M}
\eeq
with $M(\tau)$ a memory kernel that must be determined self-consistently. It will include an instantaneous---on evolutionary time scales---part from the invader's {\it ecological} feedback on its own abundance, in addition to a slow memory part that extends over many epochs~\cite{fisher2021inevitability}.
Both the memory effects and the $\ss_0-\Upsilon$ part of $\xi_0$ mean that the cavity strain will not go extinct when $\zeta_0$ goes through zero, but instead at some other time when $\phi_0(\TT)$ first goes negative. 
 
The statistics of the drives $\zeta_0(\TT)$ are determined by the extant community, but $\CF(\tau)$ is no longer a simple exponential. Nevertheless, one can write an effective stochastic equation for $\zeta_0$:
 \beq
\frac{d\zeta_0}{d\TT} = -\int_{\tinv_0}^\TT d\TT' B(\TT-\TT') \zeta_0(\TT') +\sqrt{2\JJ}\eta_0(\TT) \label{eq:B}
\eeq
with $B(\tau)$ another memory kernel and $\eta_0(\TT)$ still white noise which represents the diffusion-like effects of an invader on the other strains discussed in Section~\ref{sec:fragility}. The Fourier transform of $B$ is related, via a Weiner-Hopf factorization, to the Fourier transform of $\CF$: denoting the transforms by tildes, we have
\beq
\tilde{\CF}(\omega)= \frac{2J}{[-i\omega+\tilde{B}(\omega)][+i\omega + \tilde{B}(-\omega)]} 
\eeq
with $\tilde{B}(\omega)$ analytic in the upper half of the complex plane. 

In principle, to proceed one assumes a $\CF(\tau)$---or equivalently a $B(\tau)$ and $\JJ$---as well as $\Upsilon$ (and if $\gamma\ne0$, also $M(\tau)$. 
Then one computes $\langle\nu_0(\TT)\nu_0(\TT+\tau)\rangle$ conditioned on $\nu_0$ invading at time $\tinv_0$. By averaging over $\nu_0(\tinv_0)$, $\tinv_0$ and $\ss_0$, one can enforce the self-consistency of $\CF(\tau)$ using Equation~\ref{eq:cav_self_con}, as for the $\gamma=0$ case. Then $M(\tau)$ is self-consisted from calculating the response of $\nu_0$ to a change in its bias, and $\Upsilon$ is self-consisted to ensure $\sum_i \nu_i = 1$. In the linearized model with resources, the resources have to be handled by a cavity method as well but their effects can be integrated out to give a more complicated relation between strain abundances and $M$, with $Q$ replaced by $\kappa\sqrt\DD$ and $\gamma$ replaced by $\kappa^2$ (Appendix~\ref{app:cavity}).

 Unfortunately, even numerical analysis of the self-consistency conditions to determine the memory functions $M$ and $B$ is likely challenging~\cite{roy2019numerical}. However the above structure allows us to glean some insights. 
 
 In the absence of selective differences, all strains are statistically similar and the systematic decrease in $\zeta_0$ should make the strain lifetime distribution decay exponentially at long times. This suggests that both the memory kernels $M$ and $B$ will decay exponentially at long times, as will the correlation function of the drives, $\CF(\tau)$. To explore the effects of feedback, we carried out simulations of the $\Sigma=0$ LV model for $\gamma=-0.8$, $0$, and $0.5$, and plotted $\CF(\tau)$ in Figure~\ref{appfig:corr_func}B. We see that $\CF(\tau)$ indeed decays exponentially at long times for all these $\gamma$, with some deviations at shorter times in opposite directions for positive and negative $\gamma$. Concomitantly, as $\gamma$ increases, the timescale over which $\CF(\tau)$ decays, scaled by the overall turnover time of $\overline\LL$ epochs, grows. 
 
 There is no sign of marginality in the exact solution for $\gamma=0$, $\Sigma=0$---therefore the Red Queen phase should be robust to small perturbations, as evinced by these simulations for $\gamma\neq0$.
The heuristic arguments below suggest that the robustness extends as well to arbitrarily large $\Sigma$, and numerical results suggest that it extends up to modest---although {\it not all}---$\gamma$: see Section~\ref{sec:discussion} for more discussion of this point. 
In contrast to the LV model, simulations of the linearized resource model suggest that the Red Queen phase obtains for all $\kappa<1$.

\subsection{Effects of selective differences}
\label{sec:s_dist}

The Red Queen phase that obtains for $\Sigma>0$ in both the linearized and LV models has many properties in common with the $\Sigma=0$ case we have analyzed. Here we discuss more systematically the effects of a distribution of general fitnesses, focusing first on the distribution of strain lifetimes, $p_\tsurv(\tsurv)$, and then on the correlation function $\CF(\tau)$, whose decay at large $\tau$ is dominated by strains with large $\ss$.
We then consider large $\Sigma$ and give scaling arguments for how the community evolves into the tail of the $\ss$ distribution, which allows us to find the dependence of $\overline\Upsilon$ and $\overline{\log\pinv}$ on $\Sigma$. 
When the distribution of extant $\ss_i$ pushes into the tail of the $\sdist(\ss)$ distribution, all of the steady-state properties are determined by the form of $\sdist(\ss\to\infty)$---thus we consider non-gaussian $\sdist(\ss)$ of the form $\sdist(\ss)\sim e^{-(\ss/\Sigma)^\psi/\psi}$. We will also comment on $\sdist(\ss)$ with bounded support.

{\it Correlation function and lifetime distribution:}
From the general structure of the cavity analysis, we can attempt to understand the behavior of a strain with anomalously-large $\ss$ (we suppress the strain index for notational convenience). The behavior is simplest in the LV model with $\gamma=0$, for which the only memory is in $\zeta(\TT)$ which still has mean zero and determines the abundance of the cavity strain when in the community (Equation~\ref{eq:zero_gamma_nu}). Extinction occurs when $\xi=\zeta+\ss-\Upsilon\searrow 0$, i.e. $\zeta \searrow \Upsilon-\ss$ which is unlikely for large $\ss$ since $\xi$ will fluctuate around $\ss-\overline\Upsilon$, staying far from zero for a long time. 
The rare event for $\zeta$ to decrease to $\overline\Upsilon-\ss$ has, quite generally for well-behaved stationary gaussian processes, probability of order $\sim \exp[-(\ss-\overline\Upsilon)^2/2\sigzeta^2]$ for sufficiently large $\ss-\overline\Upsilon$, where the extinction is most likely to occur once the correlation has decayed significantly from $\CF(0)=\sigzeta^2$. However, we need to examine if the cavity {\it Ansatz} of a well-behaved gaussian process is consistent when large $\ss$ can occur. 

The effects of large $\ss$ are most clearly understood by their effects on the average strain lifetime $\langle\tsurv\rangle$, which is equal to $\overline{\LL}$ in steady state. The naively-predicted behavior is that the average lifetime of strains with large $\ss$ is at least as long as the inverse of their estimated asymptotic extinction rate. Thus, conditioned on $\ss$, the mean $\tsurv$ is $\langle\tsurv|\ss\rangle \sim \exp[(\ss-\overline\Upsilon)^2/2\sigzeta^2]$. The unconditional expected lifetime is then 
\beq
\langle \tsurv\rangle=\frac{\int {\cal P}({\rm invade}|\ss)\sdist(\ss)\langle\tsurv|\ss\rangle d\ss}{\int {\cal P}({\rm invade}|\ss)\sdist(\ss)d \ss},
\label{eq:mean_T}
\eeq
with ${\cal P}({\rm invade}|\ss)$ the invasion probability of a strain with general fitness $\ss$, which for large $\ss$ will be close to unity. The effects of large $\ss$ on the correlation function are related to their effects on the lifetime distribution, since long-lived strains have abundance $\nu \sim \ss-\overline\Upsilon$   for most of their lifetime (with coefficient $1/Q$ for the $\gamma=0$ LV model). Thus $\CF(\tau)$ is dominated at long $\tau$ by the fraction of strains whose lifetime is at least $\tau$. We thus expect that for large $\tau$, 
\beq 
\CF(\tau) \approx \int_\tau^\infty (\tsurv-\tau)\langle\nu^2|\tsurv\rangle p_\tsurv(\tsurv)d\tsurv,\label{eq:Ctau_pT}
\eeq 
with the $\nu^2$ factor likely yielding $\log(\tau)$ factors.
If $p(\ss)$ falls off sufficiently rapidly, the integral in Equation~\ref{eq:mean_T} that determines $\langle \tsurv \rangle$ is convergent, and the correlation function concomitantly decays to zero. For $\psi>2$, one finds $\CF(\tau)\sim \exp[-a\log^{\psi/2}(\tau)] $ with $p_\tsurv(\tsurv)$ decaying similarly so that $\langle\tsurv\rangle<\infty$. The gaussian case, $\psi=2$, is marginal with the integrals over $\ss$ convergent only for small enough $\Sigma <\Sigma_\times$ where the critical value is determined by $\Sigma_\times=\sigzeta(\Sigma_\times)$. For $\psi=2$ and $\Sigma<\Sigma_\times$, $\CF(\tau)$ decays as a $\Sigma$-dependent power of $\tau$.

When $\Sigma>\Sigma_\times$ in the gaussian case, as well as for any $\psi<2$, the integrals over $\ss$ that determine both $\langle \tsurv \rangle$ and the correlation function appear to diverge. Thus this naive analysis, which relied on the gaussian rarity of large $\zeta$ fluctuations, must break down---or the Red Queen steady state must cease to exist. In order to better understand what happens for $\Sigma>\Sigma_\times$, we turn to numerics. 

 In Appendix~\ref{app:corr_func} we explore numerically the effects of $\Sigma$ for the $\gamma=0$ LV model with gaussian $\sdist(\ss)$ across a range of $\Sigma$. In spite of the naive expectation from the above analysis of a loss of steady state for large enough $\Sigma$, the numerics show a steady state even for $\Sigma \gg \Sigma_\times \cong 1.4/Q$. The distributions of $p_\tsurv(\tsurv)$ appear to have a finite first moment and $\CF(\tau)$ still appears to decay to zero (Figure~\ref{appfig:lv_s_corr_func}). What are possible resolutions of this apparent discrepancy between the cavity analysis and the observed behavior? 
 
Since our simulations are for only modestly-large $\overline{\LL}$ up to $\sim100$, finite-$\overline\LL$ corrections to the cavity analysis could be substantial for the range of parameters we investigated here: maybe such corrections enhance the robustness of the steady state to large $\Sigma$.
It is also possible that fluctuations in $\Upsilon$, which are expected in any case to increase as $\Sigma$ increases (Appendix~\ref{app:corr_func}), become much larger than the expected 
$\Oh(1/\LL)$ in the regimes where the naive cavity analysis breaks down. Large positive fluctuations in $\Upsilon$ will eventually drive extinctions of strains that had anomalously large $\ss-\Upsilon$ when they invaded, although it is unclear how such large fluctuations could leave the predicted average properties unchanged. 
 
 Another possibility is that the gaussian {\it Ansatz} for the statistics of $\zeta(\TT)$ breaks down in the tail even for large $\overline\LL$---with some non-trivial scaling with $\zeta$ and $\overline\LL$---invalidating the rare event analysis for large $\ss$. This could make the rate of ``barrier" crossing of $\zeta(\TT)$ much larger than the naive gaussian estimate, perhaps associated with anomalously long-lived fluctuations in $\zeta$ that themselves arise from the persistence of large-$\ss$ strains. 
 
 We discuss further in Appendix~\ref{app:corr_func} some of these issues but for now must leave the subtle effects of longish tails of $p(\ss)$ unresolved. For the present purposes, 
we note that empirically, even for quite large $\overline\LL$, the primary conclusions of the cavity analysis, here and below, are at least semi-quantitatively good.

The above discussion has focused on the simplest case of the $\gamma=0$ LV model. 
 For $\gamma\neq0$, or in the linearized model with $\kappa$ independent of $\DD$, there are additional complications that could arise for $\Sigma>0$ when the correlation function develops a long tail, which will induce long tails to both the memory kernels. It is unclear whether or not the persistent memory in $M(\tau)$ and $B(\tau)$ (Equations~\ref{eq:M} and~\ref{eq:B}), will change the behavior qualitatively and make the simple analysis break down even when it is still valid for $\gamma=0$. However the numerics show that the naive analyses are roughly correct for average properties of the evolved communities. 

{\it Evolutionary transients and steady state for $\Sigma>0$:} 
We now analyze the evolutionary dynamics in the limit of large $\Sigma$ assuming that either $p(\ss)$ decays sufficiently rapidly for the problems with rare events not to matter, or that the analysis ignoring them is correct for averaged properties. In what follows we mainly discuss the linearized model in which the distribution of biases has $\Oh(\sqrt{\DD/\LL}) = \Oh(1)$ width---this can be translated to the LV model by recalling that there the width of the bias distribution is $\Oh(1/Q) = \Oh(1/\sqrt{\LL})$ with $\Upsilon$ scaling similarly. 

For $\Sigma\gg 1$, there is a long evolutionary transient during which the distribution of extant $\ss_i$ creeps into the tail of $\sdist(\ss)$. The tail of $\sdist(\ss)$ therefore determines the long term evolutionary dynamics, including---when it occurs---the properties of the Red Queen phase. As the population-mean $\ss$ (and therefore $\Upsilon$) increases, the extant $\ss$'s will be near to $\Upsilon$ with the scale of their variation set by the shape of $\sdist(\ss)$ near $\Upsilon$. The width of the extant $\ss_i$ distribution will be of order 
\beq \Delta(\Upsilon) \equiv \EE[\ss_\aa-\Upsilon|\ss_\aa>\Upsilon] \approx -\left[\left.\, \frac{d\log \sdist(\ss)}{d\ss}\right|_{s=\Upsilon}\right]^{-1},
\label{eq:Delta_ups}
\eeq
which decreases with increasing $\Upsilon$ for $\sdist(\ss)$ decaying faster than exponentially and increases with $\Upsilon$ for $\sdist(\ss)$ decaying slower than exponentially. At any point in the evolution, once $\Upsilon\gg\Sigma$, the distribution of extant $\ss_i$ is concentrated around $\Upsilon$ and the system behaves as if there were an exponential distribution of the $\ss_i$ with scale $\Delta(\Upsilon)$.

Given this exponential approximation to the general fitness distribution, it is helpful to understand the behavior of an {\it assembled} community with an exponential $\sdist(\ss)$ when the extant $\ss_i$ are large enough that the lower bound of $\sdist(\ss)$ does not matter---this will determine a key relationship between $\LL$ and $\Upsilon$. If we assemble a large number of strains with an exponential distribution of general fitnesses $\sdist(\ss)=\frac{1}{\Delta}e^{-\ss/\Delta}\Theta(\ss)$, for sufficiently large $\Delta$ the ecological cavity analysis (Appendix~\ref{app:exp_s_dist}) yields an assembled community with $\LL=\kappa\DD/\Delta$ in the linearized model (and $\LL = Q/\Delta$ in the LV model).
We therefore expect that during the transient evolution, the community will reach a slowly-changing quasi steady-state with diversity given, at any point in time, by $\kappa \DD/\Delta(\Upsilon)$.

We have shown above that the behavior of $\Upsilon$ determines the behavior of $\LL$. How does $\Upsilon$ depend on $\TT$ during the evolutionary transient and into the Red Queen steady state? The increase of $\Upsilon$ is tied to the increasing mean $\ss$, denoted $\langle \ss\rangle \equiv \sum_i \nu_i \ss_i$, so we are led to analyze how $\langle\ss\rangle$ changes in evolutionary time. Each successful invader $\aa$ must have $\zeta_\aa+\ss_\aa>\Upsilon$. The probability density of $\ss_\aa$ conditioned on invasion is (see Appendix of ref.~\cite{mahadevan2023spatiotemporal})
\beq
\sdist(\ss_\aa|{\rm invasion})\approx \frac{1}{\Delta}\exp\left[-\frac{\ss_\aa-\Upsilon}{\Delta}-\frac{\sigzeta^2}{2\Delta^2}\right]\Phi\left[\frac{\ss_\aa-\Upsilon}{\sigzeta}\right],
 \eeq
where $\sigzeta^2\equiv\Var[\zeta_\aa] \sim \DD/\LL$ and $\Phi$ is the CDF of the standard normal distribution. It follows that 
\beq
\EE[\ss_\aa-\Upsilon|{\rm invasion}]=\Delta(\Upsilon)-\frac{\sigzeta^2}{\Delta(\Upsilon)}.\label{eq:cond_increment}
\eeq
As a result, when $\Delta(\Upsilon) \sim \sigzeta\sim 1/\kappa$, there is a steady state and $\Upsilon$ is roughly constant in evolutionary time, with its value depending on the form of the function $\Delta(\Upsilon)$. During the transient and in steady state, $\overline\LL\sim \kappa\DD/\overline\Delta(\Upsilon)$ which saturates at $\Oh(\kappa^2 \DD)$. (Note that for $\Delta$ constant---as in the case of a pure exponential $\sdist(\ss)$---there is no special $\Upsilon$ for which $\EE[\ss_\aa-\Upsilon|{\rm invasion}]$ vanishes, so there is no natural scale for $\Upsilon$.)

We can now proceed to analyze the behavior with $\sdist(\ss)\sim e^{-(\ss/\Sigma)^\psi/\psi}$, for which $\Delta(\Upsilon)\approx \Sigma^\psi/\Upsilon^{\psi-1}$. 
For stretched exponential decay of $\sdist(\ss)$, corresponding to $\psi<1$, $\Delta(\Upsilon)$ is an increasing function---therefore $\Upsilon$ will run away to positive values with $\LL$ concomitantly decreasing. The diversity thereby crashes and eventually the community will reduce to a single strain that goes extinct when a fitter strain invades. Even if $\Sigma$ is small and seemingly unimportant, eventually a rare large $\ss$ will emerge, leading to $\Upsilon$ large enough that $\Delta(\Upsilon)\gtrsim 1$, and therefore to diversity crashing.

For $\psi>1$ (including the gaussian case $\psi=2$), initially $\Delta$ is of order $\Sigma$. From the above analysis, we expect diversity $\LL\sim \kappa\DD/\Delta$.

With $\Delta$ large, invasions will tend to increase $\Upsilon$, thereby decreasing $\Delta$ and increasing $\LL$. Eventually $\Delta$ will decrease to $\overline\Delta \sim \sqrt{\DD/\LL}\sim 1/\kappa$, independent of $\psi$, so that the widths of the extant $\ss_i$ and $\zeta_i$ distributions are comparable. Concomitantly, $\LL$ will saturate at $\overline{\LL}\sim\kappa\DD/\overline\Delta \sim \kappa^2\DD$, and $\Upsilon$ will saturate at $\overline{\Upsilon}\sim (\Sigma^\psi/\overline\Delta)^{1/(\psi-1)}$, with $\overline{\log\pinv}\sim (\Sigma/\overline\Delta)^{\psi/(\psi-1)}$.

We are now in a position to analyze the transient evolutionary dynamics that occur when $\Sigma\gg1$ and $\psi>1$. 
For large $\Delta$, as it will be initially, the change in $\Upsilon$ due to a successful invasion by strain $\aa$ is roughly $\Delta(\Upsilon)/\LL$, coming from the fitness increment multiplied by the $\Oh(1/\LL)$ typical abundance $\nu_\aa$. Since the fitness increment is $\sim \Delta$ and $1/\LL\sim \Delta/\DD$, the change in $\Upsilon$ per invasion is $d\Upsilon/d\TT\sim \Delta(\Upsilon)^2/\DD$. On average $d\Upsilon/d\TT$ is positive until the contribution of $\zeta_\aa$ becomes comparable to $\ss_\aa-\Upsilon$, at which point $\Upsilon\approx\overline{\Upsilon}$ and $\Delta\approx\overline{\Delta}$.

The transient behavior is found by integrating up the typical change per invasion: roughly $d\Upsilon/dT\sim \Delta(\Upsilon)^2/\DD \sim \Sigma^{2\psi}/\DD\Upsilon^{2\psi-2}$ until $\Upsilon$ is close to $\overline{\Upsilon}$. Thus during the transient 
\begin{align}
\Upsilon &\sim \left(\frac{\Sigma^{2\psi} T}{\DD}\right)^{\frac{1}{2\psi-1}}\\
\frac{\LL}{\DD}&\sim \Sigma^{-\frac{\psi}{2\psi-1}}\left(\frac{\TT}{\DD}\right)^{\frac{\psi-1}{2\psi-1}}
\end{align}
for a duration
\beq
\TT_{\it transient}\sim\DD\Sigma^{\psi/(\psi-1)},\eeq after which the steady state ensues.

An important observation is that some aspects of the Red Queen steady state for large $\Sigma$ are {\it independent of details} of the $\ss$ distribution and of $\Sigma$: in particular, for large $\Sigma$, $\overline\LL$ is independent of $\Sigma$, as seen in Figure~\ref{fig:LR_L} for gaussian $\sdist(\ss)$, since $\overline\Upsilon$ becomes large enough that the distribution of the extant $\ss_i$ has $\Oh(1)$ width, irrespective of $\Sigma$.

 To compare with our numerics for gaussian $\sdist(\ss)$, we specialize to $\psi=2$ in which case $\Delta(\Upsilon)\approx \Sigma^2/\Upsilon$. Here we have $\overline{\Upsilon}\sim \Sigma^2 $ and $-\overline{\log\pinv} \sim\Sigma^2$. 
This scaling is similar to the saturated assembled community (Section~\ref{sec:assembled_comm}), in which $-\log\Smax\sim\Sigma^2$ was conjectured to scale with $\overline{\log\pinv}$.
During the long initial transient, $\Upsilon\sim T^{1/3}$ and $\LL\sim T^{1/3}$ with the transient lasting for a number of epochs $\TT_{\it transient}\sim \DD\Sigma^2$, consistent with the rapid growth of transient length with $\Sigma$ in Figures~\ref{fig:LR_L},~\ref{fig:LR_Ups} and~\ref{fig:LR_pinv}. 
In Appendix~\ref{app:evo_transient}, we check that this predicted transient behavior of $\Upsilon$ agrees with numerics. Perhaps surprisingly, given the potential problems caused by large $\ss_i$'s and discussed earlier in this section, the predicted transients and steady state behavior seem to persist for gaussian $p(\ss)$ even when $\Sigma\gg\Sigma_\times$.

{\it Invasion probabilities}: Because of large $\kappa$-dependent coefficients, even moderate values of $\Sigma$ can make $\pinv$ unrealistically small at steady state. For $\Sigma=1$, $\rho=0$ and $\kappa=0.8$, the invasion probability at steady state is less than $10^{-15}$ (Figure~\ref{fig:LR_Ups}D) which suggests that $\overline{\Delta}$, which determines $\overline\Upsilon$ via $\overline{\Upsilon}\sim \Sigma^2/\overline\Delta$, has a value $\overline{\Delta}\cong 0.12$ for $\kappa=0.8$, substantially smaller than the value for the top-down assembled community, $\overline{\Delta}=(1-\kappa^2)/\kappa= 0.45$ (Equation~\ref{eq:Upsilon_hat}).

In the CR model, invasion probabilities are even smaller: as discussed earlier, the width of the general fitness terms dominates over that of the interactions by a factor of $\sqrt\DD$, corresponding to $\Sigma\sim \sqrt{\DD} $ in the linearized model. Therefore, for our gaussian ensemble, $\overline\pinv$ is exponentially small in $\DD$. Indeed we see (Figure~\ref{fig:CR_L}B) that for $\kappa=0.4$ and $\DD=50$ in the CR model, the invasion probability in steady state is $\sim 10^{-400}$, where here the distribution of general fitnesses $\UU_i$ inherits its gaussian distribution from the gaussian ensemble of the $\GG$ matrix that we studied.

Generally, distributions with faster-decaying tails but similar $\Sigma$ have higher invasion probability at steady state. However even in the limit $\psi\to\infty$ which corresponds to a double exponential decay $\sdist(\ss)\sim \exp(-\exp(\ss/\Sigma))$, $\overline\pinv$ still decays exponentially in $\Sigma$, and therefore as a stretched exponential in $\DD$ for the CR model.


For distributions that have a sharp cutoff at some maximum $\ss$, the invasion probability in steady state is not so small, even with a broad distribution of $\ss$'s. If $\sdist(\ss) \sim (\Sigma-\ss)^\lambda \Theta(\ss)\Theta(\Sigma-\ss)$ for large $\Sigma$ and some power $\lambda$, then the width of the extant $\ss_i$ distribution scales as $\frac{1}{\lambda+2}(\Sigma-\Upsilon)$. The Red Queen phase obtains when the distribution of extant $\ss$ has $\Oh(1)$ width, in which case $\pinv\sim1/\Sigma^{\lambda+1}$, which decays with $\Sigma$ much more slowly than in the case of $\sdist(\ss)\sim e^{-(\ss/\Sigma)^\psi/\psi}$. Importantly, once the Red Queen steady state is established, the population-mean $\langle\ss\rangle$ does not creep ever closer to the cutoff at the maximum $\ss$. This is in contrast to the situation with $\kappa=1$, where $\Upsilon$ would inch up as $\langle\ss\rangle$ increases towards the maximum $\ss$, since the Lyapunov function $\Upsilon$ has a part that is exactly $\langle\ss\rangle$.

We have found that there is a wide range of scalings of $\overline\Upsilon$ and $\overline\pinv$ with $\Sigma$ in the linearized model or with $\DD$ in the CR model. These scalings can be reasonably obtained with different distributions of the $\ss_i$ or the elements of $\GG$ and $\FF$. We have used gaussian distributions since they allow us to easily draw invaders conditioned on having positive invasion fitness, and to explore issues of {\it principle} regarding the Red Queen phase. However it is important to note that quantitative aspects such as the scaling of the invasion probability with $\DD$ are very non-universal. In biological systems, situations where the evolution pushes far into the tail of $\sdist(\ss)$ are anyway likely dominated by effects that we do not account for, such as the discreteness of genomes.

\section{Discussion}
\label{sec:discussion}

In this work we have explored evolution in a range of ecological models that exhibit diverse stable communities, and found that, generically, after a transient period of evolution, the community converges to a Red Queen state of constant turnover with a roughly-constant probability of new strains successfully invading.
In contrast to previous work which finds continual evolution in selection-driven host-pathogen models~\cite{yan2019phylodynamic,bonsma2023dynamics,marchi2021antigenic}, the diverse communities we find here stably coexist and do not rely on continual mutations to replenish the diversity. Here, there are no ``kill-the-winner" dynamics in which one subpopulation ``chases'' the other through trait space; nonetheless constant turnover ensues. The evolving ecology acts as a concrete manifestation of the environmental feedback that can engender continual Red Queen evolution in random landscape models without diversity~\cite{fisher2021inevitability}. The consumer resource model and a linearized approximation to it display a Red Queen phase as long as the consumption and growth are not perfectly correlated ($\kappa<1$): the perfectly correlated case often studied behaves differently, not exhibiting a Red Queen phase. The Lotka-Volterra model with random interactions also exhibits a Red Queen phase but only for symmetry parameter less than a critical value.

Importantly, the Red Queen phase persists in the presence of arbitrarily large general fitness differences $\ss_i$ provided that their distribution falls off at least exponentially for large $\ss$ (with a more rapid falloff possibly required in the limit of large $\overline\LL$ due to the issues discussed in Section~\ref{sec:s_dist}). 
Previous analyses of high-dimensional ecological models~\cite{cui2020effect,bunin2017ecological} have, as noted earlier, first assumed that in the large $\DD$ (or similar) limit, the distribution of general fitnesses (or similar quantities) has a width that is smaller than that of the interactions by a factor of $1/\sqrt{\DD}$, and then taken a well-behaved large-$\DD$ limit. We have shown that this assumption is not needed: In a long-evolved community, the width of the distribution of general fitness differences {\it within the evolved community} will become comparable to the variations between strains of the effects of ecological interactions. Even starting with a parametrically (in $\DD$) broad distribution of general fitnesses, some properties (such as the diversity) of evolved communities are roughly independent of the distribution, though other properties (such as the invasion probability of new strains) can be strongly affected.

If the distribution of the $\ss_i$ falls off with a large characteristic scale---which occurs naturally in the CR model---there is a long transient period during which evolution is dominated by increases in $\langle\ss\rangle$ and the diversity gradually increases as the population moves out into the tail of $\sdist(\ss)$. 
The transient behavior converges to a steady state when the strain-to-strain variations of $\ss_i$ and the ecological influences of other strains in the community are comparable---here the diversity of the Red Queen steady state becomes independent of $\sdist(\ss)$, but the invasion probability at steady state can be extremely small. However if there is a maximum $\ss$, the invasion probability, and hence the community turnover rate, stabilizes at a value that need not be as small. 
 
Below, we discuss some unresolved issues from this current work, and suggest future directions.

\begin{figure}
\centering
\includegraphics[scale=.6]{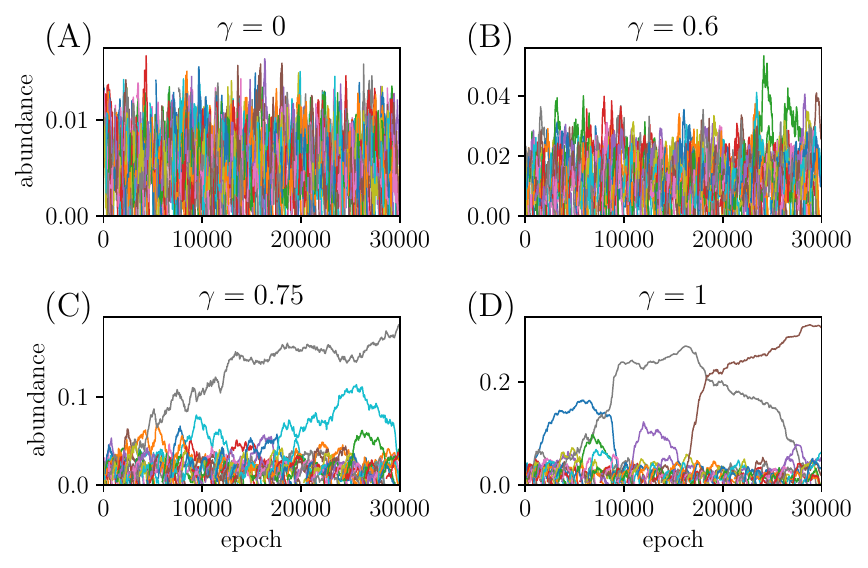}
\caption{Strain trajectories (subset of strains shown) for LV model with various $\gamma$ in Red Queen (A)--(B) and oligarch (C)--(D) phases. In oligarch phase, an $\Oh(1)$ fraction of the total abundance ``condenses" into a few strains which continue to turn over more and more slowly. $Q =20$ and $\Sigma=0$ throughout.}
\label{fig:gamma_phases}
\end{figure}

\begin{figure}
\centering
\includegraphics[scale=.6]{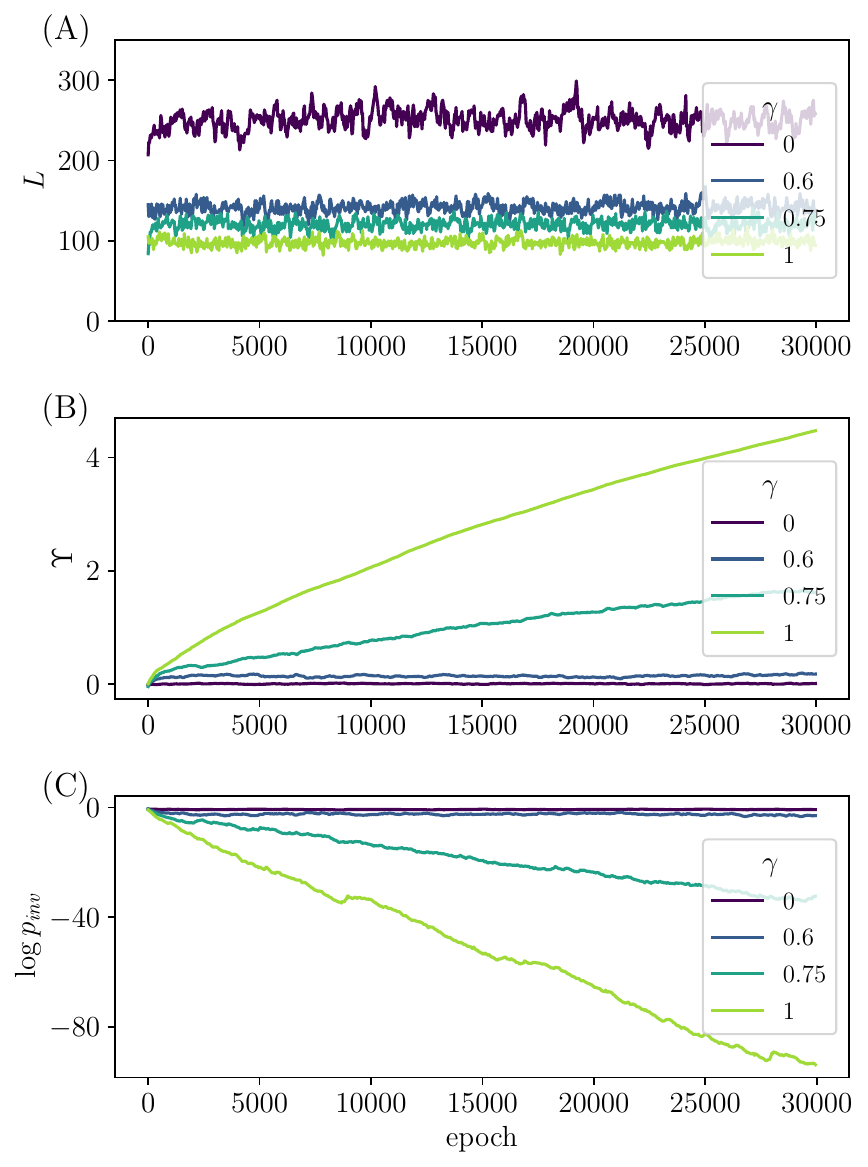}
\caption{Evolution in LV model ($Q=20$, $\Sigma=0$) undergoes a transition from Red Queen for $\gamma<\gamma_c\cong 0.65$ to oligarch phase for $\gamma>\gamma_c$. (A) $\LL$ is roughly constant in both phases. (B) $\Upsilon$ fluctuates around steady state in Red Queen phase but increases without bound in oligarch phase (C) $\pinv$ achieves a steady state in Red Queen phase but is ever-decreasing in oligarch phase.}
\label{fig:gamma_phases_L}
\end{figure}

{\it Other eco-evolutionary phases in well-mixed systems:} 
What other long term states occur naturally and robustly in eco-evolutionary models with no spatial structure or externally varying environment? For models with perfect symmetry, such as the $\kappa=1$ CR and linearized models or the $\gamma=1$ LV model, the presence of a Lyapunov function means that the community can continue to evolve and turn over but will do so more and more slowly. Does such slowing evolution occur away from special slices of parameter space? In the CR and linearized models with a convex Lyapunov function, the Red Queen phase appears to replace such a state as soon as $\kappa<1$. 
However in the LV model we have found evidence for a sharp (in the limit of large $Q$) qualitative change as a function of $\gamma$, around $\gamma_c\approx0.65$ for $\Sigma=0$. For $\gamma<\gamma_c$, the Red Queen phase obtains and there is continual turnover with the mean general fitness, $\Upsilon$, and the invasion probability, $\pinv$, roughly constant. However for $\gamma>\gamma_c$, we find no steady state. This transition is not apparent in the dynamics of the diversity, which is still roughly steady over time, but shows up strikingly in other properties.

For $\gamma>\gamma_c$, the dynamics continually slows down, with $\Upsilon$ ever-increasing and $\pinv$ thereby decreasing so that the community becomes progressively harder to invade. This behavior, which we have dubbed an {\it oligarch} phase, has surprising features that we do not yet understand.
Most strikingly, the evolutionary dynamics show that a few strains---the oligarchs---become far more abundant than the others. 

Figure~\ref{fig:gamma_phases} shows the abundance dynamics in both the Red Queen and oligarch phases, and Figure~\ref{fig:gamma_phases_L} the corresponding trajectories of $\LL$, $\Upsilon$ and $\pinv$. For $\gamma=0$ and $0.6$, the constant turnover of the Red Queen phase is evident in the narrow distribution of strain abundances and lifetimes. However for $\gamma=0.75$ and $1$, an $\Oh(1)$ fraction of abundance ``condenses" into an $\Oh(1)$ number of oligarch strains which still slowly turn over.

Qualitatively, the oligarch phase appears to occur due to feedback between the abundant strains and the invaders for larger $\gamma$: if a prospective invader, $\aa$, is to successfully enter the community it must have a positive contribution to its invasion fitness from an abundant strain $\ell$, i.e. $\VV_{\aa \ell}>0$. Then since $\gamma$ is strongly positive, the invader will tend to have $\VV_{\ell\aa}>0$ as well, thus increasing the abundance of the oligarch strain $\ell$. However this argument does not explain the turnover of oligarchs, nor their absence in the linearized model. The non-steady-state nature of the oligarch phase makes analysis difficult since correlation and response functions of strain abundances in evolutionary time depend on two times, not just time differences.

The transition between the Red Queen and oligarch phases (discussed in Appendix~\ref{app:lv_evo}) appears to be first order-like. As then expected, it persists for at least small general fitness variations with $\gamma_c$ depending on $\Sigma$ in a way that we have not carefully investigated. Hysteresis near the transition is observed when $\gamma$ is suddenly reduced through $\gamma_c$ (Figure~\ref{fig:hysteresis}). 
Strikingly, if one evolves with $\gamma>\gamma_c$ for some time, and then rapidly quenches to $\gamma<\gamma_c$, one can obtain a steady state which exhibits aspects of both the Red Queen and oligarch phases: a handful of high-abundance strains persists at roughly constant abundance without turning over, while the majority of strains turn over steadily as they would in the ``normal" Red Queen phase. In this coexisting state between the two phases, $\Upsilon$ and $\pinv$ are roughly constant although remain respectively elevated and suppressed relative to their values in the Red Queen phase, while $\LL$ does not show such a signature (Figure~\ref{fig:hysteresis}). Further analysis of such a steady---and therefore more analytically tractable---oligarch state could lead to a better understanding of the oligarch phase as well as the transition between the two eco-evolutionary phases.

What classes of models can exhibit an oligarch phase? 
Naively, the heuristic argument above would suggest an oligarch phase in the linearized model, though such a phase does not seem to exist. This difference in behaviors is likely related to the properties of the {\it totally symmetric} LV and linearized models: the linearized model always has a {\it convex} Lyapunov function whereas the LV Lyapunov function becomes non-convex for sufficiently large $\SS$~\cite{biroli2018marginally}. What other phenotype models---perhaps with more general interaction via chemicals---can display an oligarch phase of eco-evolutionary dynamics? Given aforementioned subtleties regarding the tails of the general fitness distribution and the long-lived strains with large $\ss_i$, one might also ask how the tails of the distribution of the elements of $\VV$ affect the oligarch phase.

If some fraction of the $\DD$ resource-mediating chemicals have negative correlations between the corresponding columns of $\AA$ and $\BB$, it is possible that there is a $\DD\to\infty$ limit, with carefully tuned correlations between $\AA$ and $\BB$, for which the correlations across the diagonal of the interaction matrix $\VV = - \AA\BB^\intercal$ remain of order unity, with the diagonal terms larger by a factor analogous to the LV model's $Q$. If this were the case, one would recover an LV model with $\gamma\neq0$, which could then display an oligarch phase. 
Regardless, the presence of the oligarch phase in the LV model illustrates the range of phenomenology that can occur in different models, with each distinct eco-evolutionary phase robust over a range of parameters.

{\it Parent-mutant correlations:} 
In this work, we have primarily studied a process that is not strictly speaking evolution: new strains that are introduced are unrelated to extant strains. Nevertheless, this process is crucially different from well-studied models of top-down assembly. In our pseudo-evolutionary process, strain extinctions are permanent, which qualitatively changes the shape of the abundance distribution: in steady state few strains with low abundance exist as these are most easily driven extinct by invasion-induced perturbations. 

We have shown that introducing mutant strains related to extant parent strains (with correlation $\rho>0$) results in very similar behavior if the mutant and parent are not strongly correlated phenotypically. 
However if the effects of mutations are small---surely relevant biologically~\cite{goyal2022interactions}---diversification is slower because of the increased importance of general fitness differences. This regime raises important questions: How do the existence and properties of a diverse Red Queen phase depend on the magnitude of the mutational effects when these are very small---i.e. in the limit $\rho\nearrow 1$? What happens with a distribution of mutational effect sizes? What is the interplay between general fitnesses and parent-mutant correlations? What types of phylogenies are generated by the Red Queen eco-evolutionary process? Do they resemble any known class of coalescent trees? 
We will address these issues in a future paper.

{\it Changing environments:}
The CR and linearized resource models naturally allow study of how evolving communities are affected by {\it extrinsic changes} in the environment. During the initial transient period of evolution before the Red Queen phase obtains, the community in the CR model adapts to its resource supply by aligning the growth vectors of strains (rows of $\GG$) with the resource supply vector $\vec \KK$. 
 If the supply vector $\vec \KK$ changes extrinsically after this initial period of evolution, the distribution of $\UU_i$ will broaden and the diversity will crash, only to rebuild as the community readapts to the new resource supply vector. This scenario can be studied in the linearized model if we parameterize the general fitnesses through an interaction with the environment, and define the resource vector as $\RR_\alpha = k_\alpha-\sum_j \BB_{j\alpha}\nu_j$, with $\bf k$ proportional to deviations of the resource supplies from some reference values. 
This parameterization gives rise to selective differences $\ss_i = \sum_\alpha \AA_{i\alpha} k_\alpha$: therefore changing $\bf k $ will affect the $\ss_i$.

What do the eco-evolutionary dynamics look like for a continually changing supply vector, either driven extrinsically or through some further feedback from the ecological community? How can this externally driven continual evolution be distinguished from the self-driven Red Queen phase? 
 
{\it Clonal interference:}
In this work we have taken advantage of a separation of timescales between evolution and ecology. However such a separation is not at all guaranteed to exist, particularly in host-pathogen systems~\cite{yoshida2007cryptic}, in large microbial populations with abundant beneficial mutations~\cite{good2017dynamics}, or when mutational effects are small enough that ecological timescales are slow. Investigation of the dynamics of ecological models in a ``clonal interference'' regime, with multiple ecological mutants emerging before any single one of them has reached its fixed point abundance, is an open challenge of definite relevance to microbial populations.

{\it Spatial structure:}
Spatial structure and transport remains an understudied---although crucial---feature of ecological and evolutionary models. Gut and skin microbiota associated with different individuals are realizations of distinct ecological communities coupled by person-to-person migration, with new mutants arising locally and having the potential to spread across many different people~\cite{valles2023person,zhou2020host}. 

We have found that even without spatial structure, the eco-evolutionary behavior of stable communities is quite rich. However generalizing our analyses to spatial-mean-field-like ``island" models with all-to-all migration---roughly like microbes from person-to-person---is a very interesting direction, as are extensions to two or three spatial dimensions.
Can a diverse Red Queen phase exist with only migration of strains between spatial patches, but no evolution? Or is evolution required to keep generating new types and produce turnover of biodiversity? If the potential number of niches ($\DD$ or roughly $Q^2$ in our models) is not large enough for all the strains in the system to coexist locally, then different strains could still inhabit different spatial locations. Is this heterogeneity maintained when strains migrate between spatial patches? When evolution continually produces new variants? How much persistent diversity can be maintained, and how does this grow with the spatial size or number of microbiome hosts?

{\it Other spatiotemporal eco-evolutionary phases:} 
The range of models we have studied in this work hints at the generality of the Red Queen phase in ecological models with stable communities. However, with spatial structure there is another ecologically diverse phase known: a state of spatiotemporal ecological chaos in which the spatial structure stabilizes the fluctuating ecological dynamics against global extinctions, leading to an ecological state that is exponentially long-lived in its spatial size~\cite{pearce2020stabilization}. In previous work~\cite{mahadevan2023spatiotemporal} we have studied gradual evolution and assembly in such a spatiotemporally chaotic population. In the simplest predator-prey-like models that yield this phase, there are no niches created by different resources: the model corresponds to a $Q=0$ Lotka-Volterra model with negative correlations across the diagonal ($\gamma<0$). However the dynamical nature of the ecological phase allows one to pack unlimited diversity into a single ``niche,'' even when strains are added gradually through evolution. In the presence of general fitness differences, diversification to arbitrarily large $\LL$ can still occur, but the evolution continually slows down, with $\pinv$ becoming smaller and smaller as the general fitnesses (and hence $\Upsilon$) continue to increase. 

We can understand the tendency of these host-pathogen-like models towards unlimited diversity from the phase diagram of the top-down assembled community. If one starts in the unstable phase, then increasing the diversity does not bring one closer to the phase boundary between stable and unstable phases; in fact it takes one further from this boundary. There is no natural diversity scale (like $\DD$ in the resource models) and no divergent fragility that would limit the community's diversity. 
By contrast, for the stable dynamics studied here, the presence of sufficiently strong niche interactions sets a scale for the diversity which is $\overline\LL\sim\DD$ in the CR and linearized models, and $\overline\LL\sim Q^2$ in the LV model. This niche structure qualitatively changes the outcome of the evolutionary dynamics. Nevertheless, one can ask: is it possible that a spatiotemporally chaotic phase could exist---and perhaps enable more persistent diversity---also in resource models?

{\it Diversity across phylogenetic scales:}
Experimental data have shown that strain dynamics can differ based on the resolution at which they are observed, with closely related strains undergoing larger fluctuations than family-level groupings~\cite{goyal2022interactions,martin2018high}. Further study of ecological dynamics across scales of diversity is needed. How---hopefully already in simple models---does evolution create diversity over many phylogenetic scales, and how does this diversity interact with the ecological dynamics across such scales? Can evolution robustly lead to ecologically chaotic dynamics of micro-diversity but stable dynamics at coarser levels of differentiation?

{\it Prospects}: The models and scenarios described in this work provide a simple framework and examples in which ecology produces an evolutionary Red Queen phase. The analytical tractability of the models in the simplest case---and the generality and robustness of the Red Queen phase we have found---gives hope that the study of simple models might advance our understanding of the evolution and maintenance of fine-scale biodiversity in natural microbial communities.

\acknowledgements
We thank Gabriel Birzu for illuminating discussions. This research was supported in part by the National Science Foundation (NSF) via NSF PHY-2210386 to DSF. AM and DSF acknowledge support from the Kavli Institute for Theoretical Physics (KITP) through the Gordon and Betty Moore Foundation Grant No. 2919.02, NSF PHY-1748958 and PHY-2309135, the Heising-Simons Foundation, and the Simons Foundation (216179, LB). We thank the Stanford Sherlock cluster for computing resources.


\appendix

\section{Structure of CR model}
\label{app:parameters}

{\it Lyapunov function}: The Lyapunov function of the CR model which exists when $\GG=\FF$ ($\kappa=1$) is
\begin{subequations}
\begin{align}
\Lyap &= -\sum_\alpha \left[\KK_\alpha \log\frac{\KK_\alpha}{\rr_\alpha}-(\KK_\alpha-\rr_\alpha)\right]\\
&\text{with }\rr_\alpha = \frac{\KK_\alpha}{\omega+ \sum_{j=1}^\SS \FF_{j\alpha}n_j }.
\end{align}
\end{subequations}
$\Lyap$ is a convex function of the $\{n_i\}$ and increases under the ecological dynamics of Equation~\ref{eq:CR_dyn} and hence also under the evolutionary dynamics even with extinctions, as extinct strains do not contribute~\cite{tikhonov2017collective}.

{\it Non-dimensionalizing}: In order to study the CR model in more detail, it is useful to identify the combinations of parameters that matter for its fixed points. We make the choice $\omega=0$, which does not qualitatively change the behavior, but simplifies the model. Then we define $\tilde \rr_\alpha = \rr_\alpha \mu_g/\dd$, $\tilde \sigma_g = \sigma_g/\mu_g$, $\tilde\GG = \GG/\mu_g$, $\tilde \sigma_f = \sigma_f/\mu_f$, $\tilde\FF = \FF/\mu_f$ and $\tilde \KK_\alpha = \mu_g \KK_\alpha/\dd\mu_f$, such that Equation~\ref{eq:CR_dyn} becomes 
\begin{subequations}
\begin{align}
 \frac{d n_i}{d(mt)} &= n_i\left(\sum_\beta \tilde\GG_{i\beta} \tilde \rr_\beta-1\right)\\
 \frac{d\tilde \rr_\alpha}{d(\mu_f t)} &= \tilde\KK_\alpha-\tilde \rr_\alpha \sum_j \tilde\FF_{j\alpha} n_j.
 \end{align}
 \end{subequations}
Therefore $\kappa$, $\DD$, $\tilde\sigma_g$, $\tilde \sigma_f$ and the rescaled supply vector $\tilde {\vec K}$ are the only quantities that matter for the fixed point abundances. We can set $\mu_f=\mu_g=1$ and $\dd=1$ without loss of generality. The fixed point abundances $\{ n_j^*\}$ and $\{ \rr_\alpha^*\}$ satisfy the equations
\beq \sum_\beta \GG_{i\beta} \rr_\beta^* = 1 \ \ \forall i\in\Comm;\quad \rr_\alpha^* = \frac{\KK_\alpha}{\sum_j \FF_{j\alpha}n^*_j}.\eeq
We do not expect that the variance of the components of $\vec\KK$ plays an important role, so we set $\KK_\alpha=\KK$. The value of $\KK$ matters only insofar as it determines the scale of the $\{n_j^*\}$, which can be rescaled to eliminate $\KK$. 

\section{Parent-mutant correlations in CR model}
\label{app:CR_corr}

\begin{figure*}
\centering
\includegraphics[scale=.6]{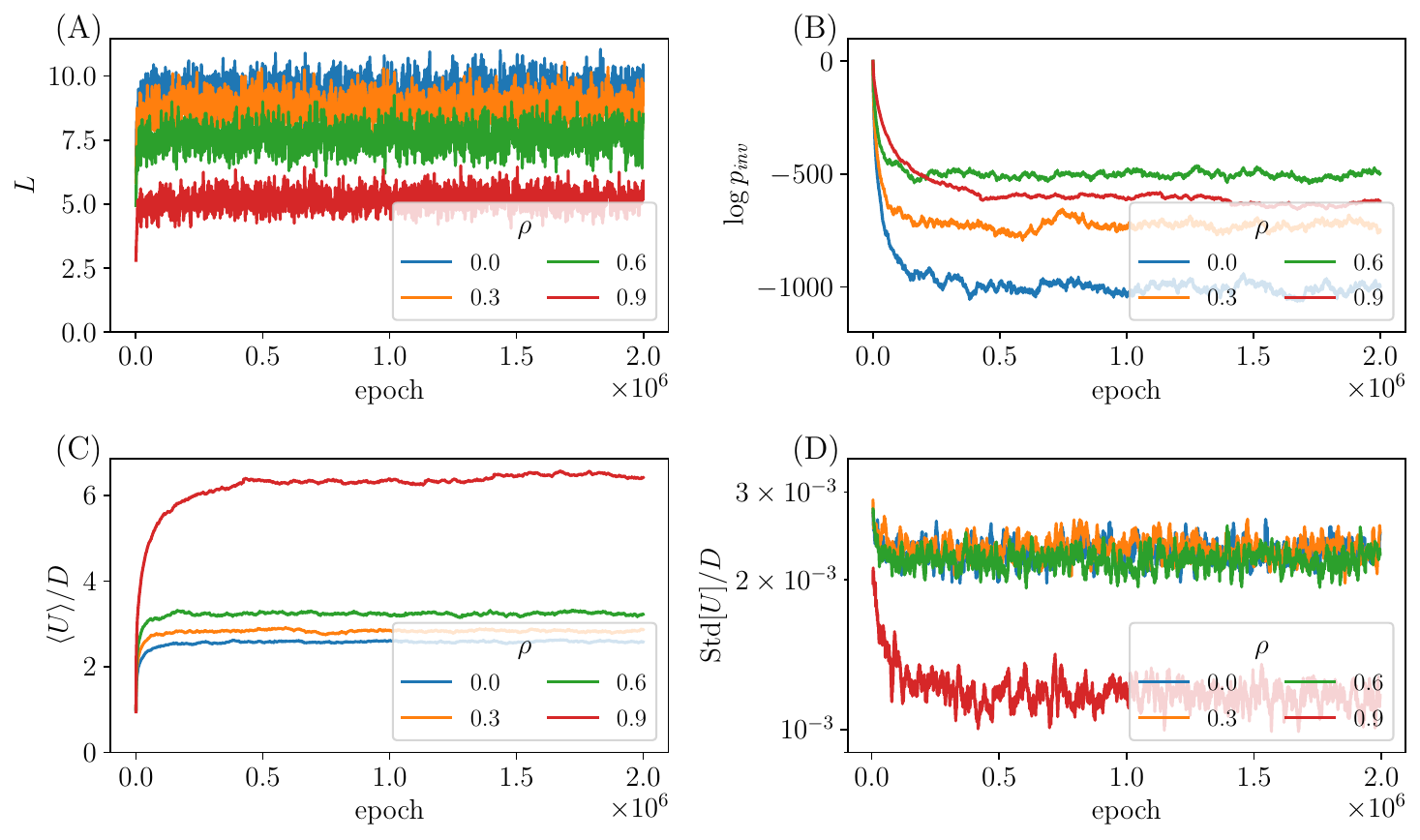}
\caption{Effects of parent-mutant correlation, $\rho$, on evolution in CR model with $\DD=50$, $\kappa=0.4$. (A) Steady state $\overline\LL$ is a decreasing function of $\rho$. (B) $\overline{\log\pinv}$ is a nonmonotonic function of $\rho$, largest for intermediate $\rho$. (C) After initial transient, community-mean general fitness $\langle\UU\rangle$ saturates at a steady state value. $\overline{\langle \UU\rangle}$ increases with $\rho$ as distribution of extant $\UU_i$ pushes farther into tail. (D) $\overline{\Std[\UU]}$ decreases with $\rho$.}
\label{appfig:resource_corr}
\end{figure*}

We can draw mutant phenotypes to be correlated with the phenotype of a parent, which is chosen with probability proportional to its abundance for each epoch. Given the parent phenotype---the vectors $\boldsymbol\GG_P$ and $\boldsymbol\FF_P$---we draw a mutant phenotype $\boldsymbol\GG_M$ and $\boldsymbol\FF_M$ which has elementwise Pearson correlation $\rho$ with the parent phenotype. We also enforce elementwise correlation $\kappa$ between $\boldsymbol\GG_M$ and $\boldsymbol\FF_M$. In Figure~\ref{appfig:resource_corr} we show numerical results from this process. The trends are similar to what we observe when we vary $\rho$ in the linearized model and, importantly, show that there is still a Red Queen steady state across a range of $\rho$, with the phase qualitatively similar to the $\rho=0$ phase studied in depth. Of particular phenomenological interest is the dependence of $\overline{\log\pinv}$ on $\rho$, which displays a nonmonotonicity. We deliberately stay away from the limit $\rho\to1$, which is left for future work as additional subtleties arise.

\section{Relation between CR and linearized models}
\label{app:consistency}

\begin{figure}
\centering
\includegraphics[scale=.6]{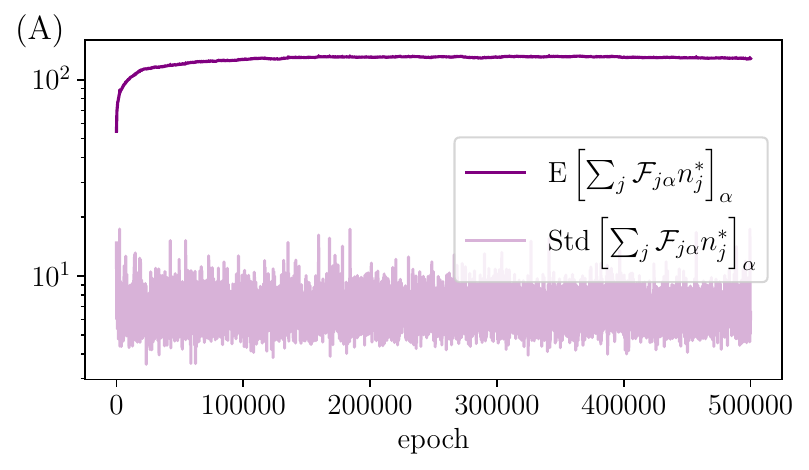}
\includegraphics[scale=.6]{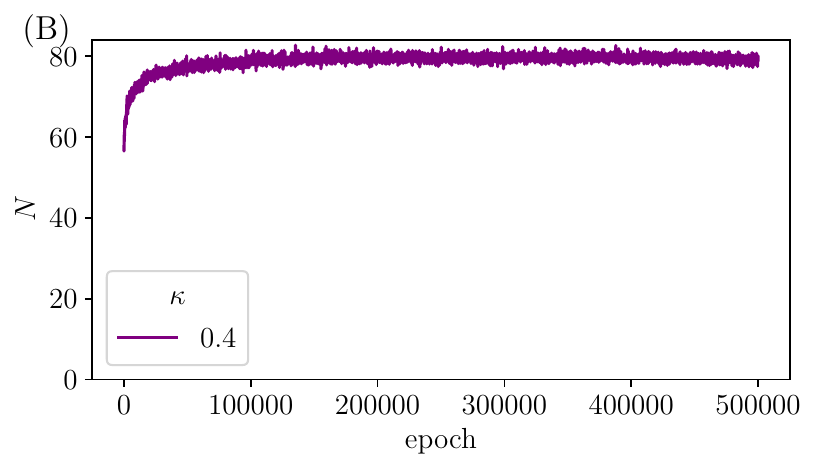}
\caption{In Red Queen phase of the CR model ($\DD=50$, $\kappa=0.4$), linearization of resource abundances is justified, and total consumer population is roughly constant. (A) At steady state, consumption rates of resources, $\sum_j \FF_{j\alpha} n^*_j$, have a narrow distribution, justifying expanding variations in this quantity out of the denominator. $\KK_\alpha=1~\forall~\alpha$, so weighting means and standard deviations by $\KK_\alpha$ does not affect results. (B) The total consumer population at the fixed point, $N =\sum_j n^*_j$, is roughly constant in the Red Queen phase, motivating switching to fractional abundances. }
\label{appfig:resource_width}
\end{figure}

We here connect the CR model to the linearized model, which we have argued captures the behavior of the CR model and is easier to analyze.
The linearization of $\boldsymbol\rr^*$ as a function of the $\{ n^*_j\}$ is justified when the mean of $\sum_j \FF_{j\alpha}n^*_j$ over $\alpha$ (weighted by the $\KK_\alpha$) is larger than its variations. 
In Figure~\ref{appfig:resource_width}A we show that this condition is met in the Red Queen steady state for $\kappa=0.4$. Figure~\ref{appfig:resource_width}B shows that the total consumer population is roughly constant in evolutionary time, which motivates a description in terms of fractional abundances for the linearized model. We carry out an expansion which allows us to separate out general fitness and interaction contributions to each strain's growth rate. 

In the Red Queen steady state the means of $\FF$ and $\GG$ are greater than $1$ due to evolutionary conditioning---for example in Figure~\ref{fig:CR_U}A, the plot for $\kappa=0.4$ shows that the mean of $\GG$ plateaus at $\langle\UU\rangle/\DD\approx2.5$. The mean of $\FF$ is smaller than that of $\GG$ but still larger than $1$. If we define $\langle\FF\rangle \equiv \frac{1}{N\Ktot}\sum_{j,\beta}\FF_{j\beta}n^*_j\KK_\beta$ and $f_{i\alpha} = \FF_{i\alpha} - \langle \FF\rangle$, then we have $\sum_j \FF_{j\alpha}n^*_j=\langle\FF\rangle N + \sum_j f_{j\alpha} n^*_j$. Defining $\NN = \langle \FF\rangle N$, we can Taylor expand the resource abundances, yielding
\beq \rr_\alpha^* \approx \frac{\KK_\alpha}{\NN} \left(1-\frac{\sum_j f_{j\alpha} n^*_j}{\NN}\right),\eeq
where our definition of $\langle\FF\rangle$ ensures that $\frac{1}{\DD}\sum_\alpha\rr_\alpha^*=\Ktot/\NN\DD$.
If we approximate this expansion as an equality, at the fixed point the condition on the extant strains that their growth rates vanish becomes
\begin{subequations}
\begin{align}
&0=\sum_\beta \GG_{i\beta} \frac{\KK_\beta}{\NN}\left(1-\frac{\sum_j f_{j\beta} n^*_j}{\NN}\right)-1\text{ for } i\in\Comm\\
\implies&0=\frac{1}{\NN}\sum_\beta g_{i\beta} \KK_\beta-\frac{1}{\NN^2}\sum_{\beta,j} g_{i\beta} \KK_\beta f_{j\beta} n^*_j\nonumber\\
&-\frac{\langle \GG\rangle}{\NN^2} \sum_{\beta,j} \KK_\beta f_{j\beta}n^*_j + \frac{\langle\GG\rangle}{\NN}\sum_\beta \KK_\beta -1\text{ for } i\in\Comm,\label{eq:CR_decomposition}
\end{align}
\end{subequations}
where we have similarly defined $\langle\GG\rangle\equiv \frac{1}{N\Ktot}\sum_{i,\beta}\GG_{i\beta}n^*_i\KK_\beta$ and $g_{i\beta} = \GG_{i\beta} - \langle\GG\rangle$ so that the average of $\sum_\beta g_{i\beta}\KK_\beta$ over $i$ (weighted by $n^*_i$) vanishes.
This form allows us to separate out the general fitness and interaction pieces in the growth rate of each strain. On the right hand side of Equation~\ref{eq:CR_decomposition}, the first term contains the general fitnesses and is proportional to $\UU_i-\langle \UU\rangle$. The second term is the contribution to the growth rate of strain $i$ from the ecological interactions, and the last three terms are $i$-independent and constitute an effective Lagrange multiplier that keeps $N$ roughly constant when $\langle \GG\rangle\gtrsim \Oh(1)$.

In the assembled community, $\NN = \Oh(\LL)$ and the $f_{j\alpha}$ and $g_{i\beta}$ have width $\sigma_f$ and $\sigma_g$ respectively. Therefore the width of the general fitness term scales as $\sigma_g \sqrt \DD/\LL$ and the width of the interaction piece scales as $\sigma_f \sigma_g \sqrt{\LL\DD}/\LL^2$.
Since $\LL = \Oh(\DD)$ in an assembled community for sufficiently large $\SS$, with each of the $n_j^*$ being $\Oh(1)$, the distribution of general fitnesses is a factor of $\sqrt\DD/\sigma_f$ wider than the distribution of drives. Since $\sigma_f$ and $\sigma_g$ are $\lesssim1$ in order to make the majority of entries in $\FF$ and $\GG$ positive, the general fitnesses dominate in the assembled community.
However when the community evolves, the $\boldsymbol\GG_i$ vectors elongate in the direction of the supply vector $\vec\KK$. As they push into their gaussian tail, they decrease the width of the distribution of extant $\sum_\beta g_{i\beta} \KK_\beta/\NN$, which is governed by the variations of the $\vec g_i$ vectors in the direction parallel to $\vec\KK$. However the width of the extant drive distribution does not decrease and this allows the distributions of general fitnesses and growth rates from ecological interactions to attain comparable widths in an evolved community with $\LL=\Oh(\DD)$ (Section~\ref{sec:CR_evo}).

\section{Algorithm to find fixed point}
\label{app:numerics}

Integrating the ecological dynamics to find their stable fixed point is too slow to observe many of the features of the long-term evolution of interest. In order to probe longer evolutionary trajectories, we have used a heuristic method to find the fixed point; we then check that it is stable and cannot be invaded by any of the extinct strains. Below we describe the algorithm for the linearized and LV models, and a modification of the algorithm for the CR model. For the LV and linearized models, given an interaction matrix, the procedure can be modified to include selective differences by adding $\ss_i$ to the $i$th row of $\VV$: for ease of notation we refer to this effective interaction matrix simply as $\VV$. The main challenge is to find the set $\Comm$ of strains in the community that has nonzero abundance at the stable uninvadable fixed point. We accomplish this with an iterative method that converges most of the time.

\begin{enumerate}[itemsep=-1ex]
\item Initialize $\Comm$ as the set of all strains in the community (including the new invader) and initialize the extinct set $\Ext$ as an empty set. $\VV^*$ is the submatrix of interactions for the extant strains, and therefore contains the rows and columns corresponding to strains in $\Comm$.
\item Invert $\VV^*$ to find a fixed point of the dynamics via $\nu^*_i = \Upsilon^*\sum_j{\VV^*}^{-1}_{ij}$ and $\sum_i\nu^*_i = 1$ which fixes $\Upsilon^*$. In general some of the $\nu^*_i$ will be negative. Remove the set $\{i|\nu^*_i<0\}$ from $\Comm$ and add it to $\Ext$.
\item Calculate the invasion fitnesses $\xi_j$ of each strain in $\Ext$ as $\xi_j = \sum_{k\in\Comm} \VV_{jk}\nu^*_k-\Upsilon^*$. Transfer the set of strains that could invsade, $\{j|\xi_j>0\}$, from $\Ext$ to $\Comm$.
\item Iterate steps $2$ and $3$ until convergence to sets $\Comm$ and $\Ext$ such that $\nu^*_i>0$ for all $i\in\Comm$ and $\xi_j<0$ for all $j\in\Ext$.
\end{enumerate}
One can then check stability of the final fixed point by computing the Jacobian of the dynamical equation. The vast majority of the time, the fixed point found in this way is stable.

When the algorithm does not converge within $10$ iterations, we stop adding extinct strains with positive invasion fitnesses back into the community, and thereby converge to a fixed point with positive abundances, which is {\it invadable} by some of the strains that were declared extinct. This adjustment is most often deployed when $\Ext$ and $\Comm$ have similar sizes at the fixed point, which can happen for an initial assembled community with large $\SS$, or for the evolved state with small $\kappa$ where $\LL$ is small and each invader can cause $\Oh(\LL)$ extinctions. In these cases the iterative algorithm does not converge reliably. However for small perturbations to the community, as in the case of primary interest, $\Ext$ is much smaller than $\Comm$ and the algorithm converges rapidly. Results from the error-prone regimes of evolution do not play a substantial part in any of our conclusions.

In the CR model, one cannot simply invert a matrix to find the fixed point for a subset of strains due to the nonlinearity in the interactions. Therefore we use a numerical solver to find the fixed point for a given $\Comm$---the rest of the algorithm is the same as in the linearized and LV cases.

\section{LV evolution across range of $\gamma$}
\label{app:lv_evo}
\begin{figure}
\centering
\includegraphics[scale=.6]{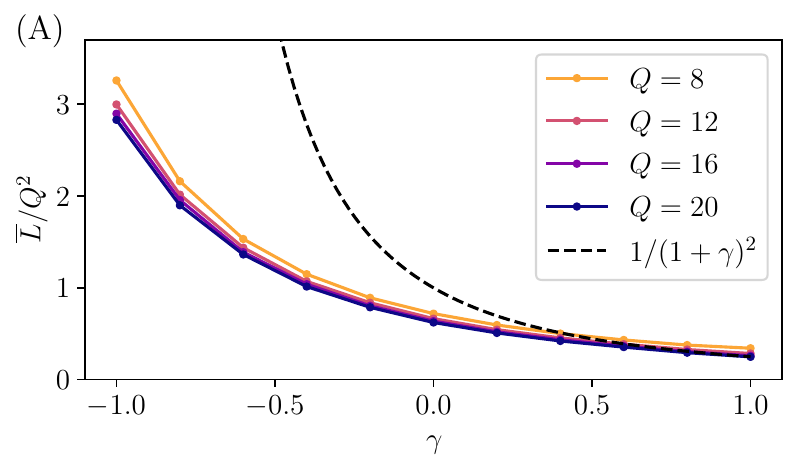}
\includegraphics[scale=.6]{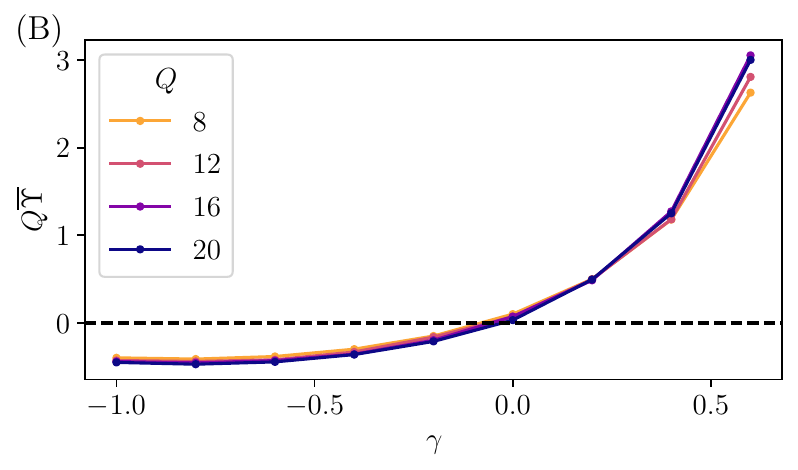}
\caption{Evolutionary dynamics of LV model (with $\Sigma=0$) depend on $\gamma$, with $Q$ setting an overall scale for both $\overline \LL$ and $\overline \Upsilon$. (A) Steady state $\overline\LL$ scaled by $Q^2$, in the LV model for $\Sigma=0$ as a function of $\gamma$. Dashed line shows $\Lmax/Q^2 = 1/(1+\gamma)^2$ from saturated assembled community~\cite{opper1992phase}. When $\gamma>\gamma_c\approx0.65$ the Red Queen phase breaks down, though $\LL$ remains roughly constant, and its average is plotted. (B) $Q\overline\Upsilon = \Oh(1)$ across a range of $\gamma<\gamma_c$ in the Red Queen phase. For $\gamma>\gamma_c$ there is no steady state, and $\overline\Upsilon$ is formally infinite. Dashed line shows $\Upsmax=0$ from saturated assembled community. }
\label{appfig:LV}
\end{figure}

As in the linearized model, the LV model (shown in Figure~\ref{appfig:LV} for $\Sigma=0$) exhibits a Red Queen steady state in which $\overline\LL$ is less than its maximal value in an assembled community, $\Lmax = Q^2/(1+\gamma)^2$, independent of $\Sigma$, as reviewed in Appendix~\ref{app:cavity}. The gap between $\Lmax$ and $\overline\LL$ diverges as $\gamma\to-1$, reflecting the fact that the fragility $\Xi$ is numerically larger for $\gamma\to-1$, and so it drives an average of one strain extinct per invasion for $\LL$ farther from $\Lmax$. For $\Sigma=0$ (as in these simulations) $\overline\Upsilon$ changes sign at $\gamma=0$, at which we have a quantitative understanding of the Red Queen phase (Section~\ref{sec:dyn_cavity}). As in the linearized model, we expect weak dependence of $\overline\LL$ on $\Sigma$, and $\overline\Upsilon\sim\Sigma^2$ for gaussian $\sdist(\ss)$ with large $\Sigma$.

\begin{figure*}
\centering
\includegraphics[scale=.6]{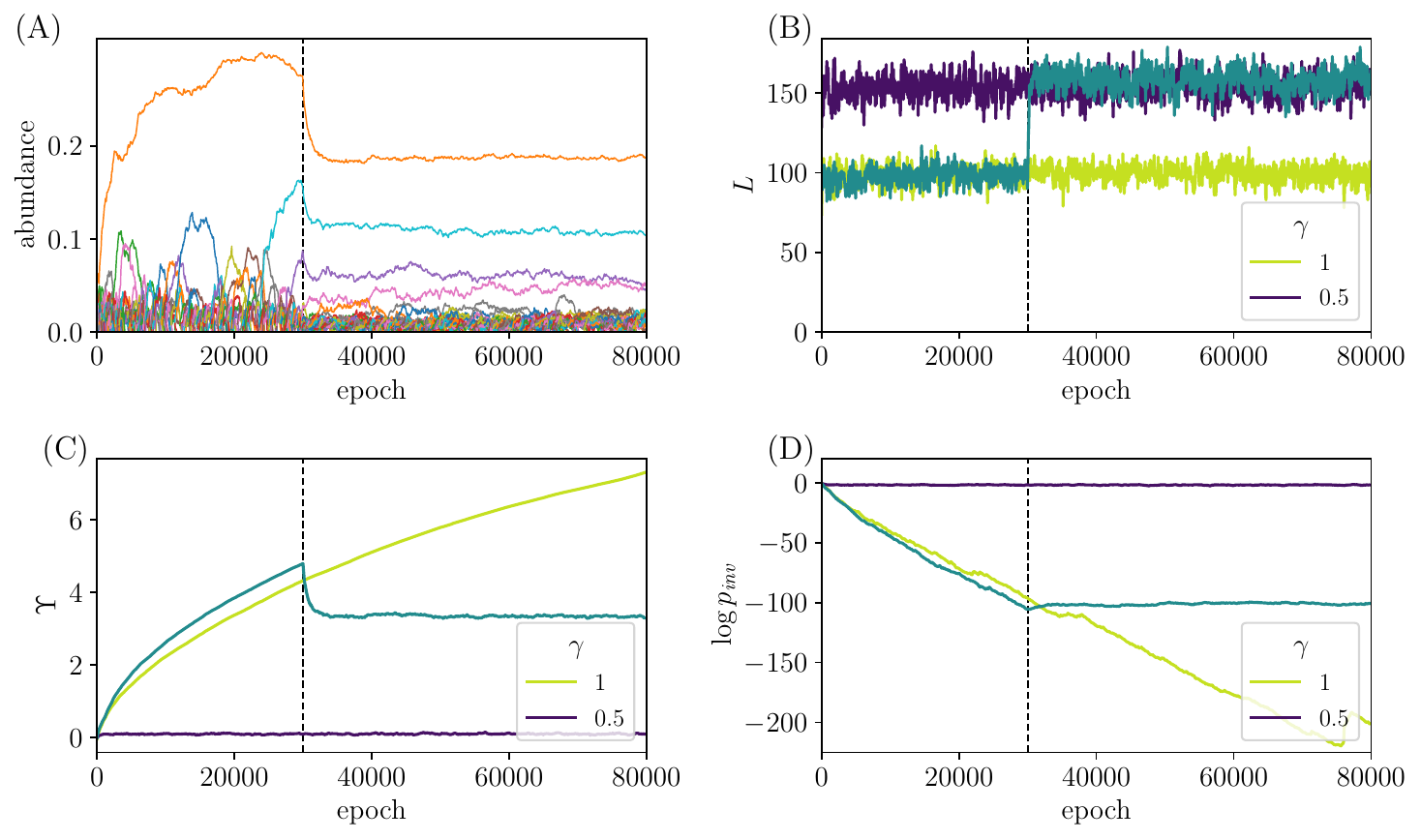}
\caption{Transition between Red Queen and oligarch phases is first order, and exhibits hysteresis and phase coexistence when the system is tuned across the transition, shown for $Q=20$, $\Sigma=0$. (A) Strain trajectories where initially $\gamma=1$, and after $3\times10^4$ epochs (at the vertical dashed line) $\gamma$ is ``quenched'' to $0.5<\gamma_c$. System does not revert back to the original steady state, instead showing several long-lived strains persisting. (B) The trajectory of $\LL$ over such an evolution (dark green), shown alongside $\LL$ for constant $\gamma=0.5$ and constant $\gamma=1$. The value of $\LL$ is indistinguishable in the post-quench phase (after a short transient) from the $\gamma=0.5$ Red Queen phase. (C)--(D) Dynamics of $\Upsilon$ and $\pinv$ show a strong departure after the quench from their values when $\gamma=0.5$ throughout.}
\label{fig:hysteresis}
\end{figure*}

In the oligarch phase, for $\gamma>\gamma_c$ with $\gamma_c\approx0.65$ for $\Sigma=0$, there are strains that grow to comprise an $\Oh(1)$ fraction of the population, and $\Upsilon$ continually increases with the community becoming harder and harder to invade. The width of the crossover between these two phases appears to sharpen for increasing $Q$ (not shown), indicating a sharp transition in the high-diversity limit $Q\to\infty$. The oligarch phase is not a steady state and $\Upsilon$ runs away to arbitrarily large values, but $\LL$ remains roughly constant, and it is this $\LL$ that is plotted as $\overline\LL$ for $\gamma>\gamma_c$ in Figure~\ref{appfig:LV}A.

The rate of increase of $\Upsilon$ jumps discontinuously up from zero through the transition. This discontinuity suggests a first order transition, motivating our investigation of hysteresis around the critical $\gamma$. One can start the evolution with $\gamma>\gamma_c$, and, after some period of time, reduce $\gamma$ to below $\gamma_c$ for the newly invading strains. Figure~\ref{fig:hysteresis}A shows strain abundance dynamics during such a ``quench.'' Here $\gamma=1$ for the first $3\times 10^4$ epochs of evolution, after which the invading strains start to enter the community with $\gamma=0.5<\gamma_c$. The strains into which an $\Oh(1)$ fraction of the community have condensed start to decrease in abundance, but instead of going extinct and turning over, they reach an apparent steady state in which they fluctuate around their still-$\Oh(1)$ mean values. This behavior becomes rarer as the post-quench $\gamma$ is decreased, indicating a region of phase coexistence around $\gamma_c$. Figures~\ref{fig:hysteresis}B--D show the dynamics of $\LL$, $\Upsilon$ and $\pinv$ through this process, compared with the behaviors for constant $\gamma=0.5$ throughout the evolution and constant $\gamma=1$. If one looks only at $\LL$, the hysteresis is not apparent, with the post-quench $\LL$ similar to $\LL$ for a community that has always had $\gamma=0.5$. Is it possible that, as for $\gamma=0$, the fragility only depends on $\LL$ but not the abundance distribution? If so, the fragility could set $\LL$ in steady state with little dependence on the history. 

In contrast to $\LL$, $\Upsilon$ and $\pinv$ show clearly that the post-quench steady state is significantly different from a Red Queen state that has been evolving with $\gamma=0.5$ from its beginning. This post-quench steady state displays properties of both the oligarch phase and the Red Queen phase, and may therefore be a useful route to better understand the oligarch phase.

\section{Cavity calculation for assembled community}
\label{app:cavity}

Here we carry out the static cavity calculation for a top-down assembled community in the linearized model: our results and approach are similar to that of ref.~\cite{blumenthal2024phase}.
An analogous calculation for the LV model proceeds in a similar manner~\cite{bunin2017ecological,pearce2020stabilization,opper1992phase}, and we summarize it at the end of this section. We consider a large community of strains at its fixed point, and add in another strain and resource, treating these as a small perturbation and then enforcing self-consistency between the statistics of this newly added strain and resource and the rest of the community. We will use the convention that repeated indices are summed over, unless otherwise stated.

{\it Linearized model:}
The strains obey the dynamical equation
\beq \frac{\dot \nu_i}{\nu_i} = s_i + V_{ij}\nu_j -\Upsilon\eeq
for $j\in\{1,\dots,\SS\}$, where $V_{ij} =- \AA_{i\alpha}\BB_{j\alpha}$ for $\alpha\in\{1,\dots,\DD\}$ and $\EE[\AA_{i\alpha}\BB_{j\beta}]=\kappa\delta_{ij}\delta_{\alpha\beta}$ with $\Upsilon = \nu_i(s_i+V_{ij}\nu_j)$. Defining resources $\RR_\alpha =-\BB_{j\alpha}\nu_j$, the fixed point abundances satisfy
\beq 0 = s_i + \AA_{i\alpha}\tilde \RR_\alpha-\Upsilon,\quad \tilde \RR_\alpha = -\BB_{j\alpha}\tilde\nu_j,\eeq
where tildes indicate the fixed point values prior to adding a new strain and resource.
Now we add in a new strain and resource with fixed point abundances $\nu_0$ and $\RR_0$ respectively.
Defining quantities $\zeta_0 = \AA_{0\alpha}\tilde \RR_\alpha$ and $\phi_0 = -\BB_{j0}\tilde \nu_j$, we have
\beq 0 = s_0 + \zeta_0 + \AA_{0\alpha}\delta \RR_\alpha-\Upsilon,\quad \RR_0 = \phi_0-\BB_{j0}\delta \nu_j.\eeq
If we define ``local susceptibilities" $\loceta_\alpha = \frac{dR_\alpha}{d\phi_\alpha}$ and $\locsusc_i = \frac{d\nu_i}{d\zeta_i}$, then the changes in $\RR_\alpha$ and $\nu_j$ are $\delta\RR_\alpha = -\loceta_\alpha \BB_{0\alpha}\nu_0$ and $\delta \nu_j = \locsusc_j \AA_{j0}\RR_0$. (Here repeated indices are not summed.)

Assuming self-averaging (distributions concentrating around their mean for large systems), the new strain and resource abundances are
\beq \nu_0 = \frac{\zeta_0+s_0-\Upsilon}{\kappa\globeta},\quad \RR_0 = \frac{\phi_0}{1+\kappa\globsusc},\label{eq:linearized_self_con}\eeq
where $\globeta = \sum_\alpha \loceta_\alpha$ and $\globsusc = \sum_i \locsusc_i$ are ``global " susceptibilities.

Since $\zeta$ and $\phi$ are weighted sums of the $\AA$ and $\BB$ matrix elements respectively, they are gaussian distributed with mean $0$. We will denote their variances as $\assembsig^2$ and $\sigma_\phi^2$ respectively.
Let us take the $s$ to be gaussian distributed with mean $0$ and variance $\Sigma^2$, and define $\sigma_\xi^2 = \assembsig^2 + \Sigma^2$ and $\YY = \Upsilon/\sigma_\xi$. We can now write self-consistency conditions for the variables $\sigma_\phi$, $\sigma_\xi$, $\globeta$, $\globsusc$ and $\YY$, which require that the distributions of the added strain and resource abundances are the same as the distributions in the original community: in other words, Equation~\ref{eq:linearized_self_con} is true for all strains, not just the cavity strain.
Defining partial moments
\beq \PM_m(\YY)=\frac{1}{\sqrt{2\pi}}\int_0^\infty {x}^m e^{-(x+\YY)^2/2}dx,\eeq
our self-consistency conditions are 
\begin{subequations}
\label{eq:gaussian_p_self_con}
\begin{align}
\globsusc&=\frac{S}{\kappa\globeta} \PM_0(\YY)\\
1&= \frac{S\sigma_\xi}{\kappa\globeta} \PM_1(\YY)\\
\sigma_\phi^2&=\frac{S\sigma_\xi^2}{\kappa^2\globeta^2}\PM_2(\YY)\\
\globeta&=\frac{D}{1+\kappa\globsusc}\\
\sigma_\xi^2&=\frac{D\sigma_\phi^2}{(1+\kappa\globsusc)^2} + \Sigma^2.
\end{align}
\end{subequations}
These equations can be reduced to 
\begin{subequations}
\label{eq:self_con_red}
\begin{align}
\globsusc &= \frac{\SS\PM_0(\YY)}{\kappa(\DD-\SS\PM_0(\YY))}\\
\globeta &= \DD-\SS\PM_0(\YY)\\
\sigma_\xi &=\kappa\frac{\DD-\SS\PM_0(\YY)}{\SS\PM_1(\YY)}.
\end{align}
\end{subequations}
along with an equation for $\YY$ alone:
\beq \frac{S \PM_2(\YY)}{\kappa^2 D} = 1-\left[\frac{\Sigma S \PM_1(\YY)}{\kappa(D-\SS\PM_0(\YY))}\right]^2.\label{eq:cavity_Z}\eeq
A solution to these equations give us the distributions of the strain and resource abundances in the community, as well as the susceptibilities $\globeta$ and $\globsusc$, and the scaled Lagrange multiplier $\YY$. This mean field solution is expected to describe the unique stable uninvadable fixed point as long as it is stable: in particular to any small random perturbation on each of the strains' growth rates.

To look at the nonlinear instability that signals the breakdown of the mean field theory, we add a random field $h_i$ to each of the biases and look at the change $\delta\xi_0$ in the bias of strain $0$. We find that
\beq \delta \xi_0 = \loceta_\alpha \AA_{0\alpha}\BB_{j\alpha}\locsusc_j \delta\xi_j + h_0,\eeq
where the sum is over both $\alpha$ and $j$.
Then in evaluating the mean square of both sides of this equation we have
\beq \EE\left[( \loceta_\alpha\AA_{0\alpha}\BB_{j\alpha}\locsusc_j \delta\xi_j)^2\right] = \sum_\alpha \loceta_\alpha^2\sum_j \locsusc_j^2\EE[\delta\xi_i^2].\eeq
Assuming that the strains are statistically identical, we can drop the strain indices to get
\begin{subequations}
\begin{align}
&\EE[\delta\xi^2] = \frac{\EE[ h^2]}{1-\sum_\alpha \loceta_\alpha^2\sum_j \locsusc_j^2}\\
\implies &\sum_i\EE[\delta\nu_i^2] =\frac{\sum_j\locsusc_j^2\EE[ h^2]}{1-\sum_\alpha \loceta_\alpha^2\sum_j \locsusc_j^2}=\Xi\times \EE[h^2]
\end{align}
\end{subequations}
where we have defined the {\it fragility} $\Xi$ as the factor relating the variance of the random fields to the sum of the squared changes in abundances that these produce. Since $\loceta_\alpha = \globeta/\DD$ and $\locsusc_j = \globsusc/\SS \PM_0(\YY)$ in the mean field solution, we can write the fragility in terms of $\LL = \SS\PM_0(\YY)$ as
\beq \Xi = \frac{\LL/\kappa^2\globeta^2}{1-\LL/\kappa^2 \DD} = \frac{\LL\DD}{(\DD-\LL)^2(\kappa^2\DD-\LL)}.\eeq

An instability signaling a phase transition occurs when the fragility diverges, at $\LL= \kappa^2 \DD$. We could have (more naturally) defined the fragility as the response of the strain abundances to adding a random field to the resource fields $\{\phi_\alpha\}$, but the form of divergence at the phase transition is the same in both cases (and in LV model, only the fragility we use here is defined).
The location of the divergence implies that the number of surviving strains is $\Lmax = \kappa^2 \DD$ at the transition. Substituting this relation into Equation~\ref{eq:cavity_Z} tells us that $\Ymax$ is the solution to
\beq \frac{\PM_2(\Ymax)}{\PM_0(\Ymax)} = 1- \left[\frac{\PM_1(\Ymax)\Sigma \kappa}{\PM_0(\Ymax)(1-\kappa^2)}\right]^2.\eeq
For $\Sigma=0$, the solution to this equation is $\Ymax=0$ and $\Smax= 2\kappa^2 \DD$. 
For $\Sigma\gtrsim1$, we have $\Ymax\approx\frac{\Sigma\kappa}{1-\kappa^2}$ and $\hat\sigma_\xi\approx\Sigma$, meaning that $\Upsmax\approx\frac{\kappa\Sigma^2}{1-\kappa^2}$ and $\Smax\approx \sqrt{2\pi}\kappa^2\DD \Ymax e^{\Ymax^2/2}\sim \exp[\frac{\kappa^2\Sigma^2}{2(1-\kappa^2)^2}]$. The reciprocal of $\Smax$ gives the invasion probability of the saturated community by an independent invader. The scaling of $\Upsmax$ and $\log \Smax$ with $\Sigma$ and $\kappa$ is reflected in the scaling forms of $\overline\Upsilon$ and $\overline{\log\pinv}$ in the Red Queen steady state.

Note that the linear susceptibility $\globsusc$ diverges when $\LL=\DD$. The nonlinear instability occurs for lower $\SS$ than the linear instability for $\kappa<1$, meaning that the cavity solution becomes invalid before it displays a linear instability. The nonlinear and linear instabilities occur in the same place for $\kappa=1$, marking the onset of a ``shielded'' phase when $\Sigma=0$~\cite{tikhonov2017collective}. With $\Sigma>0$, $\Smax$ diverges for $\kappa\to1$, indicating the marginality of the shielded phase which occurs only for $\Sigma=0$, $\kappa=1$. 

{\it LV model:}
We now summarize the cavity calculation for the LV model. In this case the cavity method yields, for the fixed point abundance of the cavity strain,
\beq \nu_0 = \frac{\zeta_0 + \ss_0 - \Upsilon}{Q-\gamma\globsusc}\eeq
where $\zeta_0 = \sum_j V_{0j}\tilde\nu_j$, and $\globsusc = \sum_i \locsusc_i = \sum_i\frac{d\nu_i}{d\zeta_i}$.
Note the similarity with the expression from the linearized model
\beq \nu_0 = \frac{\zeta_0+\ss_0-\Upsilon}{\kappa\DD-\kappa^2\globeta\globsusc},\eeq
(a consequence of Equations~\ref{eq:linearized_self_con} and~\ref{eq:self_con_red})
if one identifies $\kappa\sqrt\DD$ with $Q$, $\kappa^2$ with $\gamma$ and $\globeta\globsusc$ with $\sqrt\DD\globsusc$, which account for the integrated-out resources as well as the factor of $\sqrt\DD$ between the LV and linearized models.

The self consistency equations for the LV model with gaussian distributed $\ss_i$ of variance $\Sigma^2$, for unknowns $\sigma_\xi$, $\YY$ and $\globsusc$, are then
\begin{subequations}
\begin{align}
\globsusc&=\frac{S}{Q-\gamma\globsusc} \PM_0(\YY)\\
1&= \frac{S\sigma_\xi}{Q-\gamma\globsusc} \PM_1(\YY)\\
\sigma_\xi^2&=\frac{S\sigma_\xi^2}{(Q-\gamma\globsusc)^2}\PM_2(\YY)+\Sigma^2
\end{align}
\end{subequations}
where, as before, $\YY = \Upsilon/\sigma_\xi$. These equations can be reduced to
\begin{align}
\globsusc = \frac{Q-\sqrt{Q^2 - 4\gamma\SS\PM_0(\YY)}}{2\gamma}\\
\sigma_\xi = \frac{Q+\sqrt{Q^2-4\gamma\SS\PM_0(\YY)}}{2\SS\PM_1(\YY)}
\end{align}
along with an equation for $\YY$ alone:
\beq \left[\frac{Q+\sqrt{Q^2-4\gamma\SS\PM_0(\YY)}}{2}\right]^2= \SS\PM_2(\YY) + (\Sigma\SS\PM_1(\YY))^2.\eeq
The fragility can be calculated from local susceptibility $\locsusc_i = 1/(Q-\gamma\globsusc)$ in terms of $\LL = \SS\PM_0(\YY)$:
\beq \Xi = \frac{\LL}{(Q-\gamma\globsusc)^2-\LL}\eeq
implying a divergence of $\Xi$ at $\hat\LL = \frac{Q^2}{(1+\gamma)^2}$.

\section{Exponential distribution of general fitnesses}
\label{app:exp_s_dist}

In order to analyze the transient behavior of $\LL$ as $\Delta(\Upsilon)$ (the local scale of the $\ss_i$ distribution) decreases during the evolutionary transient, it is useful to investigate the behavior of the assembled community with an exponential distribution of positive $\ss_i$ given by $\sdist(\ss) = \frac{1}{\Delta}e^{-\ss/\Delta}\Theta(\ss)$.
We define the variable $x_i \equiv \ss_i + \zeta_i$. The self consistency conditions then deal with the distribution of $x_i$, which is
\beq p_x(x) = \frac{1}{\Delta} \exp\left[-\frac{x}{\Delta}+\frac{\assembsig^2}{2\Delta^2}\right]\Phi\left(\frac{x}{\assembsig} - \frac{\assembsig}{\Delta}\right),\eeq
where $\Phi$ is the standard normal CDF and $\assembsig^2$ is the variance of the drives in the {\it assembled} community.
If we define partial moments
\beq \Wtilde_m(\Upsilon,\assembsig) = \int_0^\infty y^m p_x(y+\Upsilon)dy\eeq
then the self consistency conditions can be written down in a similar fashion to those for the gaussian $\sdist(\ss)$ in Equation~\ref{eq:gaussian_p_self_con}.
For the LV model, these conditions are
\begin{align}
\globsusc &= \frac{\SS}{Q-\gamma\globsusc}\Wtilde_0(\Upsilon,\assembsig)\\
1&=\frac{\SS}{Q-\gamma\globsusc}\Wtilde_1(\Upsilon,\assembsig)\\
\assembsig^2 &= \frac{\SS}{(Q-\gamma\globsusc)^2}\Wtilde_2(\Upsilon,\assembsig).
\end{align}
For large $\Upsilon$, the lower limit of the exponential $\sdist(\ss)$ does not matter and the distribution $p_x(x)$ is roughly an unbounded exponential. Therefore we have
\beq \Wtilde_m(\Upsilon,\assembsig) \approx m!\times \Delta^m \exp\left[-\frac{\Upsilon}{\Delta} + \frac{\assembsig^2}{2\Delta^2}\right],\eeq
which allows us to write the self consistency conditions just in terms of the number of extant strains $\LL=\Wtilde_0(\Upsilon,\assembsig)$, regardless of $\SS$ and the normalization of $\sdist(\ss)$:
\begin{align}
\globsusc &= \frac{\LL}{Q-\gamma\globsusc}\\
1&=\frac{\Delta\LL}{Q-\gamma\globsusc}\\
\assembsig^2 &= \frac{2\Delta^2\LL}{(Q-\gamma\globsusc)^2}
\end{align}
In the large-$\Delta$ regime, we therefore have $\globsusc \approx 1/\Delta$, implying that to leading order in $\Delta$, the diversity is given by $\LL(\Delta) \approx Q/\Delta$. A similar analysis in the linearized model yields $\LL(\Delta)\approx \kappa\DD/\Delta$. 

As discussed in Section~\ref{sec:s_dist}, the analysis of the exponential $\sdist(\ss)$ also describes the steady-state behavior for $\sdist(\ss)$ decaying faster than exponentially, and $\Sigma$ large. In that case, $\Delta$ is a function of $\Upsilon$ and the distribution of $\ss_i$ has pushed far enough into the tail of $\sdist(\ss)$ to look locally exponential around $\Upsilon$.

Note that $\hat\Upsilon$ and $\hat\SS$ diverge for sufficiently large $\Delta>\Delta_c$ whether the critical $\Delta$, denoted $\Delta_c$, is proportional to---with an $\Oh(1)$ factor---$(1+\gamma)/Q$ in the LV model and $1/\kappa$ in the linearized model. This is because, as $\Delta$ increases, the maximum $\LL$---given by $\globsusc(Q-\gamma\globsusc)$ and $\globsusc\kappa\DD/(1+\kappa\globsusc)$ in the LV and linearized models respectively---becomes smaller than $Q^2/(1+\gamma)^2$ or $\kappa^2\DD$ respectively, and so $\Xi$ cannot diverge. Therefore for an exponential $\sdist(\ss)$, sufficiently large $\Delta$ eliminates the instability of an assembled community.

For understanding the evolutionary transients in models with faster-than-exponential tails (Section~\ref{sec:s_dist}), it is important to note that for $\Delta>\Delta_c$, as more strains are added, $\Upsilon$ continues to increase without the community otherwise changing.

\section{Dynamical cavity analysis in simplest case}
\label{app:dyn_cav}

Here we carry out the dynamical cavity analysis for the evolutionary steady state of the LV model with $\gamma=0$, $\Sigma=0$. In particular we calculate the temporal correlation function of the cavity strain in Equation~\ref{eq:cav_self_con} and derive self-consistency conditions to find the statistical properties of the evolution of strain abundances. Throughout, we suppress the index $0$ of the cavity strain.

{\it Solution for correlation function:}
Because of the absorbing boundary condition at $\nu=0$, we can use the solution of the Ornstein Uhlenbeck process (Equation~\ref{eq:drive_sde}) together with the method of images to write down the distribution of $\nu(\TT)$ given that it started at some initial value $\nu(\tinv)$ at invasion time $\TT=\tinv$. For simplicity we set $\tinv=0$: integrating the correlation function over $\tinv$---as we will later do---is the same as integrating over $\TT$. We use the fact that, for $\gamma=\Sigma=\overline\Upsilon=0$, we have $\nu = \zeta/Q$. Therefore we can write the distribution of $\nu(\TT)$ conditioned on $\nu(0)$ as 
\begin{widetext}
\begin{align}
&p^*(\nu(\TT)|\nu(0)) = \nonumber\\
&\frac{1}{\sqrt{2\pi}\sigma(\TT)} \left[\exp\left(-Q^2\frac{[\nu(\TT)-\nu(0) e^{-\bb \TT}]^2}{2\sigma(\TT)^2}\right)-\exp\left(-Q^2\frac{[\nu(\TT)+\nu(0) e^{-\bb \TT}]^2}{2\sigma(\TT)^2}\right)\right],\text{ where } \sigma^2(\TT) =\frac{\JJ}{\bb}(1-e^{-2\bb \TT}),
\end{align}
\end{widetext}
where we have used the notation $p^*$ to denote the probability density with an absorbing boundary condition at $\nu=0$: the integral of $p^*$ over $\nu(\TT)$ is the survival probability of the strain as a function of time $\TT$.
We calculate the expectation $\langle \nu(\TT)\nu(\TT+\tau)\rangle$ over the joint distribution $p^*(\nu(\TT),\nu(\TT+\tau)|\nu(0)) = p^*(\nu(\TT)|\nu(0))\times p^*(\nu(\TT+\tau)|\nu(\TT))$. We then integrate over all times $\TT\in[0,\infty)$ and over all $\nu(0)\in[0,\infty)$ distributed according to
\beq p(\nu(0))=\frac{r}{\sqrt{2\pi}\sigzeta}\exp\left[-\frac{Q^2\nu(0)^2}{2\sigzeta^2}\right]\Theta[\nu(0)],\eeq
where $\sigzeta = \sqrt{\JJ/\bb}$ is the standard deviation of the incoming drive distribution, and $r$ is the rate of attempted invasions. $r=2$ in our case since time is in units of successful invasions, and $\overline\Upsilon=0$ so each invader has probability $1/2$ of successfully invading. The integrals over $\TT$ and $\nu(0)$ correspond to summing over all the strains in the community and can be done exactly, yielding
\begin{subequations}
\begin{align}
\CF(\tau) &= \EE[\langle\nu(\TT)\nu(\TT+\tau)\rangle]_{\TT,\nu(0)} \\
&= \frac{r\sigzeta^2}{4Q^2\bb}(1 + 2\log2)e^{-\bb\tau}.
\end{align}
\end{subequations}
We now assert that the dynamics of the cavity drive are the same as those of the drives in the community. Since the cavity drive can fluctuate in sign, its correlation function is $\CF(\tau)= \sigzeta^2e^{-\bb\tau}$, from an unconditioned Ornstein Uhlenbeck process at stationarity. Equating our two expressions for $\CF(\tau)$ gives
\beq 1 =\frac{r}{4Q^2 \bb}(1 + 2\log2).\eeq
We can get another self-consistency condition by enforcing $\sum_j \nu_j=1$. Note that $\langle\nu(\TT)|\nu(0)\rangle = \nu(0)e^{-\bb \TT}$, where the expectation is taken with an absorbing boundary condition at $\nu=0$. Integrating this expression over the past and over the half-gaussian distribution of starting positions (as before) gives $r\sigzeta = Q \bb \sqrt{2\pi}$.

 The number of strains in steady state, $\overline\LL$, is determined by the fact that the survival probability up to time $\TT$ is $\langle \Theta[\nu(\TT)]|\nu(0)\rangle = \erf[Q\nu(0) e^{-\bb \TT}/\sqrt 2 \sigma(\TT)]$. Integrating over all past times and over the distribution of $\nu(0)$ gives $\overline\LL = \frac{r}{2\bb}\log2$. Now we have two self-consistency conditions for $\bb$ and $\sigzeta$, along with an equation for $\overline\LL$:
\begin{align}
r\sigzeta &= Q \bb \sqrt{2\pi}\\
4Q^2\frac{\bb}{r} &= 1+2\log2\\
\overline\LL &= \frac{r}{2 \bb}\log2.
\end{align}
We can solve these equations in terms of $Q$. The most easily testable prediction from numerics is
\beq \overline\LL =\frac{2\log 2}{1+2\log 2}Q^2 \cong 0.58 \times Q^2,\eeq
which agrees well with numerics for $\overline\LL$ across a range of $Q$ (Figure~\ref{fig:theory}A). We also find that $\sigzeta = \sqrt{2\pi}\lambda/Q$, and $\bb= \lambda r/Q^2$ where $\lambda = \frac{1+2\log 2}{4}$, implying that $\JJ = 2\pi r\lambda^{3}/Q^4$. Therefore the theory prediction for the correlation function of the cavity drive is
\beq \CF(\tau) = \frac{2\pi\lambda^2}{Q^2} \exp\left[-\frac{\lambda}{Q^2}r\tau\right].\eeq
As stated previously, $r=2$ since this makes the integral of the half-gaussian incoming drive distribution (and therefore the rate of incoming strains) equal to $1$.

{\it Lifetime distribution:}
The distribution of strain lifetimes can be computed using known results from the theory of Ornstein Uhlenbeck (OU) processes---which can also be derived from the method of images. A useful result~\cite{ricciardi1988first} is that the distribution of survival times $\tsurv$ for an OU process $\dot x = -\bb x +\sqrt{2\JJ}\eta(\TT)$ starting at position $x_0>0$ and ending when it hits the origin is
\beq
p_\tsurv(\tsurv|x_0) = 
\frac{2\bb^{3/2}x_0}{\sqrt{2\pi\JJ}}\frac{e^{2\bb\tsurv}}{(e^{2\bb\tsurv}-1)^{3/2}}\exp\left[-\frac{\bb x_0^2}{2\JJ(e^{2\bb\tsurv}-1)}\right].
\eeq
We can then integrate over the distribution
\beq
p_{x_0}(x_0) = \frac{2\Theta(x_0)}{\sqrt{2\pi}\sigzeta}e^{-x_0^2/2\sigzeta^2}, \text{ where } \sigzeta = \sqrt{\JJ/\bb}.
\eeq
We obtain $p_\tsurv(\tsurv) = 2\bb/\pi\sqrt{e^{2\bb\tsurv}-1}$, as quoted in Equation~\ref{eq:lifetime_dist}.

{\it Abundance distribution:}
The steady state distribution of abundances can also be computed from our self consistent solution.
The abundance distribution obeys a Fokker Planck equation (with the appropriate boundary conditions) that captures the dynamics of the drives from Equation~\ref{eq:drive_sde} and includes a {\it source} term for incoming invaders. Therefore in the steady state, the condition for stationarity of the distribution $P(\nu)$ is
\beq \frac{\bb \sigzeta^2}{Q^2} P''(\nu) + \bb[P(\nu) + \nu P'(\nu)] + \frac{rQ}{\sqrt{2\pi} \sigzeta} \exp\left(-\frac{\nu^2Q^2}{2\sigzeta^2}\right) = 0\label{eq:FP_drive}\eeq
with boundary conditions $P(0) = P(\nu\to\infty) = 0$.
Equation~\ref{eq:FP_drive} contains effects of the systematic and random parts in $\delta\nu$ as well as the input from invading strains.

If we change variables to $z = Q^2\nu$ and $\dens(z) = \frac{1}{Q^4}P(\nu)$ so that $\overline\LL = Q^2 \int dz \dens(z)$, we find that the number density $\dens(z)$ of the rescaled abundances $z$ obeys the $Q$-independent differential equation
\beq 2\pi \lambda^3 \dens''(z) + \lambda [\dens(z) + z\dens'(z)]+\frac{1}{2\pi\lambda} \exp\left(-\frac{z^2}{4\pi \lambda^2}\right)=0 \label{eq:FP_rescaled} \eeq
with $\lambda = \frac{1+2\log2}{4}$.
The solution to this equation with boundary conditions $\dens(0)=0$, $\dens(\infty)=0$ is plotted in Figure~\ref{fig:abund_dist} as the theoretical prediction for the distribution of scaled abundances in the evolved community.

\section{Fluctuations in steady state: $\Sigma=0$}
\label{app:L_fluct}

\begin{figure}
\centering
\includegraphics[scale=.6]{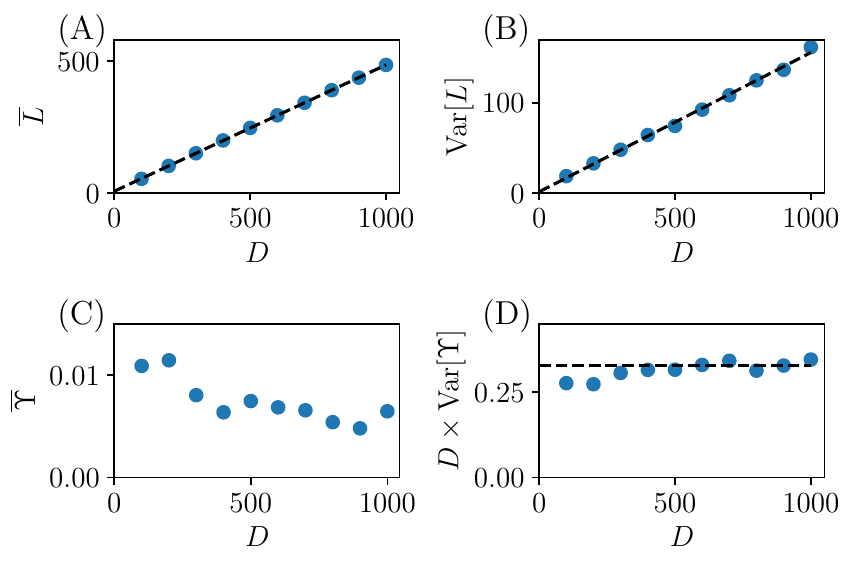}
\caption{Fluctuations of $\LL$ and $\Upsilon$ in the Red Queen phase of the linearized model with $\kappa=0.8$, $\rho=0$ and $\Sigma=0$. (A)--(B) Both the mean and variance of $\LL$ scale roughly linearly with $\DD$, so that fluctuations are smaller than the mean by a factor of $\sqrt\DD$. (C)--(D) The mean $\Upsilon$ in steady state, expected to be $\Oh(1)$, is rather small: unclear whether it goes to zero as $\DD\to\infty$. The variance of $\Upsilon$ scales, as expected, with $1/\DD$.} 
\label{appfig:L_fluct}
\end{figure}

Figures~\ref{appfig:L_fluct}A and B show that in the Red Queen steady state for the linearized model, $\LL$ fluctuates around a mean value $\overline\LL$ which is proportional to $\DD$, with fluctuations that scale with $\sqrt\DD$. This scaling is consistent with expectations from the finite fragility of the Red Queen steady state---with evolutionary perturbations to the community not causing large changes and fluctuations therefore down by a factor of $1/\sqrt{\DD}$. Figures~\ref{appfig:L_fluct}C and D show the same statistics for $\Upsilon$. We expect $\overline\Upsilon$ to be $\Oh(1)$ but is anomalously small in simulations (since $\Sigma=0$), and it is possible that the coefficient of its $\Oh(1)$ part is zero, with a $1/\sqrt\DD$---or some other---correction showing up in numerics. For $\Sigma=0$, we know that $\overline\Upsilon=0$ in the limit $\kappa\to0$, $\DD\to\infty$, $\kappa^2\DD=Q^2$ (which reduces to the $\gamma=0$ Lotka-Volterra model), as well as when $\kappa\to1$, since in the latter case $\Upsilon$ becomes nonpositive and asymptotes to $0$ from below. It is therefore possible that $\overline\Upsilon=0$ for intermediate $\kappa$ as well, but our results are inconclusive. Note that fluctuations in $\Upsilon$ have size $\Oh(1/\sqrt\DD)$ as expected.

\section{Evolutionary transients with $\Sigma>0$}
\label{app:evo_transient}

\begin{figure}
\centering
\includegraphics[scale=.6]{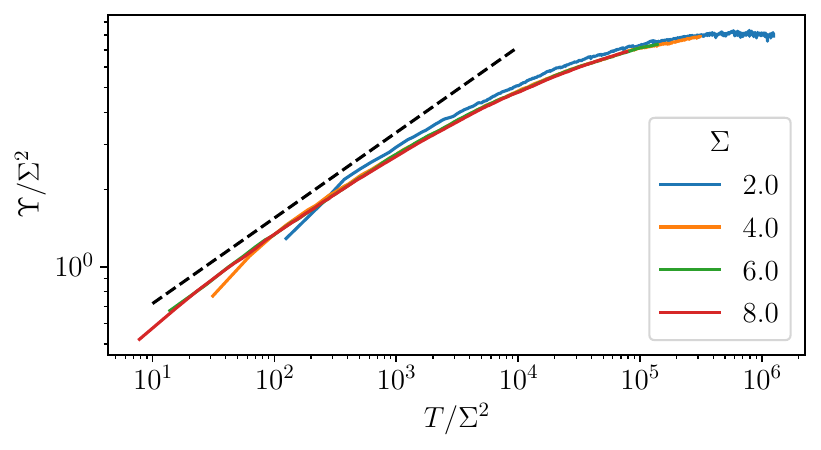}
\caption{Transient behavior of $\Upsilon$ during the evolutionary dynamics with large $\Sigma>\Sigma_\times\approx1.9$ in the linearized model. Parameters are $\DD=200$, $\kappa=0.8$ and $\rho=0$. Dashed black line has the predicted slope of $1/3$. Each simulation was run for $5\times10^6$ epochs.} 
\label{appfig:evo_transient}
\end{figure}

In Section~\ref{sec:s_dist} we discussed the transient behavior of the evolution as the distribution of extant $\ss_i$ pushes into the tail of $\sdist(\ss)$. For a gaussian distribution of the $\ss_i$, we expect that during the transient,
\beq \frac{\Upsilon}{\Sigma^2} \sim \left(\frac{\TT}{\Sigma^2}\right)^{1/3}.\eeq
In Figure~\ref{appfig:evo_transient} we find that this prediction agrees well with numerics for $\Upsilon$ as a function of evolutionary time, and the collapse of the curves scaled by $\Sigma$ verifies the predictions. The curves collapse even for $\Sigma>\Sigma_\times$, beyond which the cavity analysis would naively predict that the mean strain lifetime diverges. The data indicate that, in agreement with observation in Figure~\ref{fig:LR_Ups}C, for our choice of parameters, the simulations with $\Sigma\geq4$ have not yet reached their steady state by the end of $5\times10^6$ epochs. The dynamics of $\LL$ during the transient (not shown) are less well-fit by a simple power law since $\LL$ varies over a smaller range and has a larger constant correction.

\section{Correlations and lifetimes for nonzero $\gamma$ and $\Sigma$}
\label{app:corr_func}
\begin{figure}
\centering
\includegraphics[scale=.6]{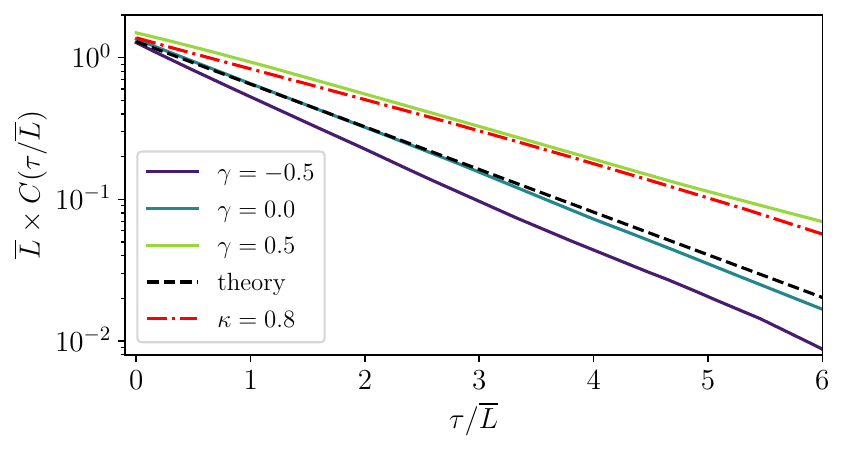}
\caption{In the LV model with $Q=20$ and $\Sigma=0$, $\CF(\tau/\overline\LL)$ decays more slowly with increasing $\gamma$ (normalization of $\tau$ by $\overline\LL$ eliminates the trivial scaling of the decay timescale with $\overline\LL$). Dashed black line is the simple-exponential prediction for $\gamma=0$, with deviations from this result due to finite-size effects. Red dash-dotted line shows $\CF(\tau)$ for linearized model with $\DD=400$, $\kappa=0.8$.} 
\label{appfig:corr_func}
\end{figure}

\begin{figure}
\centering
\includegraphics[scale=.6]{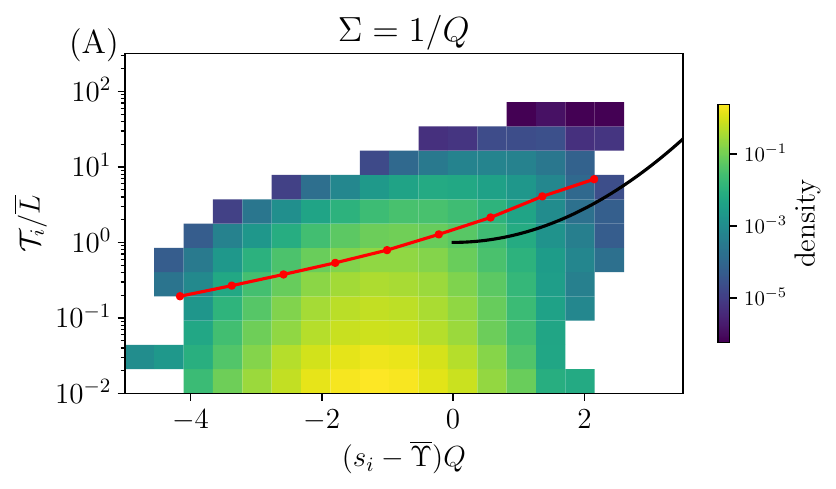}
\includegraphics[scale=.6]{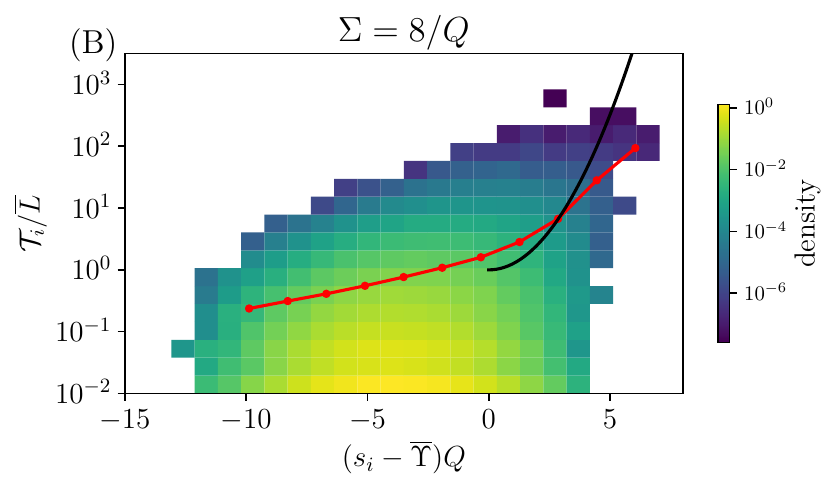}
\caption{Joint distribution of strain lifetimes $\tsurv_i$ and general fitnesses $\ss_i$ in $\gamma=0$ LV model ($Q=10$, yielding $\overline\LL\approx70$). (A) Data for $\Sigma Q = 1$. Red line shows the mean lifetime $\langle \tsurv|\ss\rangle$ conditioned on $\ss$ within a given range. Black line shows naive theoretical expectation $\langle\tsurv|\ss\rangle\sim \bar{\LL}\exp[(\ss-\overline\Upsilon)^2/2\sigma_\zeta^2]$ (up to an unknown coefficient). Data shows $\langle\tsurv|\ss\rangle$ increasing more slowly than this expectation. (B) Same plot for $Q\Sigma=8$, also showing empirical lifetime shorter than expectation from gaussian steady state {\it Ansatz}.}

\label{appfig:lv_s_T}
\end{figure}

\begin{figure}
\centering
\includegraphics[scale=.6]{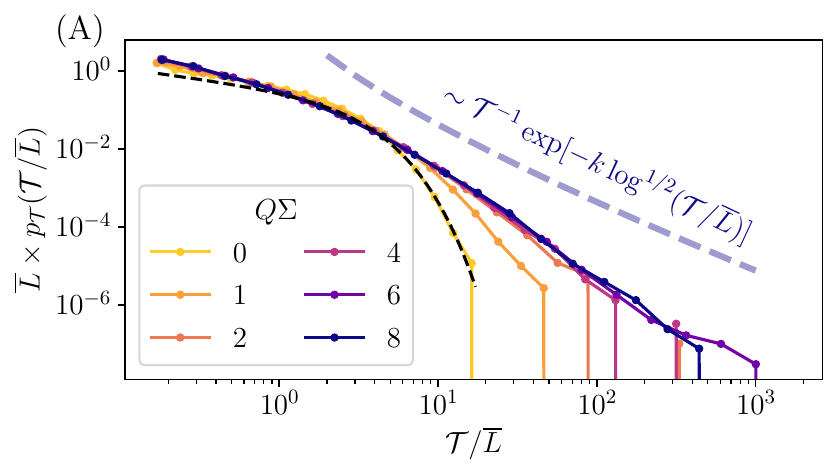}
\includegraphics[scale=.6]{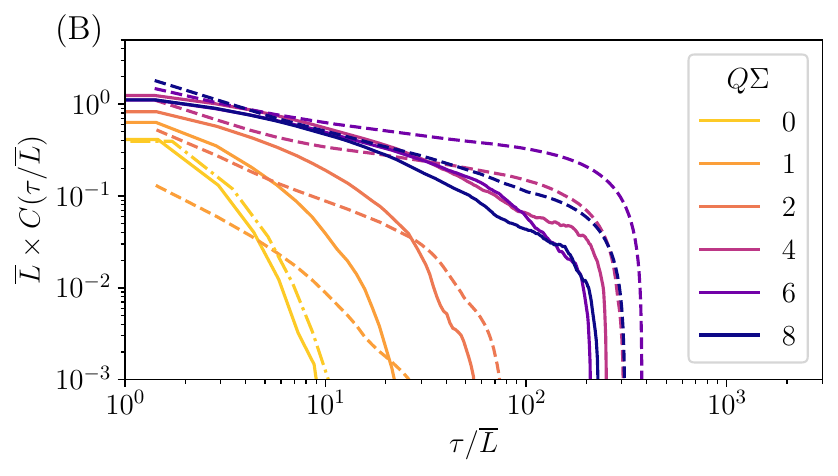}
\caption{(A) Distribution of lifetimes $\tsurv_i$ for various values of $\Sigma$ in $\gamma=0$ LV model ($Q=10$, yielding $\overline\LL\approx70$). $p_\tsurv(\tsurv)$ appears to decay sufficiently rapidly to yield finite $\langle\tsurv\rangle$ even for large $\Sigma$. Thin dashed line shows theoretical prediction for $\Sigma=0$ (Equation~\ref{eq:lifetime_dist}), which decays exponentially for large $\tsurv$. Thick dashed line shows poor agreement with naive prediction (for $\Sigma=8/Q$) for transient in $p_\tsurv(\tsurv)$ of form  shown with $k = \sqrt{2}\sigma_\zeta/\overline\Delta = \sqrt 2 \sigma_\zeta \overline\Upsilon/\Sigma^2\cong3.6$, and unknown multiplicative coefficient. 
(B) Correlation function $\CF(\tau)$ for various $\Sigma$. Yellow dash-dotted line shows theoretical result for $\Sigma=0$ (Equation~\ref{eq:corr_func}), with deviations from this prediction due to finite $Q$. Dashed lines show predictions of $\CF(\tau)$ computed from $p_\tsurv(\tsurv)$ using Equation~\ref{eq:Ctau_pT}, with a similar number of strains averaged in $p_\tsurv(\tsurv)$ as in $\CF(\tau)$. Note that time ranges in (A) and (B) are not aligned between panels.
} 
\label{appfig:lv_s_corr_func}
\end{figure}

\begin{figure}
\centering
\includegraphics[scale=.6]{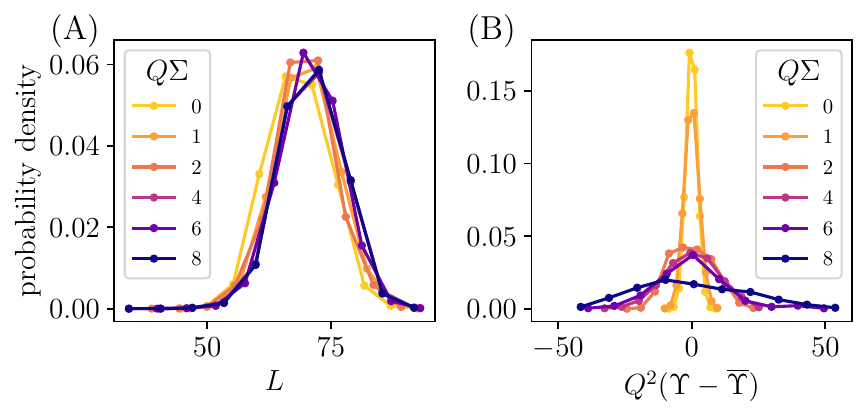}
\includegraphics[scale=.6]{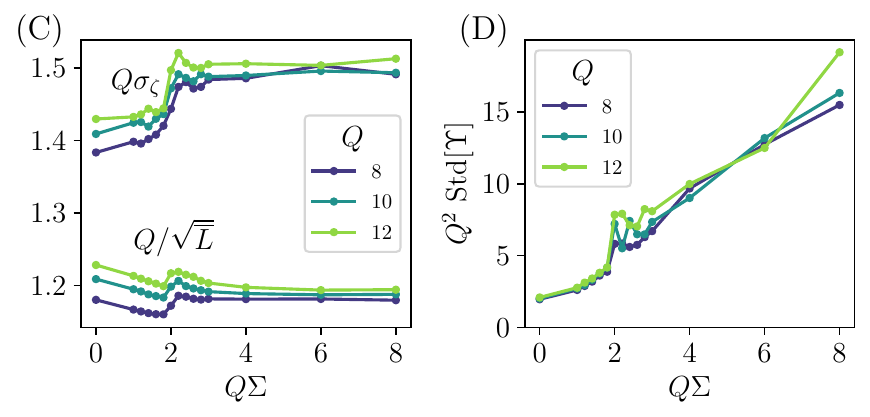}
\caption{Fluctuations in $\gamma=0$.LV model as function of $\Sigma$.   (A) Distributions of $\LL$ showing that  $\overline\LL$ and the fluctuations of $\LL$ depend only weakly on $\Sigma$. (B) Distributions of fluctuations in $\Upsilon$ (whose mean scales as $Q\Sigma^2$) increase with $\Sigma$. The scale of these fluctuations is $Q^2$ at $\Sigma=0$. $Q=10$ in panels (A) and (B). (C) Scaled $\sigma_\zeta$ and scaled $1/\sqrt{\overline{\LL}}$ as function of scaled $\Sigma$ (different sets of curves are labeled with colors indicating $Q$). The naive critical $\Sigma_\times$, defined by $\Sigma_\times = \sigzeta(\Sigma_\times)$, is roughly $\Sigma_\times \approx1.4/Q$ near which $1/\sqrt{\overline\LL}$ and $\sigma_\zeta$ both exhibit a small increase. (D) $\Upsilon$ fluctuations increase dramatically as $\Sigma$ passes through $\Sigma_\times$. For smaller $\Sigma$, ${\rm Std}(\Upsilon)$ scales as $1/Q^2$, while for larger $\Sigma$ it appears to decrease more slowly with $Q$.} 
\label{appfig:lv_ups_fluct}
\end{figure}


Here we examine the lifetime distribution, $p_\tsurv(\tsurv)$ and the correlation function $\CF(\tau) = \sum_{j} \nu_j(\TT)\nu_j(\TT+\tau)$ in the LV model for various $\gamma$ with $\Sigma=0$, and for gaussian $\sdist(\ss)$ with various $\Sigma$, restricting to $\gamma=0$. 
We find that in all cases---even when the cavity analysis suggests otherwise (Section~\ref{sec:s_dist})---a steady state appears to obtain in numerics, and the lifetime distribution decays sufficient rapidly that the relation $\langle\tsurv\rangle=\overline\LL$ between the average strain lifetime and the average diversity holds. 

{\it Nonzero $\gamma$ with $\Sigma=0$:} 
 In Figure~\ref{appfig:corr_func} we show $\CF(\tau)$ for the $\Sigma=0$ LV model, with time rescaled by the average strain lifetime, equal to $\overline\LL$. Defining rescaled time $\tilde\tau = \tau/\overline\LL$, we see that $\CF(\tilde\tau)$ decays exponentially for large $\tilde\tau$, with the rate of this exponential depending on $\gamma$. Larger $\gamma$ results in slower turnover and slower decay, presumably because with positive $\gamma$, invading strains are likely to increase the abundance of already-abundant strains, which therefore decay to extinction more slowly. Before the exponential regime it appears that for $\gamma<0$, $\CF(\tilde\tau)$ decays slightly faster for small $\tilde\tau$ than for large $\tilde\tau$, with the opposite trend for $\gamma>0$.

{\it Nonzero $\Sigma$ with $\gamma=0$:}
In the LV model, due to the interactions being $\Oh(1)$ (instead of $\Oh(\sqrt\DD)$ in the linearized model), the natural scale for the width of the extant $\ss_i$ distribution at steady state is $\Oh(1/Q)=\Oh(1/\sqrt{\LL})$, rather than $\Oh(1)$. We therefore consider $\Sigma = \Oh(1/Q)$ in the following analysis.

For $\Sigma>0$, subtleties arise for gaussian $\sdist(\ss)$ as discussed in Section~\ref{sec:s_dist}. For $\Sigma$ less than some crossover scale $\Sigma_\times$, $p_\tsurv(\tsurv)$ is predicted to decay as a $\Sigma$-dependent {\it power law}, while for $\Sigma>\Sigma_\times$ (or for broader-than-gaussian $\sdist(\ss)$) one naively expects that large-$\ss_i$ strains cause $\langle\tsurv\rangle$ to diverge, disallowing a steady state. However in contrast to this expectation, we find in numerics that a seemingly well-behaved steady state occurs even for large $\Sigma$. 

In order to understand this behavior it is useful to examine deviations from the cavity analysis exhibited by our numerics. As discussed in Section~\ref{sec:s_dist}, extinction for large $\ss-\overline\Upsilon$ requires a rare fluctuation of $\zeta_i$ away from zero (its mean) by an amount $\ss_i-\overline\Upsilon$. If the stochastic process $\zeta(\TT)$ has a {\it gaussian} steady state even out into the rare tails of large negative $\zeta$, the rate of extinction (and therefore inverse lifetime) will be roughly $\sim\exp[-(\ss_i-\overline\Upsilon)^2/{2\CF(0)}]$, where $\CF(0) = \sigma_\zeta^2$ is the variance of the stationary state distribution of $\zeta(\TT)$.
Figure~\ref{appfig:lv_s_T} shows the joint distribution of $\tsurv_i$ and $\ss_i-\overline\Upsilon$, and compares the mean lifetime for a given $\ss$, denoted $\langle\tsurv|\ss\rangle$, to the predicted inverse-gaussian-large dependence. The numerics show a slower increase of $\langle \tsurv|\ss\rangle$ with $\ss$ than suggested by a gaussian steady state, indicating wider fluctuations of $\zeta(\TT)$ and shorter lifetimes.

These shorter-than-expected lifetimes can also be seen directly from $p_\tsurv(\tsurv)$. Under the gaussian cavity assumption, strains which last for time $\tsurv$ will have typical general fitness $\ss\approx \overline\Upsilon + \sigzeta \sqrt{2}\log^{1/2}(\tsurv/ \overline{\LL})$. If $\sdist(\ss)\sim e^{-(\ss/\Sigma)^\psi/\psi}$, this yields an estimate for the large-$\tsurv$ decay of $p_\tsurv(\tsurv)$: for $\psi>2$, we have $p_\tsurv(\tsurv) \sim \tsurv^{-1}\exp[-k\log^{\psi/2}\tsurv]$, while for gaussian $p(\ss)$ we have, at asymptotically long times, 
 \beq
 p_\tsurv(\tsurv) \sim \tsurv^{-1-\beta} \quad \text{with}\quad \beta=\sigzeta^2/\Sigma^2, \label{eq:p_T_decay}
\eeq
which yields a finite $\langle \tsurv\rangle$ only for $\beta>1$, i.e. $\Sigma< \Sigma_\times=\sigzeta(\Sigma_\times)\cong 1.4/Q$.
At intermediate times, especially for larger $\Sigma$, $p_\tsurv(\tsurv)$ is expected to decay as $\sim \tsurv^{-1}\exp[-k\log^{1/2}(\tsurv/ \overline{\LL})]$ with $k=\sqrt{2}\sigzeta/ \overline{\Delta}$, due to the roughly exponential decay of $p(\ss)$ (with slope $1/ \overline{\Delta}$) near $\overline\Upsilon$.

In Figure~\ref{appfig:lv_s_corr_func}A $p_\tsurv(\tsurv)$ is plotted on a log-log scale for gaussian $p(\ss)$ with various $\Sigma$. As $\Sigma$ goes from $0$ to $8/Q$, $\sigzeta$ increases modestly from $1.4/Q$ to $1.5/Q$, and $\overline\LL$ barely changes (Figures~\ref{appfig:lv_ups_fluct}A and C). Even for $\Sigma=1/Q$, which is less than $\Sigma_\times$, the decay of $p_\tsurv(\tsurv)$ is considerably faster than the predicted $\beta+1\cong 3$. For $\Sigma\gg \sigzeta$ the difference between numerics and cavity analysis is even larger: $\beta$ appears to saturate at $\cong 1.7$ in contrast to the predicted $\beta\cong 0.06$ for $\Sigma=8/Q$, and $p_\tsurv(\tsurv)$ is not well fit by the intermediate-$\tsurv$ transient either (Figure~\ref{appfig:lv_s_corr_func}A). Thus something is making the rare fluctuations to large negative $\zeta$---assuming the basic cavity structure in terms of $\zeta$ is still valid---much larger than gaussian.

Given the difference between naive expectations and numerics for $p_\tsurv(\tsurv)$, we investigate whether these differences carry over to the correlation function, $\CF(\tau)$, which, as discussed in Section~\ref{sec:s_dist}, is determined by strains that survive for at least time $\tau$. $\CF(\tau)$ is plotted in Figure~\ref{appfig:lv_s_corr_func}B for gaussian $p(\ss)$ with the same range of $\Sigma$ between $0$ and $8/Q$. With the joint distribution of $\ss$ and $\tsurv$, and approximating the contribution to $\CF(\tau) $ of a strain with $\tsurv>\tau$ by $ (\tsurv-\tau)\nu^2$, $\CF(\tau)$ can be estimated from the {\it observed} $p_\tsurv(\tsurv)$ (Equation~\ref{eq:Ctau_pT}). The result of this estimation is shown in Figure~\ref{appfig:lv_s_corr_func}B along with the empirically calculated $\CF(\tau)$, and we find rough agreement.

As discussed in Section~\ref{sec:s_dist}, a possible source of the discrepancy between naive cavity analysis and numerics is the emergence of long-tailed memory kernels that increase the rate of extinction by allowing anomalously long-lasting negative $\zeta_i$. Such dynamics could drive strains with large $\ss_i-\Upsilon$ extinct at a higher rate than the naive ``barrier-crossing" estimate.  But as long as the $\zeta(\TT)$ is a stationary gaussian process, this cannot happen: the correlations between $\zeta(\TT)$ and the typically-small $\zeta$ at the invasion time can only decrease the probability of a large deviation. 
Instead, a more viable scenario is that the large-$\overline\LL$ limit is subtle with large deviations from the gaussian {\it Ansatz} of the $\zeta(\TT)$ statistics making extinction events of large $\ss-\overline\Upsilon$ strains much less unlikely, as seen in Figure~\ref{appfig:lv_s_T}.
 
 In general, one does not expect that the limits of large $\overline\LL$ and behavior in the tails of distributions will commute: the central limit theorem is only for the non-too-rare center of the distribution. Likewise, the long-time limit and large $\overline\LL$ limit need not commute. Concretely, if large fluctuations that drive extinctions are only exponentially rare in $\zeta$, and hence lifetimes ``only" exponentially long in $\ss-\overline\Upsilon$, $p_\tsurv(\tsurv)$ would decay roughly as a power of $\tsurv$ even for very large $\Sigma$ where the distribution of extant $\ss$ is close to exponential. This behavior is roughly consistent with the apparent power-law decay in this regime (Figure~\ref{appfig:lv_s_corr_func}A). Exponential---rather than gaussian---tails often occur, for example if the rare events that drive extinctions result from the sum of effects of a series of roughly independent events. Thus this putative resolution of the apparent paradox may well be viable. 
 
Another effect that could cause extinctions is large fluctuations in $\Upsilon$, particularly for exponential $\sdist(\ss)$, which plays a key role in the behavior for large $\Sigma$. With exponential $\sdist(\ss)$ (in contrast to $\psi>1$) there is no special value of $\Upsilon$ in steady state. 
Instead there is a special value of $\Sigma$ for which a steady state occurs, equal to the steady-state $ \overline{\Delta}$ for all $\psi>1$ in the limit of large $ \Sigma$ (Section~\ref{sec:s_dist}). Since any $\Upsilon$ in an exponential tail is equivalent (conditioned on invasion of strains) there are likely to be at least diffusive---and hence unbounded---fluctuations in $\Upsilon$. Upwards fluctuations in $\Upsilon$ would eventually drive extinctions of strains that had anomalously large $\ss-\Upsilon$ at their invasion time. Additionally, persistent large-$\ss$ strains will themselves cause substantial increases in $\Upsilon$, perhaps providing sufficient feedback for these to limit their own lifetimes. 

 The fluctuations in $\Upsilon$ about the steady state for gaussian $p(\ss)$ can be estimated by noting that in the LV model, an invader changes $\Upsilon$ by an amount of order $\LL^{-3/2}$, the typical amount needed to keep $\sum_i\nu_i=1$ for the random changes in $\nu_i$ caused by the invasion (Section~\ref{sec:fragility}). 
Combining these diffusive fluctuations with the $\Oh(1/\LL^2\Sigma^2)$ linear deterministic push towards $ \overline{\Upsilon}$ (which can be seen by expanding Equation~\ref{eq:cond_increment} for $\Upsilon$ near $\overline\Upsilon$) yields a prediction: $\Var(\Upsilon)\sim \Sigma^2/Q^2$. Simulations for modest $Q=10$ in Figure~\ref{appfig:lv_ups_fluct} show that while the distribution of $\LL$ in steady state depends only weakly on $\Sigma$, the distribution of $\Upsilon$ in steady state appears consistent with this dependence on $\Sigma$.
We expect that the diffusion coefficient of $\Upsilon$ would increase---and perhaps diverge---for large $\Sigma$, due to long-time correlations of the fluctuations in $\zeta$ and hence in $\Upsilon$. In order to play a role in the extinctions of strains with large $\ss-\overline\Upsilon$, fluctuations of $\Upsilon$ would have to be larger than expected by a factor of $\sqrt{\LL}\sim Q$, and would likely need to have non-gaussian tails as well. While we have not explored the tails, Figures~\ref{appfig:lv_ups_fluct}B and D indeed show increased fluctuations in $\Upsilon$ as a function of increasing $\Sigma$, with a sharp increase as $\Sigma$ passes through $\Sigma_\times$ at which the naive analysis predicts a divergence of $\langle\tsurv\rangle$. For larger $\Sigma$, the dependence of $Q^2 \Std[\Upsilon]$ on $Q$ becomes stronger, perhaps indicating a crossover to a regime with the fluctuations scaling as $1/Q$ and thus comparable to $\sigma_\zeta$.
 
 At this point it is unclear what kind of non-gaussian fluctuations, or other subtleties of large but finite $\LL$, cause the persistence of a Red Queen steady state even when large $\ss$'s can occur. However our numerics indicate that the eco-evolutionary steady state obtains even for large $\Sigma$ with gaussian $\sdist(\ss)$. Further investigation of $\psi<2$ is needed, but it is possible that a similar mechanism preserves the steady state there as well.

\section{Comparison with results of de Pirey and Bunin}
\label{app:dPB}

Recent work of de Pirey and Bunin~\cite{arnoulx2024many} (dPB for short) has explored a Red Queen-like phase in a different context. dPB study {\it ecological dynamics} in a $\gamma=0$ Lotka-Volterra model in the limit of small migration for each species, such that there is a clear separation between extant species which have substantial abundance, and species which would be extinct if not for the small migration. 
They find that if the diversity is high enough, the ecological dynamics are chaotic such that there are species which go nearly extinct before rising again to high abundance and rejoining the extant subset which is almost at a stable fixed point. Due to a separation of timescales arising from the very small migration rate (analogous to our separation of evolutionary and ecological timescales), the ecological dynamics consist of a series of almost-fixed points punctuated by changes that occur when a near-extinct species rises up to non-negligible abundance. 

When a species transitions from near-extinct to extant, the process is similar to an independent invader joining the community, as the abundance of the invader can jump up to a nonzero value determined by its ecological interactions with the extant community. 
This entrance to the community causes a change in the biases of the other strains, which can drive low abundance strains out and produces a steady state in which low-abundance strains are depleted relative to the truncated gaussian distribution of the assembled community. Indeed dPB find that their distribution at steady state is not well approximated by a truncated gaussian, and the number of extant species at any given time is less than the total capacity of an assembled community.

However, in the model studied by dPB, the steady state abundance distribution does not vanish at zero abundance, in contrast to our evolved communities. This is because when a species goes from extant to nearly-extinct, its probability of rejoining the extant community due to an ecological fluctuation is a decreasing function of time as ecological fluctuations drive its abundance farther from the boundary between extant and near-extinct. Therefore the boundary condition at zero abundance is not an absorbing one but rather a permeable boundary where species that hit zero abundance can be resurrected by a small change in the community and come back into the community soon after their extinction. dPB are able to write dynamical mean field equations for the abundance dynamics and solve these equations numerically to get the steady state abundance distribution. Our case turns out to be simpler in this respect since with evolutionary dynamics the extinction boundary is strictly absorbing which, fortuitously, makes the self-consistent correlations simple and enables an exact solution,  with the equation satisfied by the abundance distribution reduced to a boundary value problem (Equation~\ref{eq:FP_rescaled}) that can be solved numerically (Figure~\ref{fig:abund_dist}).

\bibliography{comm_evo_draft}

\end{document}